\documentclass[AMA,STIX1COL]{WileyNJD-v2}
\usepackage{amsmath, amssymb, fontenc,  physics, mathtools, dsfont,bbm,dsfont, cases, float}
\usepackage{moreverb,url, multicol, multirow, soul,  lscape, longtable,booktabs,threeparttable, makecell, bm,sublabel}
\usepackage{caption}
\usepackage{subcaption}
\newcommand{\bigCI}{\mathrel{\text{\scalebox{1}{$\perp\mkern-10mu\perp$}}}}
\DeclareMathOperator{\logit}{logit} 

\usepackage{color, soul, threeparttable, multirow, multicol, setspace}
\usepackage{amsmath, amssymb, amsfonts, tikz, subcaption, booktabs,
	colortbl, dcolumn, wrapfig, bbding}
\articletype{Article Type}%

\received{}
\revised{}
\accepted{}

\begin{document}
	\title{Causal inference in absence of positivity: the role of overlap weights}
	\author[1,2]{Roland A. Matsouaka}
	\author[1,3]{Yunji Zhou}
	
	\authormark{Matsouaka 
		and Zhou}

	\address[1]{\orgdiv{Department of Biostatistics and Bioinformatics}, \orgname{Duke University}, \orgaddress{\state{Durham}, \country{North Carolina, USA}}}
	\address[2]{\orgdiv{Program for Comparative Effectiveness Methodology}, \orgname{Duke Clinical Research Institute}, \orgaddress{\state{Durham}, \country{North Carolina, USA}}}
	\address[3]{\orgdiv{Duke Global Health Institute}, \orgname{Duke University}, \orgaddress{\state{Durham}, \country{North Carolina, USA}}}
	\corres{Roland A. Matsouaka\\ Duke Clinical Research Institute, \\300 W. Morgan Street, Durham, NC 27701 , USA\\\email{roland.matsouaka@duke.edu}}

\everymath{\displaystyle} 

\abstract[Summary]{
		What do we do when violations of the positivity assumption are expected? Several possible solutions exist in the literature. We  consider propensity score (PS) methods that are commonly used in observational studies to assess causal treatment effects when the positivity assumption is violated. We focus on  and examine four specific alternative solutions to the inverse probability weighting (IPW) trimming and truncation: matching weight (MW), Shannon's entropy weight (EW), overlap weight (OW) and beta weight (BW) estimators.
		
		In this paper, we demonstrate that these estimators target the same population of interest, the population of patients for whom we have sufficient PS overlap, i.e., where lies clinical equipoise. Then, we establish the nexus among the different corresponding weights (and estimators); this allows us to highlight the share properties and theoretical implications of these estimators. Finally, we introduce their augmented estimators that take advantage of estimating both the propensity score and regression models to enhance the treatment effect estimators in terms of bias and efficiency.  We also elucidate the role of OW estimator as the flagship of all these methods that target the overlap population.
		
		Our analytic results demonstrate that OW, MW, and EW are preferable when there is a moderate or extreme violations of the positivity assumption. We then evaluate,  compare, and confirm their finite-sample performance of the aforementioned estimators via Monte Carlo simulations. Finally, we illustrate these methods using two real-world data examples marked by expected violations of the positivity assumption. 
	}

	\maketitle
	\section{Introduction}
	A well-designed and well-executed randomized clinical trial (RCT)---with a perfect randomization, large sample size, full adherence, no attrition and no measurement error---is gold standard for causal inference since it leads to comparable treatment groups with respect to both measured and unmeasured covariates. Randomization ensures the comparability between the treatment groups where participants are enrolled based on clearly defined and objective eligibility criteria, which are embedded in the study protocol. Any difference in patients' baseline covariates between treatment groups is random  and their distributions coincide on average, which affords the study its internal validity.\cite{fonarow2016randomization} This allows a straightforward assessment of the treatment effect: association is causation. More often physicians and investigators  in such a trial operate, at least in principle, under the ideal of clinical equipoise, i.e., they would be a priori genuinely  uncertain (from a historical or general practice perspective) about the therapeutic merits and the effectiveness of each treatment option.\cite{freedman2017equipoise}
	As a result, randomization ensures positivity, i.e., each study participant has a chance of being assigned to either treatment option.  
	
	Unfortunately, it is not always feasible to conduct a randomized clinical trial for practical reasons, ethical concerns, or financial constraints. Furthermore, when feasible, strict eligibility criteria of an RCT do not always allow investigators to adequately enroll patients that are representative of the whole target population to which the treatment of interest, based on its intended purposes, will eventually be applied. For instance, it is well documented that older patients, higher risk patients, women, and patients from racial/ethnic minority groups are often underrepresented  in RCTs.\cite{fonarow2016randomization} Finally, despite their high quality, data from an RCT may not reflect those from a real-world evidence practice since the available hospital resources,  logistics, and personnel as well as the treatment effectiveness may be drastically different in tightly-controlled confines of an RCT  than in  regular day-to-day clinical practice. Therefore,  for a number of questions of interest arising in biomedical research,  we often resort to  observational studies for answers. 
	
	Unlike in RCTs, the treatment assignment is not random and usually beyond the investigators' control in observational studies. Treatment options are either self-selected by patients (via some health-seeking behaviors) or recommended by their physicians following specific diagnosis or outcome prognosis. In most clinical practice, the treatment is assigned based on a range of deciding factors, including patient's demographics, diagnosis, lab values,  prognosis, provider's preference, practice patterns, and even hospital characteristics (or capabilities). Hence, observational studies can be potentially marred by confounding, which is susceptible to jointly influence the treatment a patient can received and his or her final outcome. In addition, there is a number of potential study-altering factors (e.g., endogeneity, selection bias, collider bias, compliance) we need to sleuth on if we want to conduct a rigorous and reliable study.\cite{delgado2004bias} Nevertheless, principled design and analysis of observational studies are conducted with the objective of emulating a randomized clinical trial. \cite{dorn1959some,fanaroff2020randomized,rubin2006matched,hernan2011great,hernan2016using, hernan2016does,dahabreh2018randomization} The goal of causal inference methods is thus to make the treatment groups comparable with respect to the distributions of their baseline covariates. Therefore, estimating the effects of treatment  requires that we rule out any confounding due to systematic difference in the distributions of the baseline covariates that may explain the difference in outcome between treatment groups. 
	
	For many observational studies,  propensity score (PS) methods are used to balance the distributions of the measured covariates, reduce or eliminate selection bias, emulate an RCT-like design, and evaluate the treatment effect. The propensity score is the conditional probability of treatment assignment, given the baseline covariates. It contains all measured information relevant for the treatment--outcome relationship. Furthermore, the PS has the balancing property, i.e., condition on the propensity score, the underlying distributions of covariates of the treatment groups are similar: the PS restores balance between these groups.  \cite{rosenbaum1983central,goetghebeur2020formulating}  At the heart of all propensity score methods are two fundamental assumptions: unconfoundness and positivity. For the unconfoundness assumption to be deemed acceptable, we assume that an extensive set of relevant and important covariates that are related  to the treatment or the outcome of interest has been measured based on subject-matter expertise and can be used to remove or at least reduce confounding. Therefore, condition on the measured covariates, the study is considered as good as if it were a randomized clinical trial (with respect to the treatment assignment).
	
	Once the selection of covariates is done, an important task is thus to ensure that the positivity (or experimental treatment assignment) assumption is satisfied: the propensity score must be bounded away from 0 and 1, given the smallest subset of covariates that make the uncounfoundness assumption valid. When satisfied, the positivity assumption guarantees that there is an appropriate overlap in the measured covariate distributions across the treatment groups. It ensures comparability of treated and control participants, i.e.,  some measured covariates do not deterministically lead to a specific treatment group (but not the other). Moreover, it helps avoid extrapolation beyond the common overlap region of the covariate distributions where we can reliably estimate the treatment effect. Along with the unconfoundness assumption, the positivity assumption implies that participants from one treatment group can always be used as good proxies (to infer potential outcomes) for the other group.   This allows us to leverage the outcomes of treated (resp. untreated) participants to infer the potential	outcomes of untreated (resp. treated) participants had they been treated (resp. untreated).	Whenever the distributions of the covariates between the two treatment groups  are drastically different, this may result in limited overlap of the distribution the propensity scores and violations (or near violations) of the positivity assumption. We will refer to violations or near violations of the positivity assumptions as {\it lack of adequate positivity}, as defined by Cheng et al.\cite{ju2018adaptive} 
	\subsection{Lack of adequate positivity in observational studies}
	Different propensity score methods are affected differently by insufficient overlap in the distribution of the propensity scores or by the lack of adequate positivity and, hence, have different requirements. Note that in practice, overlap refers to how well the histograms of propensity scores in each treatment group cover (i.e., overlap with) each other. As opposed to propensity score matching and propensity stratification methods where a good overlap is a must to the success of their implementations,  inverse probability weighting (IPW) methods do not impose strict requirements for good overlap of the distributions of the propensity scores per se. The main idea behind weighting methods is to create a pseudo-population, i.e., the target population of interest where, on average, the treatment groups are balanced with respect to their baseline covariates.\cite{cole2008constructing,hernan2018causal}  They usually lead to adequate covariate balance, sometimes even when there is limited overlap of the distribution of the propensity scores (see, for instance, the Lalonde data set in Ridgeway et al. \cite{ridgeway2017toolkit}). Nevertheless, a scrupulous evaluation of the common support has a fundamental implication: it allows us to speculate whether causal inference is advisable, i.e. whether the treatment groups overlap sufficiently well enough to provide reliable and stable estimate of the treatment effect. \cite{schafer2008average,khan2010irregular,canavire2021outliers} 
	
	For inverse propensity score weighting (IPW) methods, lack of adequate positivity or insufficient overlap in the distribution of propensity scores may result in observations with propensity scores near 0 or 1 and lead to extreme weights. When that is the case, the results of statistical analyses may  be largely influenced by a few observations, which can then bias the treatment effect estimate and  inflate its estimated variance considerably. \cite{schafer2008average,khan2010irregular,canavire2021outliers,crump2006moving,crump2009dealing, fogarty2016discrete} To estimate the average treatment effect (ATE), for instance, the positivity assumption requires that each participant has a positive probability of being assigned to all available treatment options as the target population corresponds to the whole population of participants: every single participants must be  eligible (hypothetically) to received either treatment. The ATE targets the average treatment effect in the hypothetical scenario where everyone in the population is assigned to the treatment versus everyone is assigned to the control. Hence the need to assess whether the distribution of the propensity scores in both treatment and control groups overlap well enough to yield reliable and efficient estimate of the treatment effects. 
	Nevertheless, the positivity assumption can be relaxed if we want to estimate the average treatment effect on the treated (ATT) or on the control (ATC) participants. For ATT (resp. ATC), the positivity assumption means that the control (resp. treated) participants have propensity score below 1 (resp. strictly positive). As argued by Heckman and Vytlacil, sometimes, the ATT (resp. the ATC) may be preferred or more appropriate for policy-decision making than the ATE since the latter may include the effect of a  treatment on people from whom the treatment under study was never intended. \cite{heckman2001policy,imbens2004nonparametric}  
	
	Lack of adequate positivity can occur for several reasons including happenstance, the choice of covariates included in the propensity score model (which may lead to its misspecification), and design flaws or limitations.  A plot of the distributions of the propensity scores  in both treatment groups may reveal an instance of lack of adequate positivity. The literature distinguished two specific violations of the positivity assumption: random and structural (or deterministic) violations.\cite{cole2008constructing} Random violation of the positivity assumption occurs in studies where we objectively know from experts' subject-matter knowledge that every individual in the population of patients are eligible to the different treatment options, but the positivity assumption happened to not be satisfied in our specific data sample.  Because the true propensity scores are unknown in observational studies and must be estimated, one may have propensity score estimates that are equal to zero or are near zero by chance (especially in small and moderate sample sizes) or due to a  (mis)specification or an overfitting of the propensity score model. In practice overfitting leads to a random violation of the positivity assumption if, for example,  we decide to include in the propensity score model covariates that are instrument of of the treatment, i.e., strongly associated with the treatment but  completely independent from the outcome of interest or those that are mediators of the treatment--outcome relationship. \cite{einbeck2009design,kang2016practice} 
	
	While we can remedy either lack of adequate positivity by increasing the sample size or by re-appraising the propensity score model with the input from  substantive area expert knowledge, a third reason for the lack of positivity pertains to the design or the nature of study and requires our careful consideration. 
	Indeed,  such structural lack of adequate positivity may arise due to an inherent structural nature of our study of interest and we cannot improve or correct it by increasing the sample size alone, readjusting the propensity model, or knowing the true propensity scores (if it were possible). In this context, there will be regions of the distribution of the patients' covariates where corresponding propensity scores are equal to (or near) 0 or 1 even as the sample size increases. This lack of adequate positivity signals  the presence of individuals in our data set who, based on their baseline covariates, have deterministic treatment  choices. In standard practice, clinicians would have a clear indication as to which treatment is most suitable, appropriate, and beneficial for these patients. 
	
	Examples of such deterministic treatment choices are common in clinical practice. In disease areas,  e.g., in cancer,  a well-established treatment regimen may be counter-indicated for certain patients, based  on   a specific protocol. Thus, for these patients, clinicians will consistently and overwhelmingly assign treatments according to the protocol.  Therefore, these individuals would not be among the participants for whom there is a genuine clinical equipoise. In comparative effectiveness studies of newly marketed medications, for instance, it is common for physicians  to  channel some patients towards these new drugs for factors related to both their expected effectiveness and tolerability. This explains why the first patients to use a new statin agent are those dissatisfied with the existing agents (e.g., they have suffered side effects or  did not achieve sufficient low-density lipoprotein cholesterol control); those with atrial fibrillation who uptake a new direct thrombin inhibitor are probably those for whom warfarin treatment was not optimal; and finally patients with rheumatoid arthritis  trying a new immunomodulating drug are probably those who have experienced little or no benefit from existing drugs.\cite{schneeweiss2011assessing}
	
	Because  patients for whom positivity assumption is violated do not have comparable counterparts in the alternative treatment, the causal effects of treatment for these patients are not really identifiable. Extrapolating results on this subpopulation has little to no additional value; treatment effect estimates based on data that include these patients may be biased and thus questionable. \cite{schafer2008average,kang2016practice,busso2014new,goetghebeur2020formulating} This is the case, for instance, if one considers  a study on the effect of breastfeeding that includes in the target population women for whom breastfeeding is precluded (due to preexisting or pregnancy-related conditions): the effects of breastfeeding on the subpopulation of infants whose mothers cannot breastfeed is irrelevant.\cite{goetghebeur2020formulating}  If we put ourselves in the perspective of analyzing observational studies that emulate ideal randomized clinical trials, the presence of such participants annihilates the principle of equipoise (had we conducted an RCT), which results in a type of infeasible intervention.  The corresponding ATE would lack a meaningful or informative clinical interpretation since it is estimated on a population that does not reflect a population of genuine scientific interest. More generally, since the choice of the estimand is governed by the underlying scientific question of interest, choosing the ATE as estimand may not adequately answer such a question.

	Since the overlap region of the distributions of the propensity score characterizes the amount of common information carried by both treatment groups, the treatment effect estimated on the support of the covariate distributions intersection "would reflect the effect that would be attained in a typical clinical trial population" according to Vansteelandt and Daniel. \cite{vansteelandt2014regression} Thus, for internally valid inference in observational studies, it is important to have a substantial overlap of the covariate distributions between the treatment groups. 	
	In this context, with limited overlap or lack of adequate positivity, there is no point in estimating the average treatment effect (ATE)  for the whole population since such an estimate will be based purely on extrapolating its value(s) in the regions of the covariate distributions without overlap. \cite{frolich2004finite, busso2014new} To preserve internal validity  and obtain a meaningful assessment of the treatment effect, it is recommended to limit the estimation of the treatment effect on the region of sufficient overlap of the distribution of the propensity scores, while acknowledging that doing so alters both the population and the estimand(s) of interest.\cite{crump2006moving,crump2009dealing, schafer2008average,fogarty2016discrete,petersen2012diagnosing}  
	
	"To make causal inference in situations with nonoverlapping densities, we must therefore eliminate the region outside of common support \dots or attempt to extrapolate to the needed data \dots such as by using a parametric model \dots ", recommend King and Zeng. \cite{king2006dangers} Furthermore, generalization and transportability of the treatment effect cannot be made on participant outside of the  overlap region or for whom there is lack of adequate positivity.\cite{schneeweiss2011assessing}    Thus, the need to "move the goalpost" by focusing on the treatment effect in regions of common support.  \cite{crump2006moving,crump2009dealing} 
	
	\subsection{Mitigating the lack of adequate positivity: two schools of thought}
			 Several methods have been proposed in the literature to mitigate the lack of adequate positivity. Two different schools of thought exist in the causal inference literature on how to appropriately handle lack of adequate positivity, beyond increasing the sample size or judiciously selecting the variables to include the propensity score model. The first one evolves around the idea of "fixing" the propensity score estimation as to obtain propensity score estimates exempt of any lack of positivity or focus directly on estimating weights that lead to covariate balance. On a technical level, the ultimate goal of using propensity score methods is to achieve covariate balance between the treatment groups in finite sample since the propensity score is a balancing score that stochastically leads to comparable covariate distributions.\cite{rosenbaum1983central} Therefore, these methods impose {\it a priori} the covariate balancing conditions, which modify  the propensity score model (see Graham et al. \cite{graham2012inverse} and Imai and Ratkovic \cite{imai2014covariate}) or they directly derive sample weights 
			from the data under some covariate constraints using direct optimization techniques (see Hainmueller \cite{hainmueller2012entropy},  Zubizarreta \cite{zubizarreta2015stable}, Wong and Chang \cite{wong2017kernel}, or Hirshberg and Zubizarreta \cite{hirshberg2017two}).  Such empirical methods to estimate weights can substantially outperform parametric methods (e.g., logistic regression) in reducing bias and variance. However, when there is lack of adequate positivity, especially the lack due to underlying structural nature of the data, these methods can still lead to unstable estimates of treatment effect with larger variances. \cite{zhao2016covariate}  In several real-world situations, "fixing" the propensity score model or imposing some extraneous constraint to reach covariate balance will not solve the issue introduced by the lack of adequate positivity entirely. Such lack of adequate positivity is sometimes expected, which often necessitates a priori subject-matter or expert knowledge (and not just statistical prowess) to adequately incorporate such an information into the causal analysis, without the need to fix lack of adequate positivity by brute force.\cite{hernan2011great,hernan2019comment} As one might say "if it ain't broke, don't fix it" since there are many instances where  a deterministic allocation of treatment is expected for some patients, if not welcomed, as we illustrated in  Sections \ref{sec:alt_positivity}  and \ref{sec:eg_equipoise}. Thus,  "fixing" the propensity score estimation might not be the right solution. Instead, acknowledging the inherent limitations of the data, the study design,  (or the nature of the target population) that lead to lack of adequate positivity and finding ways around it constitute  the acceptable alternative. 
			

			The second school of thought aims at building methods that estimate the treatment effect on a subpopulation of interest, without having to "fix" the propensity score estimation  or imposing additional covariate constraints.  Some, such as the stabilized weights where the weights of treated (resp. controlled) patients are multiplied the proportion of treated (controlled) patients can  result in consistent estimates and help improve inference. \cite{robins2000marginal,hernan2000marginal} Others, such as truncation (downweighting extreme weights to fixed thresholds, e.g., capping extreme weights to the 5th and 95th percentiles) and trimming (discarding observations with weights outside of the 5th and 95th percentiles) are ad hoc methods since they are somewhat arbitrary (mostly guided by the analyst's insights, habits, and intuition), often strongly driven by the sample size and the data at hand.\cite{ma2020robust,zhou2020propensity} Although they may help improve efficiency by discarding observations with large weights, they can also introduce or increase biases since the choice of specific thresholds varies widely and remains arbitrary, \cite{cole2003effect, cole2008constructing,sturmer2006insights,kurth2005results,sturmer2010treatment,ju2018adaptive} which obviously leaves room for cherry-picking.  
		
		Crump et al.  formalized trimming by providing an optimal thresholds that minimizes the variance of the ATE estimate.\cite{crump2006moving,crump2009dealing} Using beta-distributed propensity scores,  Crump et al. showed that a good (optimal) threshold can be approximated by 0.1 and suggested as rule-of-thumb to discard observations whose propensity scores lie outside of the interval $[0.1, 0.9]$.  However, the rule-of-thumb can still result in inefficient estimates and is not always used consistently in practice. \cite{lee2011weight,li2018addressing, yang2018asymptotic,zhou2020propensity}
		Moreover,  trimming result in loss of sample size and this can be substantial, especially when the trimming threshold is large. \cite{yang2018asymptotic,zhou2020propensity} Finally, there is no clear guideline as to whether we should use the trimmed data set to derive de novo  the propensity score estimates. While some researchers such as Crump et al. \cite{crump2009dealing} recommend refitting the propensity model after trimming, other (including Sturmer et al. \cite{sturmer2010treatment}) do not recommend such an additional step. 
		
		To circumvent the need to discard observations,  Li and Green \cite{li2013weighting} and Mao et al., \cite{mao2018propensity} propose the use of matching weights, whereas Li et al. \cite{li2018balancing, li2018addressing} recommend using overlap weights. Trimming, truncating, matching weights as well as overlap weights target subpopulations of the overall population of interest. Their respective estimands deviate from the (overall) average treatment effect (ATE), the same way targeting the subpopulation of treatment (resp. untreated)  patients leads to the average treatment effect on the treated (ATT) (resp. untreated or controls (ATC)). \cite{hirano2003efficient,crump2006moving,crump2009dealing} 		
		For both trimming and truncation of  weights, unfortunately, it is not always clear how this subpopulation---to which the corresponding estimands apply---can be described succinctly to a layman as it corresponds to a subpopulation of patients enclosed between hyperplanes of the covariate space defined by the propensity scores. \cite{yang2018asymptotic}  Although, Traskin and Small \cite{traskin2011defining} and \cite{fogarty2016discrete} provide a way to (approximately) determine and describe such a subpopulation for trimming using a regression tree approach or a discrete optimization, it is not always feasible, especially when there is a large mixture of discrete and continuous covariates. 
		
		Li and Green \cite{li2013weighting} as well as Mao et al. \cite{mao2018propensity} describe the matching weight estimand as an analogue to the one-on-one propensity score matching estimand, while Li et al. \cite{li2018balancing,li2018addressing} define the overlap weights as those targeting the subpopulation of patients with the most overlap in observed baseline characteristics between the treatment groups. Nonetheless, the last two methods have not been compared directly even though they, very often, lead to similar results.\cite{li2018balancing,li2018addressing} Li and Greene \cite{li2013weighting} were the first to attempt to derive doubly robust overlap weight and matching weight estimator. However, even though they do improve efficiency, their augmented estimators were shown not to be doubly robust by Mao et al. \cite{mao2018propensity} and confirmed later by Tao and Fu. \cite{tao2019doubly} All these aforementioned weights are components of what constitute the "balancing weights" as they often balance, on average, the distributions of the covariates between treatment groups.  
		\subsection{The objectives of this paper}
		The goal of this paper is four fold. First, we provide a set of conditions for inverse probability weights to target the subpopulation of patients for whom there exists a clinical equipoise. Using these conditions, we explain why the trimming and truncating weights cannot target such a subpopulation of patients in an observational study. Second, we establish the relationship between matching weights and overlap weights. In fact, we demonstrate that they both target the same subpopulation and thus the same estimand.  This means, as a propensity weighting scheme, the matching weights estimator is not the analogue to the one-one-one propensity score matching that has ATT as target of inference, rather an analogue to the overlap weights. We also establish the intrinsic connections that exist among the matching weight (as well as the overlap weight) estimand and the ATE, ATT, or ATC, depending on some empirical conditions that are related to the data at hand. Third,  using the beta distribution family of functions we show that the overlap weights are  member (and a very special case) of this family of functions. This provides us with the tools to provide a mathematical proof as to why the matching weights are asymptotically less efficient than the overlap weights both in terms of point and variance estimates. Lastly, as they draw from familiar concepts of balance and clinical equipoise that are inherent to randomized clinical trials, the beta distribution family of function appear to be the panacea to the lack of positivity: the further away from the boundaries $\{0, 1\}$ of the propensity score range we can get, the better. We demonstrate that there is a diminishing return, in terms of bias-variance trade-off, if we use functions of the beta family other than the overlap weights.  In a clinical context, this answer the question as to why, in a search for a perfect overlap population, we cannot limit or aim for the range of eligible participants' propensity scores  in the neighborhood of 0.5.
		 
		 The rest of this paper is structured as follow. In the next section, we present the statistical characteristics of  covariate balancing weights currently used in the propensity score literature and highlight their differences.   Section \ref{sec:balancing_wgts_examples} presents some examples of such balancing weights, their estimands, and the rationale for selecting one we deem appropriate to use based on their target populations. Then we introduce, in Section \ref{sec:wgt_eqp}, the conditions that versions of (inverse) probability  weights must satisfy to target the subpopulation of patients for whom there is clinical equipoise. They allow us to specify which are the weights that are good candidates to aim for such a subpopulation. We also introduce the entropy weights and the beta family of weights. Using the beta family, we establish the relationships between the different weights that aim at estimating the treatment effect for the subpopulation of interest and derive the most efficient. Leveraging the semiparametric augmentation  framework, we demonstrate in Section \ref{sec:augmentation} that, while efficient, the weights for clinical equipoise do not satisfy the  double robustness property. Nevertheless, their augmented estimators can improve estimation as the regression models play a preeminent role in dealing with bias and efficiency.  
		 In Section \ref{sec:simulations}, we conduct Monte-Carlo simulation studies to compare the performances and validate the theoretical results of IPW, matching, entropy, overlap, and some beta weights under diverse scenarios of propensity score overlap.  Finally, we illustrate the different methods in Section \ref{sec:illustration}, before we conclude with a brief discussion and some recommendations. Outlines of proofs, technical details as well as additional simulation results are provided in the Appendix .
	\section{Balancing weights}
	\subsection{Overview}\label{sec:balancing_overview}
	 Let $Z=z$ be the treatment indicator ($z=1$ for treatment and  $z=0$ for control), $Y$ a continuous outcome, and ${X}=(X_0, X_1, \ldots, X_p)$ a matrix of baseline covariates, where $X_0=(1,\ldots, 1)'$.  The observed data $O=\{ (Z_i, X_i, Y_i): i=1\dots, N \}$ are a sample of $N$ participants drawn independently from a large population of interest.    We adopt the potential outcome framework of Neyman-Rubin \cite{neyman1923applications, imbens2015causal} and assume that  for a randomly chosen subject in the population there is a pair of random variables  $(Y(0), Y(1))$,  where $Y(z)$ is the  potential outcome, i.e., the outcome that would been observed if, possibly contrary to fact, the individual were to receive treatment $Z=z.$ Potential outcomes are related to observed outcomes via 
	$Y=ZY(1)+(1-Z)Y(0),$ i.e., for each individual, the potential outcome  $Y(z)$  matches their observed outcome  $Y$ for the treatment  $Z=z$ they indeed received, by the consistency assumption.

	We  assume the stable-unit treatment value assumption (SUTVA), i.e., there is only one version of the treatment and the potential outcome $Y(z)$ of an individual does not depend on another individual's received treatment, as it is the case when participants' outcomes interfere with one another. \cite{rosenbaum1983central} 
	To identify causal estimands, we also assume that $Y(0)$ and $Y(1)$ are conditionally independent of $Z$ given the vector of covariates ${X}$, i.e., $E[Y(z)|X]=E[Y(z)|X,Z=z],$ $z=0,1$ (unconfoundness assumption). \\
	More often, the goal is to estimate the average treatment effect (ATE) across all members of a given population $\tau=\displaystyle E[\tau(X)]$ from the data, where  $\tau(x)=E[Y(1)-Y(0)|X=x]$ is the conditional average treatment effect (CATE), conditional on covariate values $X=x$. Note that when the CATE is constant, then ATE =  CATE and the treatment is said to be homogeneous. Otherwise, it is considered heterogeneous. In the latter case, we may focus our interest on the effect of treatment on  specific subgroups of the covariate distributions. Therefore, our goal is to estimate the weighted average treatment effect (WATE) 
	\begin{align}\label{eq:G-estimand}
		\tau_{g}&=\displaystyle \frac{\displaystyle E[g(X)\tau(X)]}{\displaystyle E(g(X))}
		=C^{-1}\!\displaystyle {\displaystyle\int \tau(x)f(x)g(x)dx},~~\text{with}~~ C={\displaystyle\int g(x)f(x)dx}
	\end{align} 
	where $f(x)$ represent the marginal density of the covariates with respect to a base measure 	 $\mu,$ which  we have equated to the Lebesgue measure, without loss of generality.
	The expectation is taken over the population of interest and $\tau_{g}$ simplifies to $\tau$ when $g\equiv 1$. Thus, $\tau_{g}$ differs from $\tau$ through the variation in $g(x)$ across levels of $X=x.$ The product $f(x)g(x)$ represents the target population density, where the function $g(X)$, which we refer to as the selection function, is a known function of the covariates \cite{hirano2003efficient} and can be modeled as $g(X; \beta)$ with  parameters  $\beta$.	It can be fully specified without unknown parameter $\beta$, for instance when $g$ specifies a population of women who are Medicare--Medicaid beneficiary or a population of physicians with a given specialty. The function $g(x)$ is often defined as function of the propensity score.\cite{zhou2020propensity}

	 We define the propensity score $e({x})=P(Z=1|{X=x})$, i.e., the conditional probability of treatment assignment  given the observed covariates. Under unconfoundness assumption, Rosenbaum and Rubin demonstrate that the propensity score is a balancing score since $X\bigCI Z|e({X})$. This implies that,  for participants with the same propensity score, the distributions of their corresponding observed baseline covariates ${X}$ are similar regardless of their treatment assignment.
	\cite{rosenbaum1983central,rosenbaum1984reducing,rubin1997estimating} Therefore, instead of controlling for the whole vector of multiple covariates $X$ to estimate treatment effects, one can leverage this property of the propensity score $e({X})$ to derive unbiased estimators of the treatment effect.  
	
	Since the propensity score $e(X)$ is usually unknown (except in randomized experiments); it must be estimated. The commonly-used estimation approach  is via a logistic regression model 
	\begin{equation} \label{eq:logistic}
	\logit\left(e({X};{\beta})\right)=\log\left(\frac{e({X};{\beta})}{1-e({X}; {\beta})}\right)={c(X)}\beta,
	\end{equation}
	where $e({X};{\beta})=P(Z=1|{X}; {\beta})=\left\{ 1+\exp(-{c(X)}{\beta})\right\}^{-1}$ and ${c(X)}$ is a function of ${X}$ that may include interactions and higher-order terms of the components of $X$. Henceforth, for ease of notation, we write ${c(X)}$ as just $X$. We say that the model \eqref{eq:logistic} is correctly specified if $P(Z=1|X)=e({X};{\beta})$, for some parameter vector $\beta$. The parameters $\beta$ can be estimated by maximum likelihood estimator $\widehat{\beta}$ solution to the estimating equation
		\begin{equation} \label{eq:logistic.esteq}
	\displaystyle\sum_{i=1}^{N} \psi_{{\beta}}({X}_i, Z_i)=\displaystyle\sum_{i=1}^{N}[Z_i-e({X_i};{\beta})]X_i=0
	\end{equation}
	 Other methods to estimate $e(X)$ exist, including the generalized additive models \cite{woo2008estimation} and machine learning techniques such as the generalized boosted regression models, random forest, classification and Bayesian additive regression trees. \cite{lee2010improving, kern2016assessing, stuart2017generalizing}

	  Consider the weights 
	$w_{z}(x) =g(x)\displaystyle e(x)^{-z}(1-e(x))^{z-1}, $  $ z =0, 1,$ 
	we showed in Appendix \ref{appendix1} that  
   \allowdisplaybreaks	
   \begin{align}
	E[Zw_{1}(X)] &= E[(1-Z)w_{0}(X)]=E[g(X)];\label{bal_zyg.wgts}\\
	E[Zw_{1}(X)Y] &= E[g(X)E(Y(1)|X)]~~\text{and} ~~E[(1-Z)w_{0}(X)Y] = E[g(X)E(Y(0)|X)].\nonumber
	\end{align} 
	Therefore, $\tau_{g}$ can be estimated by	a H\'ajek-type estimator
	\begin{gather}\label{eq:li_est}
	\widehat\tau_{g}= \displaystyle\sum_{i=1}^{N}\left[{Z_i\widehat  W_1(x_i)}-{(1-Z_i)\widehat W_{0}(x_i)}\right] Y_i,\\
	\text{where}~~ W_z(x)=\widehat w_z(x) /N_{\widehat w_z},~~\widehat w_z(x) =\widehat g(x) \widehat e(x)^{-z}(1-\widehat e(x))^{z-1}~~\text{and}~~N_{\widehat w_z}=\displaystyle\sum_{i=1}^{N} Z_i^z(1-Z_i)^{1-z}\widehat w_z(x_i),\nonumber
	\end{gather} 
 with  $\widehat g(x)=g(x; \widehat \beta)$, $~\widehat e(x) =e(x;\widehat \beta),$ for an  an estimator $\widehat \beta$   of  $\beta$ (see Li and Greene \cite{li2013weighting} as well as Li et al.\cite{li2018balancing}). 
	

	The WATE $\tau_g$ generalizes a large class of causal estimands.\cite{crump2006moving,crump2009dealing,hirano2003efficient,li2018balancing}  The selection function $g$ delimits and specifies the target subpopulation defined in terms of the covariates $X$ as well as the treatment effect estimand  of interest and helps define the related weights.  Suppose $f({x})$ is the marginal distribution of the covariates ${X}$ and consider  $f_{x|Z}(x|z)=P(X=x|Z=z)$ the density of the covariates $X$ in the treatment group $Z=z$. Because $f_1(x)w_1(x)=f_0(x)w_0(x)=f(x)g(x),$ as shown in \eqref{bal_zyg.wgts}, the weights $w_z(x)$  balance the distributions of the covariates between the two treatment groups. \cite{li2018balancing} Thus, the name {\it balancing weights}. 
	
	\subsection{Importance sampling perspective}
	To better understand the aforementioned different estimand of treatment effect, we use the importance sampling perspective (via a change of the underlying probability measure) to provide insights on the role the selection function $g(\cdot)$ plays in targeting the subpopulation of interest.
	
	First, using the pseudo-population analogy, \cite{cole2008constructing,hernan2018causal}  the weights $w_z(x)$ indicate that two different operations are happening  simultaneously. We create a pseudo-population via the inverse probability weights $e(x)^{-z}(1-e(x))^{z-1}, ~ z \in \{0, 1\},$  where the treated and control patients are exchangeable, i.e., the probability of  selection  of treatment $Z$ is independent of the covariates $X$. Then, from that pseudo-population and thanks to the selection function $g(x)$, we select a sub-pseudo-population that represents the targeted subpopulation of interest. 
	Therefore, as causal contrast, the WATE represents the potential outcomes difference in the pseudo-(sub-)population. \cite{crump2006moving} 
	
	The estimator $\widehat\tau_{g}$ in equation \eqref{eq:li_est} uses normalized weights   ${\widehat W_z(x_i)}$  that sum up to 1 in each treatment group. For the ATE, the use of normalized weights have been first proposed by H\'ajek \cite{hajek1971comment} to stabilize estimators. They have been also advocated by Imbens \cite{imbens2004nonparametric} as they seem to work better and improve the performance of non-normalized weights. \cite{kang2007demystifying,  busso2014new,ju2018adaptive} Sometimes, in lieu of normalized weights, people also  use stabilized weights, especially in marginal structural models, where $\widehat\tau_{g}$ is defined as $
	\widehat\tau_{g}=\frac{1}{N}\displaystyle\sum_{i=1}^{N}\left[ P(Z=1)Z_i{\widehat  e(x_i)}^{-1}Y_i-P(Z=0)(1-Z_i)(1-\widehat  e(x_i))^{-1}Y_i\right]$ instead.\cite{robins2000marginal,robins2000marginals}	
	Large-sample properties of the estimator \eqref{eq:li_est} are provided by Hirano et al., \cite{hirano2003efficient} Crump et al., \cite{crump2006moving, crump2009dealing} and Li et al. \cite{li2018balancing} 

	The idea of using a selection function to sample a subpopulation deemed to provide a better estimator espouse that of the (self-normalized) importance sampling commonly-used in our introductory Bayesian theory courses. Note that 
	\allowdisplaybreaks	\begin{align*}
		\tau_{g}^{z}&=\displaystyle \frac{\displaystyle E[g(X)E(Y(z)|X)]}{\displaystyle E(g(X))}
		=\displaystyle \frac{\displaystyle \int yf_{Y|X,Z}(y|x,z)w_z(x)f_{X|Z}(x|z)f(z)dydxdz}{\displaystyle \int w_z(x)f_{X|Z}(x|z)f(z)dxdz},
	\end{align*} 
	since $E[g(X)E(Y(z)|X)]=\displaystyle \displaystyle \int g(x)yf_{Y|X,Z}(y|x,z)f(x)dydx$ and $g(x)f(x) = \displaystyle \int w_z(x)f_{X|Z}(x|z)f(z)dz$. The denominator $E(g(X))$ is the special case of $E[g(X)E(Y(z)|X)]$, where we set $E(Y(z)|X) =  1$. Hence, we can view an estimation of $\tau_{g}^{z}$ (and thus of $\tau_{g}=\tau_{g}^{1}-\tau_{g}^{0}$) as a frequentist estimation of the quantity sampled from data using a form of Monte-Carlo procedure where we sample  from the original data with respect to the conditional joint density $f_{Y|X,Z}(y|x,z)f_{X|Z}(x|z),$ $z=0, 1.$ Therefore, the Monte-Carlo estimator  $\widehat \tau_{g}$ is thus from a sample from the original data---as these are samples for which $Z=z.$  This generalizes the importance sampling perspective for ATT presented by Moodie et al.\cite{moodie2018doubly}
	
	Since for a given selected function $g(x)$, the weights $w_z(x)$ re-scale the original study population to create a pseudo-population---for which the treatment groups are balanced on average, with respect to their baseline covariates to obtain a reliable treatment effect---this  may result in a loss of precision for the estimator $\widehat \tau_{g}$ due to the influence of extreme weights.\cite{cole2008constructing,hernan2018causal}  In the context of importance sampling, the behavior of the selection function at the tails of the propensity score spectrum is crucial and must be investigated.\cite{geweke1989bayesian,robertmonte2010} The (per group) effective sample size, as measure of the efficiency of the resampling procedure,  
	\begin{align} \label{eq:effssize}
		\widehat{ESS}_z= \left( \displaystyle\sum_{i=1}^{N} \widehat W_z(x_i)^2\right) ^{-1} \left( \displaystyle\sum_{i=1}^{N} {\widehat W_z(x_i)}\right)^{2}.
	\end{align}  
	provides the approximate number of independent observations drawn from a simple random sample needed to obtain an estimated with a similar sampling variation than that of the weighting observations. 	 It helps characterize the variance inflation or precision loss due to weighting. \cite{mccaffrey2004propensity} 	 
	Alternatively, one can also estimate directly the variance inflation based on  a “design effect” approximation of Kish \cite{kish1985survey} 
	\begin{align*}
		\widehat{\text{VI}}=(1/N_1+1/N_0)^{-1}  \sum_{z = 0}^1 \left[ \left( \displaystyle\sum_{i=1}^{N_z} {\widehat W_z(x_i)}\right)^{-2}\displaystyle\sum_{i=1}^{N_z} \widehat W_z(x_i)^2\right], ~~\text{where}~~N_z=~~\text{number of participants in group }~Z=z \in \{0,1\}.
	\end{align*}

		\section{Examples of balancing weight estimators}\label{sec:balancing_wgts_examples}
			\subsection{The average treatment effect (ATE)}
	The choice of $g$ dictates the nature of the estimand and characterizes the underlying population of interest. Which function $g$ to use depends on several considerations including subject-matter knowledge, the targeted causal estimand of interest, and the structure of the data at hand.   For example, choosing $g(x)\equiv 1$  yields the average treatment effect (ATE) in a population where the distribution of the covariates is similar to that of all the study participant. Such an estimand helps answer the question "what would have happened had all participants in the population been treated compared to none of them had been treated?"  Thus, the target population is the entire sampling population of interest with weights $e(x)^{-z}(1-e(x))^{z-1}$, i.e., the ATE is the familiar inverse probability weight estimator for the whole population.\cite{hirano2003efficient} To estimate the ATE, via this so-called inverse probability weighting (IPW) method, it is primordial that the positivity assumption be satisfied, i.e., $P(\{x: \zeta_0 <e(x)<1-\zeta_0\})=1$, for some $\zeta_0>0$. Otherwise, the estimate will be subject to the "tyranny of the minority", \cite{lin2013agnostic} especially when the ratio $[e(x)(1-e(x))]^{-1}$ is highly variable. \cite{robins2007comment} That is, on the one hand, the weights will be extremely (and unreasonably) large for few treated participants with $e(x)\approx 0$ or control participants with $e(x)\approx 1$, while on the other hand, the presence of few treated participants with $e(x)\approx 1$ or control participants with $e(x)\approx 0$ may unduly influence the overall treatment effect or its efficiency (even though their weights $w_z=1$). In both scenarios, the ATE estimator often leads to numerical instability, relatively high variance, and non-standard rate of convergence (its fastest rate being slower than  $N^{1/2}$). \cite{khan2010irregular,imbens2004nonparametric,ma2020robust}

	\subsection{Moving the goalpost}
	A choice of $g$ other than  $g(x)\equiv 1$ changes the estimand of interest from the ATE to an estimand that is defined by the distribution of the propensity score. 	
	As argued by Hirano et al. \cite{hirano2003efficient} and Crump et al., \cite{crump2009dealing} using the weighted average treatment effect based on the function $g(\cdot)$ shifts our  focus to a subpopulation whose covariate distributions allows a  large concentration of observations in both treatment groups. Nevertheless,  this  "moves the goalpost" from the ATE toward an estimand based on the subpopulation and improves its internal validity.\cite{crump2006moving} Such an emphasis on interval validity over external validity is crucial in many studies, especially when we intend to apply the results of a study  to a different environment or population (transportability). \cite{stuart2017generalizing,westreich2017transportability}	Furthermore, deviation from the ATE is common in practice, from the exclusion of subjects without adequate matches in matching methods \cite{stuart2010matching} to truncation or pruning of observations with extreme inverse probability weights;  the objective is to reach a more inclusive subpopulation that provide a better  bias-variance trade-off. \cite{crump2006moving, traskin2011defining,yang2018asymptotic,ma2020robust}

		\subsubsection{Two alternatives to the average treatment effect}
	
	In many studies, the average treatment effect on the treated (ATT) is the more relevant estimand, when the interest lies solely on a subpopulation of participants who actually received the given treatment. For instance, one may evaluate the impact of a smoking cessation intervention on improving health outcomes among those who actually received the intervention.   Thus, when $g(x)=e(x),$ $\tau_g$ is the average treatment effect on the treated (ATT), with weights   $(1,~ e(x)/(1-e(x)))$.  Its targeted population  is a population of participants whose the covariate distribution is similar to that of the the subpopulation of treated study participants.   The goal is to weight the control  participants  as to reach a balanced distribution of their covariates with that of the treatment group, using $g(x)=e(x)$.\cite{hirano2003efficient,moodie2018doubly} 
	Alternatively, the average treatment effect on the controls (ATC) can be of interest if one evaluates the impact of introducing a new treatment (or withholding a harmful exposure) on an outcome if we switch from the standard of care (or control treatment) to the new treatment.\cite{tao2019doubly}  To this regards, $g(x)=1-e(x)$ and the estimand $\tau_g$ becomes the average treatment effect on the controls (ATC), i.e., the treatment effect in a population where the distribution of the covariates $X$ is similar to that of the study participants who are in the control group and the corresponding  weights are   $((1-e(x))/e(x),~ 1)$.    
	
	Note that the for these targets of inference we only need weaker versions of the positivity assumption. When we estimate the ATT, the weights of all treated participants is 1; thus we only need $P(\{x: e(x)<1-\zeta_0\})=1$, for some $\zeta_0>0$ for control participants since for control participants with $e(x)=0,$ their weights is equal to 0. Similarly, to estimate the ATC, we are only required that $P(\{x: e(x)>\zeta_0\})=1$, $\zeta_0>0$, for treated participants. 
			\subsubsection{Alternatives under lack of adequate positivity}\label{sec:alt_positivity}
	To deal with the lack of adequate positivity, i.e. $P(\{x: \zeta_0 <e(x)<1-\zeta_0\})<1$, for any $\zeta_0>0$, other alternatives to ATE have been proposed in the literature. 
	These alternative methods (and estimands) include IPW trimming and IPW truncation, with $g(x)$ equal to, respectively, $\mathds{1}({\{\alpha\leq e(x)\leq 1-\alpha\}})$,  and $ \alpha^{-1} e(x)\mathds{1}({\{\alpha> e(x)\}}) + \mathds{1}({\{\alpha\leq e(x)\leq 1-\alpha\}}) + (1- \alpha)^{-1} (1- e(x))\mathds{1}({\{e(x)> 1-\alpha\}})$, where $\alpha\in[0,~0.5]$ and the indicator function $\mathds{1}({\{\alpha\leq e(x)\leq 1-\alpha\}})=1$  if  $\alpha\leq e(x)\leq 1-\alpha$ and 0 otherwise. \cite{robins2000marginal,cole2008constructing,zhou2020propensity} They weaken lack of adequate positivity, reduce variability by making the weights less extreme, and have a positive impact on the effective sample size or the variance inflation. Nevertheless, they are not a panacea since reduce variability comes at a price---in addition to changing the estimand of interest\cite{khan2010irregular}---the choice of a threshold is often subjective and both methods are sensitive to the data at hand, its size, and the selected threshold, which can still lead to biased estimators. \cite{ma2020robust}
	
	It is important to remember that there are many versions of trimming,\cite{li2018addressing} along with (or not) a re-run of the propensity score model, which play into this subjectivity of the IPW trimming. Furthermore, the rule-of-thumb of choosing $\alpha=0.10$ is not always followed \cite{yang2018asymptotic} and when it is, nothing guarantees that it works since it often leads to a substantial sample reduction and possibly limited information. \cite{zhou2020propensity} The ideal trimming or truncation threshold should be purely data-driven, far from any analyst subjectivity. \cite{ju2018adaptive,ma2020robust} Finally,  truncation, in addition to arbitrary altering  the weights of participants with initial extreme IPW weights, does not change the weights of participants with initial $\widehat w_z(x) = 1$, i.e., treated participants with $e(x) = 1$ or control participants with $e(x)=0.$ While their weights are not extreme, their corresponding outcomes can be drastically different from those of participants for whom $P(\{x: \zeta_0 <e(x)<1-\zeta_0\})=1$, for some $\zeta_0>0$. In aortic valve disease, for instance, these are often patients for whom transcatheter aortic valve replacement (i.e., $z=1$) is the only option because of older age, frailty,  multiple comorbidities or prohibitive surgical risk while surgical aortic valve surgical (i.e., $z=0$) is mostly administered to younger or less frail patients.\cite{brennan2017transcatheter} 
	
	Finally, a number of estimators that do not require the positivity assumption include the overlap weight (OW),  matching weight (MW), Shannon's entropy weight (EW),\cite{zhou2020propensity} and beta weight (BW) estimators. Their respective selection function are $e(x)(1-e(x))$,  $\min\{e(x), 1-e(x)\}$, $-[e(x)\log(e(x))+(1-e(x))\log(1-e(x))]$, and $[e(x)(1-e(x))]^{\nu-1}, \nu\geq 2$.  When $e(x)\rightarrow  0$ or 1, these estimators lead to $w_z(x)\rightarrow 1$, i.e., they smoothly and gradually downweight the influence of participants whose $e(x)$ is at both ends of the propensity scores spectrum. Thus, with extremely limited influence of participants with $e(x)=0$ or 1 in estimating $\tau_{g}$, i.e.,   no extreme weights, there is no need to exclude participants arbitrarily (as with trimming). These estimators  belong to the family of the estimands of the treatment effect on the subpopulation for whom there is clinical equipoise, which we focus on in the next sections.

			  	\begin{table}[ht] 
			  	\caption{Examples of selection function, targeted (sub)population, causal estimand, and corresponding weights} \label{tab:wgts_summary}
			  	\begin{center} 
			  		\begin{threeparttable}[]
			  			\def\arraystretch{1.4}
			  			\begin{tabular}{rccccccccccccccccccccc} 
			  				\toprule
			  				Target (sub)population   & $g(x)$ & Estimand  &   Method\\ \cmidrule(lr){1-4} 
			  				overall &   $1$  &  ATE  &   IPW          \\
			  				treated &  $e(x)$   &  ATT  & IPWT \\
			  				control &  $1-e(x)$  &       ATC  &       IPWC \\ 
			  				\multirow{1}{*}{\makecell{restricted}} &   $I_{\alpha}(x)=\mathds{1}({\{\alpha\leq e(x)\leq 1-\alpha\}})$     &  OSATE  & IPW Trimming\\ 
			  				\multirow{1}{*}{\makecell{truncated}} &   $I_{\alpha}(x) + J_{\alpha}(e(x))^zJ_{\alpha}(1-e(x))^{1-z}$     &     &       IPW Truncation\\
			  				\addlinespace
			  				trapezoidal &  $\min\left\{1, Ku(x)\right\},$  $K\geq 1.$ &            &   \\\addlinespace\cmidrule(lr){2-3}  
			  				\multirow{3}{*}{equipoise}  
			  				 &  $u(x)=\min\{e(x), 1-e(x)\} $  &          &      MW\\
			  				&  $e(x)\left(1-e(x)\right)$  &       ATO   &       OW\\ 
							&	 $[e(x)(1-e(x))]^{\nu-1}, \nu\geq 2$  &        &       BW\\ 
									  				\bottomrule
			  			\end{tabular} 
			  			\begin{tablenotes}
			  				\footnotesize
			  				
			  				\item The related weights are $ w_z(x)=g(x)[e(x)^{-z}(1-e(x))^{z-1}]$, $z\in \{0,1\}$. $~g$ is the selection function;  $\mathds{1}(\boldsymbol{\mathcal{.}})$ is the standard indicator function and 
			  				$J_{\alpha}(e(x))= e(x)\big[{\alpha}^{-1}{\mathds{1}({\{ e(x)<\alpha\}})}+{(1-\alpha)}^{-1}\mathds{1}({\{ e(x)>1-\alpha\}})\big]$, where $ \alpha \in (0, 0.5),$  $z\in \{0,1\}$.
			  				%
			  			\end{tablenotes}
			  		\end{threeparttable}
			  	\end{center} 
			  \end{table}

	 	\section{Weights for clinical equipoise}\label{sec:wgt_eqp}
	 	Estimating causal effect in observation studies through a specific selected estimand requires that we specify the target population. As we have alluded to, the selection function plays a key role in determining the estimand and the targeted subpopulation. In this section, we focus on the subpopulation of patients for whom there is a clinical equipoise. Examples of such selection function have been used in the literature by Li and Greene, \cite{li2013weighting} Mao et al., \cite{mao2018propensity,mao2020flexible} Li et al.\cite{li2018balancing} and Zhou et al. \cite{zhou2020propensity} The goal of this section is to characterize the different selection functions that can be used to target such a population and demonstrate that most of these functions are related. In particular, we will present and compare the characteristics of the matching, entropy, and overlap weights functions as equivalent selection functions that target the subpopulation of patients for whom there is equipoise. Henceforth, we will refer to these estimators as equipoise treatment effect estimators. In the next section, we describe through some examples what characterizes their target subpopulation, contextualize related estimand(s),  and elucidate what that means in practice.
	 	
	 			\subsection{Subpopulations of participants for whom there is clinical equipoise}\label{sec:eg_equipoise}
	 	While we have become accustomed to ATE, ATT, and ATC and their target populations, it is not always clear how to define  {\it a priori} the target populations of the other WATEs as they may intrinsically depend on the data at hand.  However, the subpopulation we target via the OW  corresponds to the subpopulation where treated and control patients overlap the most in terms of their respective $e(X)$ and for whom  there is equipoise. Such a subpopulation is easily described, characterized, and presented using descriptive statistics table from  patient population. \cite{thomas2020overlap}	
	 	Although Crump et al. \cite{crump2006moving} think the change of target of inference from ATE (or ATT) to  WATE  is "\dots not  motivated,  {\it per se}, by an intrinsic interest in the subpopulation for which we ultimately estimate the average causal effect", the WATE produced by OW and MW target a genuine subpopulation of interest in many areas of scientific investigation. 
	 	\cite{mao2018propensity,thomas2020overlap}

	 	For instance, in clinical settings, the subpopulation targeted by  overlap weights corresponds to the subpopulation of participants with great similarities as it comes to their covariate distributions for which clinicians have a genuine uncertainty in deciding over which treatment  can be the most beneficial since they appear to be good candidate for either treatment. Zhou, Matsouaka, and Thomas mentioned clinical studies where patients are counter-indicated for some treatment options, frail, or not eligible for surgical procedures (often referred to as exclusion-restriction criteria) and  the focus is then to assess the effect of treatment on groups of eligible patients. \cite{zhou2020propensity} Their example on the study of uterine fibroid---in which women of childbearing age decline hysterectomy and opt for myomectomy, a uterine-sparing treatment option, keeping alive their chances to bear children---is such an example where it is important to judiciously aim at an appropriate subgroup of women if we want to better evaluate the health benefits of either procedures.   
	 	
	 	In social  sciences, this population corresponds to the group of participants for whom a new policy (or a new political campaign), if efficient, can produce a greater shift in policy implementation. In marketing, this population will consist of shoppers who can be persuaded to buy a newly launched product if offered a coupon (or discount) and not those who stick to a specific brand (or despise it) regardless of the incentive. Had all people been influenced  equally by all advertisements, the advertisement targeting would not be necessary. Of course, this is not true, and advertisers generally target people who they believe would be most influenced by an advertisement campaign.
	 	Finally, in political science, the population targeted by overlap weights corresponds to the group of swing voters, whom a political campaign would like to reach out to, inform or persuade about their policies, and sway them to vote for their candidate  to change the political landscape or win an election, beyond their own political strongholds and their own (often increasingly polarized) die-hard fan base.\cite{lockhart2020america,butters2020polarized} The US 2020 presidential election is a great illustration of such a realistic approach to targeting voters: candidates chose some key states or even specific cities to aggressively campaign in (eithr in person or via advertisements) to reach as many  swing voters as possible.  \cite{lockhart2020america} 
	 		 	
	 	In all these examples, patients (or participants) for whom the positivity assumption is violated and do not have adequate comparable counterpart in the alternative treatment group, the causal treatment effect is not only unidentifiable (or very uncertain, at best), the treatment effect estimated based on data that include them may be biased and thus questionable.\cite{schafer2008average,kang2016practice,busso2014new,barsky2002accounting} The choice of estimands based on a subpopulation of  participants for whom the treatment effect is more relevant and better estimated (in term of bias and efficiency) is in fact a must.	 	
	 	Li and Greene \cite{li2013weighting}, Fan Li et al. \cite{li2018balancing} and Mao et al. \cite{mao2018propensity} provide additional reasons (along with several examples)  on why  such a target subpopulation  can be of intrinsic substantive interest. Walker et al. \cite{walker2013tool} introduce the concept of empirical equipoise and propose an algorithm to systematically identify settings in which there is empirical equipoise, i.e., where clinicians (collectively) seem evenly divided regarding the best treatment option(s) for a population of patients. Even better, Thomas et al. recommend that authors provide the descriptive statistics for the participants that compose the overlap population, in addition to the traditional "Table 1", to better understand which randomized clinical trial is thus emulated.\cite{thomas2020overlap} 
	 	Yoshida et al. extend the algorithm to multiple treatments \cite{yoshida2019tool}.  Finally, Vansteelandt and Daniel argue that the WATE derived from this targeted population represents the effect that would have been observed in a randomized trial where participants are selected into the study with  probability (proportional to) $e(x)(1-e(x))$. \cite{vansteelandt2014regression} To generalize how one can target such a subpopulation for clinical equipoise, we henceforth expand on these selection functions and their corresponding estimators.

	 	\subsection{Selection functions for clinical equipoise}\label{sec:select_eqp}
	 	 To simplify our notation, in this section, we denote $t=e(x)$ and consider as selection function $\phi: [0,1]\mapsto \mathbb{R}^{+}=\{x \in \mathbb{R}~ |~ x\geq 0\}, t\longmapsto \phi(t)$. 
	 	 A selection function $\phi(t)$ is good candidate in targeting the subpopulation of participants for whom there is clinical equipoise if it satisfies the following conditions:
	 	\begin{enumerate}
	 		\item $\phi$ is defined at $t=0$ and $t=1$ and is equal to 0; \label{cond.1a}
	 		\item $\phi$ is symmetric around the vertical line $t=0.5$;\label{cond.1c}
	 		\item $\phi$ is strictly quasi-concave and reaches its maximum at $t=0.5$ ; \label{cond.2}
	 			\item $\phi$ is non-negative, continuous and differentiable. \label{cond.3}

	 	\end{enumerate}
	 
	  In theory,  participants with propensity score near 0 and 1  lack adequate positivity and whose treatment allocation is deterministic while  patients in the common region of the propensity score distributions reinforces the positivity assumption and guaranties the internal validity of the findings. \cite{imbens2004nonparametric} 
	   
	  Condition \eqref{cond.1a} allows us to include all the participants, but assigns a value $\approx$ 0 to those who based on their covariates are either certain to receive the active treatment or certain to received the control treatment.  These participants for which  $\phi(t)=0$ are precisely those with a lack adequate positivity, whose contribution is not needed in estimating the treatment and whom the results we will obtain should not apply to.    Thus,  condition \eqref{cond.1a} enforces the positivity assumption by focusing on the regions of the covariate distributions where it is plausible to draw causal inference (see the geometric interpretation of Section \ref{sec:interpret} and its implications on both matching- and overlap-weight estimators). 
	  
	  In a two-treatment randomized clinical trial, for instance, the true propensity score $t=e(Z=1|X)$ is known and constant, usually equal to 0.5, irrespective of the covariate distributions. Similarly, conditions \eqref{cond.1c} and \eqref{cond.2} ensure that we equitably target  the subpopulation of participants whose propensity scores are around 0.5, i.e. those with high chance of being randomized if we had conducted a randomized clinical trial--- where the covariate distributions of participants in one group overlap best with those for participants in the other group.  This guarantees that we over-represent observations with propensity scores around 0.5 and thus target the subpopulation of participants those whose treatment assignment is poorly explained (or deemed unpredictable) given their measured covariates. 
	  
	   Condition \eqref{cond.2}   smoothly downweights (in combination with \eqref{cond.1a}) the unduly influence observations with extreme weights (propensity scores near 0 or 1) may exert on the average treatment effect.   	 In other words, the choice of a function  $g$ for clinical equipoise allows us to assign more weights to participants in the target subpopulation, while annihilating the impact of participants at the extreme tails of the distributions of the propensity scores of the treatment groups  without resorting to an arbitrary decision to discard  these outliers.	 
	 The quasi-concavity of condition \eqref{cond.2} and  condition \eqref{cond.3} ensure that regularity conditions for sandwich variance estimation are satisfied, along variance reduction and improved efficiency. 	 Conditions \eqref{cond.1a}--\eqref{cond.3} allow us to focus on the average treatment effect on the subpopulation of patients (defined  on the common support of the joint covariate distribution\cite{thomas2020overlap})  for whom there is clinical equipoise and estimate it more precisely, while preserving a good effective sample size.\cite{crump2006moving}

	 Several functions $\phi(t)$ satisfy  the conditions \eqref{cond.1a}--\eqref{cond.3}, including the sinus function $\sin(\pi t)$, the (scaled) Shannon's entropy function $\Omega(t)=-[t\ln(t)+(1-t)\ln(1-t)]/\ln(2)$ and the family of (scaled) beta  functions $ B_{(\nu_1, \nu_2)}(t) = 2^{\nu_1+\nu_2-2}t^{\nu_1-1}(1-t)^{\nu_2-1},$ for which $\nu_1=\nu_2=\nu\geq 2.$   Note that the selection functions for the ATE, ATT, ATC and OW are all special cases of the beta selection function (up to a multiplying constant), with $(\nu_1,\nu_2)$, respectively, equal to $(1,1)$, $(2,1)$, $(1,2)$, and $(2,2)$.  According to our definition and criteria \eqref{cond.1a}--\eqref{cond.3},  functions $ B_{(\nu_1, \nu_2)}(t)$ that are good candidate for selecting the subpopulation of participants for whom there is clinical equipoise are those for which  $\nu_1=\nu_2=\nu.$  Hereafter, we adopt the notation $ B_{\nu}(t)$ in lieu of the symmetric functions $ B_{(\nu_1, \nu_2)}(t)$, i.e., whenever $\nu_1=\nu_2=\nu.$	 
	 Strictly speaking, the (scaled) matching function $\Lambda(t)=2\min\{t, 1-t\} $ is neither concave nor differentiable  (i.e., it does not satisfy the  condition \eqref{cond.3}), as the other the functions we  consider for equipoise hereafter. Nevertheless, Li et al. \cite{li2013weighting}  as well as Mao et al. \cite{mao2018propensity} have used a smooth version of  $\Lambda(t)$ (smoothed in the neighborhood of $t=0.5$) to obtain the point and variance  of the matching weight estimators.   For the sake of exposition, we exceptionally use $\Lambda(t)$ to establish the connection between the matching estimator and the overlap weight estimator in Section \ref{sec:overlap.estimators}. However, when we implement the matching weight estimator, especially for variance estimation, we use its smoothed version. Hence, hereafter, we will also refer to such a smoothed version as the matching weight selection function, without loss of generality. 
	 %
	 	\begin{figure}[h]
	 		\begin{center}
	 			\caption{Examples of selection functions for equipoise} \label{fig:overequipoisematch}
	 			\includegraphics[trim=30 60 19 105, clip, width=0.550\linewidth]{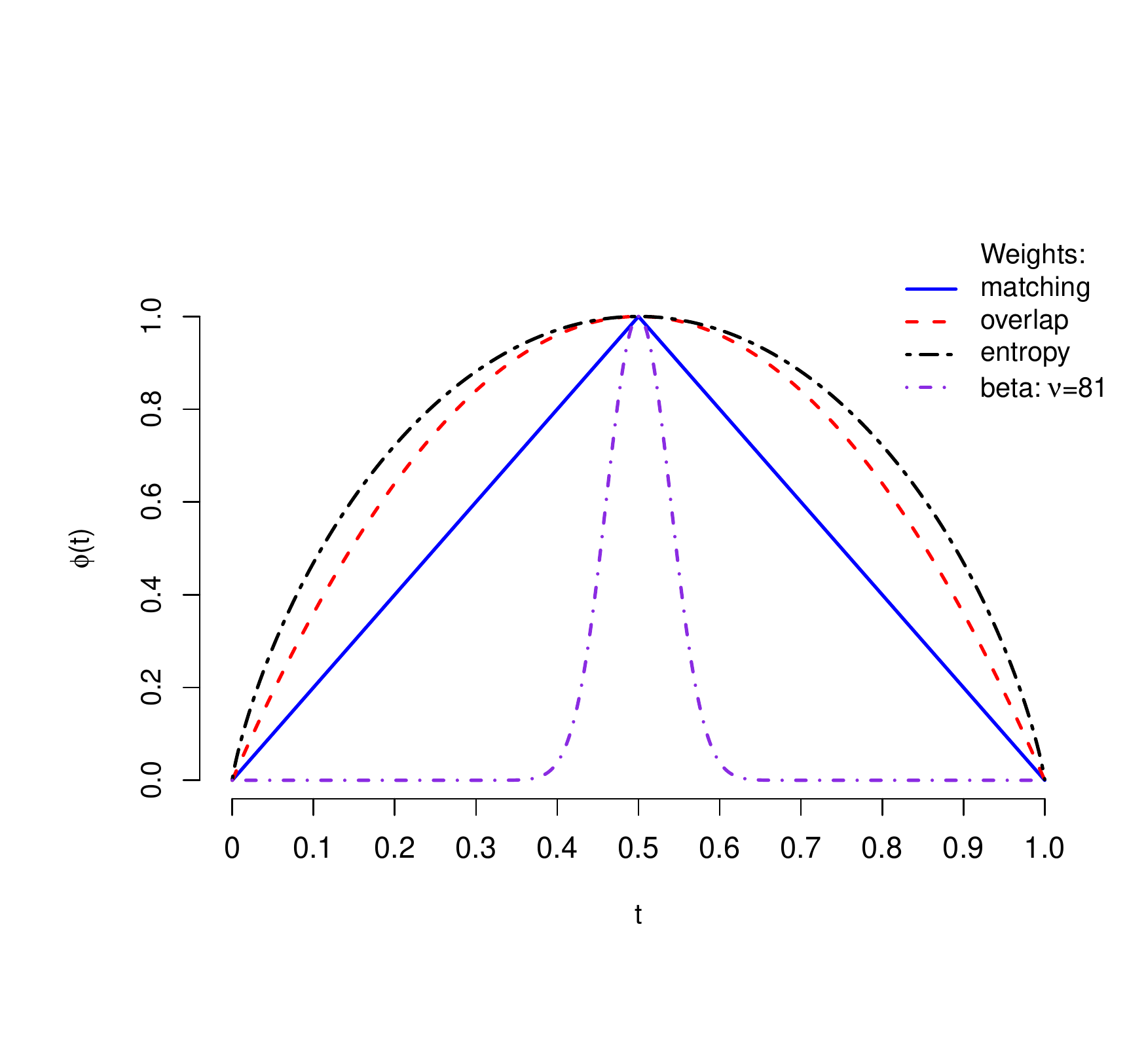}
	 		\end{center}
	 	\end{figure}
 	 
	 Figure 	\ref{fig:overequipoisematch} shows some of the selection functions for clinical equipoise. We scaled the selection functions such that the maximum $\phi(0.5)=1,$ for better visualization.  It can shown that the beta distribution function $B_{1.7}(t)$ approximate the entropy function  very well  	 and that the beta functions  $B_\nu(t)$, where  $3\leq \nu\leq 4$, can be good (and smooth!) approximations of the matching weight function. 	 
	 Clearly, the choice of $\nu$ has a direct impact on the treatment effect estimate. 	While $\nu>4$, seem to be an interesting choice to target patients with  $t$ around 0.5, such an apparent advantage only  helps reduce bias when the treatment effect is constant, but with less precision (especially when $\nu$ is large)---as shown by our simulation results. When the treatment effect is heterogeneous, it often leads to both biased and inefficient estimates as shown by our simulations.
	 For instance,  with $\nu=81$,   $B_\nu(t)$ move faster to 0 as $t$ gets further away from 0.5. Most noticeable weights are concentrated among few observations with  $t$  around 0.5 and these tends to drive the treatment effect estimate, which unfortunately will have a large  variance.  
	 
	 More formally,  we can show that $B_{1.8}(t)\leq \Omega(t)\leq B_{1.7}(t)$ and $B_{16}(t)\leq \Lambda(t)\leq B_{2.2}(t)$. This implies that, when the outcome is homoscedastic, the estimators based on the entropy  and "matching" weight  selection functions  $-[t\ln(t)+(1-t)\ln(1-t)]$ and $\max(t, 1-t)$ are always asymptotically less efficient that the estimator from the overlap weight selection function $t(1-t)$ since the latter is has the smallest asymptotic variance, under homoscedasticity, among all the selection functions $g$ (see Li et al.,  \cite{li2018balancing} Corollary 1). In addition, from a practical standpoint, the entropy weights can still be extremely large or undefined whenever the propensity scores for some participants are near (or literally equal to) 0 and 1  while the imperative to smooth out the matching weight function around the maximum 1/2 can be a hindrance to its implementation  , especially when dealing with multiple treatments.\cite{li2018addressing}

	 While the selection functions for the ATE, ATT, and ATC are also special cases of beta functions, they are not (according to our definition and criteria \eqref{cond.1a}--\eqref{cond.3}) good candidate functions to target the subpopulation of patients for whom there is clinical equipoise in our context. In addition, the indicator selection function $\mathds{1}({\{\alpha\leq t\leq 1-\alpha\}})$, $0<\alpha<0.5$ of Crump et al., \cite{crump2009dealing} the truncation selection function, and the trapezoidal function of Mao et al. \cite{mao2018propensity} defined as 
	 $\min\left\{1, K\!\min\{t, 1-t\} \right\},$ for  $K>1$ do not  satisfy some of the above conditions. 	 
	 
	 %
 	
	Table  \ref{tab:wgts_summary.equip}  summarizes the list of selection functions that target the subpopulation of participants for whom there is equipoise we will consider for the remaining of this paper when we evaluate the equipoise treatment effect.

		 \begin{table}[h] 
			\caption{Selection functions and weights for clinical equipoise} \label{tab:wgts_summary.equip}
			\begin{center} 
				\begin{threeparttable}[]
					\begin{tabular}{cccccccccccccccccccccc} 
						\toprule
						&  &  \multicolumn{2}{c}{Weights} \\ \cmidrule(lr){3-4} 
						Name  & $g(x)$ & $w_1(x)$& $w_0(x)$ \\ \cmidrule(lr){1-4} 
						
						matching &  $u(x) $  &   $\min\left\{1,~{(1- e(x))}e(x)^{-1}\right\}$ & $\min\left\{1,~{e(x)}{(1- e(x))^{-1}}\right\}$\\				\addlinespace
						beta &  $b_{\nu}(x)$ &   $e(x)^{\nu-2}(1-e(x))^{\nu-1}$ &      $e(x)^{\nu-1}(1-e(x))^{\nu-2}$\\ \addlinespace
						overlap &  $b_{2}(x)$  &     $1-e(x)$  &      $e(x)$\\ \addlinespace
						\multirow{1}{*}{entropy} &  \multirow{1}{*}{$\xi(x)$}  &   ${\xi(x)}e(x)^{-1} $	&   $\xi(x)(1- e(x))^{-1}$\\
						\bottomrule
					\end{tabular} 
					\begin{tablenotes}
						\footnotesize
						\item $u(x)=\min\{e(x), 1-e(x)\};$ $~b_{\nu}(x)=\left\lbrace e(x)\left(1-e(x)\right)\right\rbrace^{\nu-1} $, $\nu\geq 2$;
						\item  $\xi(x)=-\left\{e(x)\ln(e(x))+ \left(1-e(x)\right)\ln\left(1-e(x)\right)\right\}$. 
					\end{tablenotes}
				\end{threeparttable}
			\end{center} 
		\end{table} 
		
		\subsection{A geometric interpretation}\label{sec:interpret}
		
		In the same vein as the importance sampling perspective, conditions \eqref{cond.1a}--\eqref{cond.3} specify the subpopulation of patients we target by using the selection functions for equipoise. 
		Let $\mathcal{S}_1(X)$  (resp. $\mathcal{S}_0(X)$) be the support of $X$ for the treatment (resp. control) group, i.e., the smallest closed set such as $P(X\in \mathcal{S}_z(X)|Z=z)=1$, for $z=0,1.$ Define the intersection $\mathcal{S}(X)=\mathcal{S}_0(X)\cap\mathcal{S}_1(X)$; every $x\in \mathcal{S}(X)$ belongs to the common support of the distribution of the covariates of participants in both groups. Therefore, $\mathcal{S}(X)$ represents the set of "matchable" participants. Ideally, to estimate ATE we expect to have $\mathcal{S}(X)=\mathcal{S}_0(X)=\mathcal{S}_1(X),$ which is not always guaranteed. When that is not realistic, we often fix our choice on the common support $\mathcal{S}(X)$ such that $\mathcal{S}(X)\subset \mathcal{S}_z(X),$ $z=0, 1$ where the estimand $\tau_{g}$ is well defined for some $g$.

		It is not always easy to specify  $\mathcal{S}(X)$ in terms of the different baseline variables. Nevertheless, suppose that $\mathcal{U}(X)$ is the topological complement of  the adherence of the set $\{X=x:g(x)=0\}$ where $g$ satisfies the conditions \eqref{cond.1a}--\eqref{cond.3}, then   $\mathcal{U}(X)$ is the above common support. Naturally, $\mathcal{U}(X)$ includes $\mathcal{S}(X)$ since, by Bayes theorem,  \begin{align}\label{eq:bayes_thm}
         f_{X|z}(x)=\left[ p^{-1} e(x)\right] ^z \left[ (1-p)^{-1} (1-e(x))\right] ^{z-1}f_X(x),~~ \text{where}~~f_{X|z}(x)=P(X=x|Z=z)~~\text{and}~~p=P(Z=z).
		\end{align} As a consequence, the  functions $g$ in Table \ref{tab:wgts_summary.equip} are good candidate functions to target the set of matchable participants. Hence, the above selection functions re-weights treatment and control groups as to potentially provide a large set of participants whose distributions $f_{X|z}(x)$ lead to adequate positivity.  Therefore,  as long as we can find a function $g$ and a related set $\mathcal{V}(X)$ such that $\mathcal{S}(X)\subset \mathcal{V}(X)\subset \mathcal{U}(X)$ and the influence of the points in $\mathcal{V}(X)\setminus \mathcal{S}(X)$ is minimal, then $\mathcal{V}(X)$ is a viable set to work with. We use  the terms "influence" and "minimal" loosely and we will consider the subpopulations targeted via the matching weights function of Li and Greene \cite{li2013weighting}, the overlap weights function of Li et al. \cite{li2018balancing}, and the above entropy function as good examples of such sets $\mathcal{V}(X)$, where we can estimate the treatment effect on a subpopulation of participants for whom positivity (and thus overlap of propensity scores) holds.
		
		Using a geometric description, Shafer and Kang \cite{schafer2008average} demonstrated that patients in the treatment (resp. control) group are localized inside concentric ellipses with the highest concentration of treated (resp. control) individuals found around the center $E(X|Z=1)$ (resp. $E(X|Z=0)$). The further away we move from these center, the more sparse the number of observation becomes. Furthermore, the overall covariate average, $E(X
		)$, lies along the line segment that connects  $E(X|Z=0)$ and $E(X|Z=1)$ with its position depending on the relative sizes of the two treatment groups. Since,  $E(Y(z)|X)$, $z=0, 1$ can be estimated most reliably for values of the covariates $X$ in the vicinity of $E(X|Z=z)$, better estimation the ATE depends on how well $E(Y(z)|X)$, $z=0, 1$ can be predicted near the overall mean $E(X)$. Therefore, when $g(x)=1$  the two distributions  $f_{X|z}(x)$ of the covariates are far apart in the sense that few individuals from both treatment groups are in the common region, the estimate of the ATE may require some extrapolation, \cite{schafer2008average} which may render the estimation of the ATE unstable and prone to bias. \cite{rubin1997estimating,dehejia1999causal} In Section \ref{sec:theory_example}, we demonstrate  how and when, for a given function $g,$ the related estimand is closer to an ATT, an ATC or an ATE depending on  the sample size of the treated participants and the variability of the propensity scores among the treatment groups.
		
		\subsection{Beta, matching, overlap, and Shannon's entropy weights: close connections}\label{sec:overlap.estimators}
%
	\subsubsection{The target population of the matching weights}\label{sec:matchwgts}
	The matching weights function was proposed by Li and Greene \cite{li2013weighting} as a so-called analogue to pair matching without replacement. However, it is not clear 
	whether the  estimand of interest is a deviation from the ATT or from the ATE. 	
	Although there are some similarities between the matching weight estimand and ATT (or ATC) estimand obtains via matching \cite{busso2014new},  we  demonstrate in this section that  the matching weights by their nature do not solely target these estimand. Instead, they target an ATE-like estimand using the subpopulation of patients for whom there is equipoise. 
	
	%

	The selection function for the matching weights of Li et al. \cite{li2013weighting} is the triangular distribution function $u(x)=\min\left(e(x), 1-e(x)\right)$. This corresponds to the weights  
	\allowdisplaybreaks\begin{align}
	w_{1}(x)&=\displaystyle u(x)e(x)^{-1}=\min\left\{1,~e(x)^{-1}(1-e(x))\right\} ~\text{and}~
	w_{0}(x)=\displaystyle u(x)(1-e(x))^{-1}=\min\{1, ~e(x)(1-e(x))^{-1}\}\label{eq:matc_ow} 
	\end{align} 
	Using the above equations and the indicator $\kappa_i=\mathds{1}({\{\widehat e(x_i)\leq 0.5\}})$,  the estimator $\widehat\tau_{g}$ becomes 
	\allowdisplaybreaks	\begin{align}
	\widehat\tau_{g}
	&=\displaystyle\sum_{i=1}^{N}\left[\displaystyle \frac{Z_i}{N_{\widehat w_1}} \min\left\{1,~\frac{(1-\widehat e(x_i))}{\widehat e(x_i)}\right\}- \frac{(1-Z_i)}{N_{\widehat w_0}}\min\left\{1,~\frac{\widehat e(x_i)}{(1-\widehat e(x_i))}\right\}\right] Y_i\label{wgt.atc}\nonumber \\
	&=\displaystyle\sum_{i=1}^{N} \left[ {\kappa_i}\left( \frac{Z_i}{N_{\widehat w_1}}-\frac{\widehat e(x_i)}{(1-\widehat e(x_i))}\frac{(1-Z_i)}{N_{\widehat w_0}}\right)+ (1-{\kappa_i})\left(\frac{(1-\widehat e(x_i))}{\widehat e(x_i)}\frac{Z_i}{N_{\widehat w_1}}-\frac{(1-Z_i)}{N_{\widehat w_0}}\right)\right]Y_i. 
	\end{align}
	
	We can easily recognize the weights inside the square brackets in \eqref{wgt.atc} . Whether a matching weight estimate differs from the one-to-one PS matching ATT estimate
	depends on how much treatment effects vary with the covariates X and on the range of values for the propensity scores $e_i(x).$
	When  $\widehat e_i(x)\leq 0.5$, the weights in \eqref{eq:matc_ow}  are proportional to the ATT weights  whereas these weights are proportional to the ATC weights whenever all $\widehat e_i(x)>0.5$. Thus, the above $\widehat\tau_{g}$ is a combination of the estimator of the ATT and that of the ATC: 
	it resembles an ATT estimator if all $e_i(x)\leq 0.5$  and an ATC estimator for  $e_i(x)> 0.5$. 
	Namely, whenever $e_i(x)\leq 0.5$, matching weight algorithm assigns a weight equal to 1 to all treated participants,  but weights control participants as to make them look like treated participants. However, when  $e_i(x)> 0.5$, the matching "weight" algorithm assigns to treated participants---who are usually overrepresented in this propensity score bracket---weights that make them look like those in the control group, just as we do when we estimate the ATC. Finally, when all $e_i(x)\approx 0.5$ (which is really unexpected in practice!), $\tau_g$ approximates simple mean outcome differences between the treatment groups. 
	

	In general, the estimator $\widehat\tau_g$ is a weighted average of the ATT and ATC. It does not target an ATT-like (or ATC-like) estimand  per se as one-on-one propensity score matching does.  Therefore, it does not correspond to an analog  of an estimator obtained through propensity score "pair matching", as stated by Li and Greene.\cite{li2013weighting}

	So what does the estimand $\tau_g$ evaluate in this case and what does it represent? Based on equation \eqref{wgt.atc}, the matching weight algorithm revolves around group representation in terms of the corresponding propensity scores. Usually, treated patients are underrepresented in the lower half of the propensity score where $\{e_i(x)\leq 0.5\}$ and overrepresented when $\{e_i(x)> 0.5\}$. Therefore, just like when we estimate the ATT, the matching weights \eqref{eq:matc_ow} assign a weight equal to 1 to all treated participants in this lower range of the propensity scores,  but weights control participants as to make them look like treated participants. Similarly, when  the propensity scores is greater than 0.5, the matching weight algorithm assigns to treated participants, who are overrepresented in this propensity score bracket, weights that make treated participants to look like those in the control group and assign to control participants a weight equals to 1, just as we do when we estimate the ATC. In sum, the  matching weights triage the participants by selecting only those for whom there is clinical equipoise and thus the underlying estimand is target the equipoise treatment effect, just like the overlap weights.  
	Our illustrative example in Section \ref{sec:illust} and the simulation results in Section \ref{sec:simulations} demonstrate these features of the matching weights.

	\subsubsection{Unexpected similarities}\label{sec:closeconx}
	The aforementioned connections between the matching weights and overlap weights from Section \ref{sec:matchwgts} or among the selection functions of Section \ref{sec:select_eqp}  are not the only unexpected similarities among their corresponding estimands.	It is worth mentioning that when  $e(x)\in [0.2, 0.8],$ for all $x$, the OW estimator is close to the ATE, even when the treatment effect is heterogeneous since the selection function $g(x)=e(x)(1-e(x))\in $   $[0.16, 0.25],$ i.e., nearly constant.	
	Moreover, despite their apparent differences, connections between the overlap weight and the matching weight selection functions can be established. Several simulation studies and data applications have shown that matching weights and overlap  weights yield similar point estimates for $\tau_g$ (see Li and Greene \cite{li2013weighting} and Mao et al. \cite{mao2018propensity}). This can be explained by the inequality 
	\allowdisplaybreaks	\begin{align}\label{eq:ow_ineq}
		\min\{t, 1-t\}\leq 2t(1-t)\leq 1/8+\min\{t, 1-t\}, ~~\text{for all }~t\in [0,1]
	\end{align} and by the fact that 
	\allowdisplaybreaks	\begin{align}\label{eq:ow_mw}
		t(1-t)=\displaystyle  {\min\{t, 1-t\}}\times {\max\{t, 1-t\}}, ~~\text{for all}~t\in [0,1].
	\end{align}
	Therefore, the results at the end of Section \ref{sec:matchwgts} also applies to overlap weights, i.e., the OW estimator is a combination of the ATT and the ATC. Its estimate is close to ATT when most propensity scores $e(x)$ are less than 0.5 and close to the ATC if a large number of  the propensity score $e(x)$ are greater than 0.5. This shed lights on the role of the matching weight estimand and, more generally, on all the estimand that based on selection function for whom there is clinical equipoise as we demonstrate in Section \ref{sec:theory_example}.
	
	As we mentioned, the overlap weight function is part of the family of beta functions $g(x)=b_\nu(x)=[e(x)(1-e(x))]^{\nu-1},$ for which $\nu=2$.  While some of these functions do not satisfy all the conditions \eqref{cond.1a}--\eqref{cond.3} of section \ref{sec:wgt_eqp} when $\nu \leq 1$,  those for which $\nu>1$ do. However, these functions are concave and strictly concave only when $1<\nu\leq 2$. When  $\nu$ gets larger than 2, these functions are no longer concave and not as roughly proportional to $e(x)(1-e(x))$ as we would expect, in order to obtain results that are similar to those from the overlap weight function. 

		Just as with the matching weight selection function, the maximum distance between the curves $t(1-t)$ and $-[t\ln(t)+(1-t)\ln(1-t)]$ is small and equal to $-1/4+\ln(2)\simeq 0.44.$ Case in point, using Taylor's theorem with Cauchy remainder term, we can show that for $t\in [0,1],$
	\begin{equation}\label{eq:taylor_owent}
		-\left[t\ln(t)+(1-t)\ln(1-t)\right]=t(1-t)\!\left(2+\frac{\rho-t}{2\rho^2}\right),~~\rho\in (0, 1).
	\end{equation}
	This implies that, in general, the point estimates from both the overlap weights and the entropy weights will also be very similar. In light of the results \eqref{eq:ow_mw} and \eqref{eq:taylor_owent}, we conclude that when the variances of the potential outcomes are homoscedastic, the overlap weight estimator is asymptotically more efficient than the matching weight, the entropy weight, and the beta weight estimators.\cite{li2018balancing} When there is heteroscedasticity, we do not guarantee such an efficiency, but expect the estimated variances from these estimators to be fairly similar or in favor of the overlap weights. This question is assessed in the simulation study in the context of finite samples.
	
	That selection functions  that are roughly proportional to $e(x)(1-e(x))$ will exhibit the same behavior as the overlap weight function is not fortuitous. In fact, the resemblance of the results from the matching, entropy, and overlap weights functions have been profusely demonstrated in classification and regression trees as well as in information theory where these functions are called, respectively, misclassification or Bayes error, cross-entropy or deviance, and Gini (impurity) index (see Breiman et al. \cite{breiman1984classification} or Hastie et al. \cite{hastie2009elements}) The last two functions are the most commonly-used and, for practical reasons, the Gini index is the most preferred when dealing with categorical categorical variables. \cite{breiman1984classification} Finally, we can easily establish the relationship between the beta weights and the other three weights. For the matching weights,  we just need to realize that the selection function $u(x)$ is also the so-called  triangular distribution function; we can use the relationship between the beta  and the triangular distributions to establish the connections between the beta and the matching weights. \cite{johnson1997triangular}

	What separates the overlap weight function from the beta functions when $\nu>2$ is that the latter group of functions move faster to 0 as $e(x)$ moves away from the mode 0.5. Most observations will have very small normalized weights $w_z(x)/N_{wz}$ and the non-negligible weights will be concentrated on a very few partipants, with a spike around $e(x)=0.5$.  This have two direct consequences in the estimation of the treatment effect using the beta weights when $\nu>2$: 
	(1) the treatment effect is mostly driven by few  observations with  propensity scores around 0.5; (2) its estimated variance will tend to be larger compared to the variance from the other methods because of the presence of few observations  with large weights (i.e., $e(x)\approx 0.5$) and a great number of observations with smaller weights, as shown in the illustrative example \ref{sec:illust}. 	While the selection function for $\nu>2$ seem to be a interesting choice when one needs to target participants with propensity scores around 0.5 and reduce bias, such an apparent advantage helps only in reducing bias when the treatment effect is constant. However, even then, it lacks precision and when the treatment effect is heterogeneous, it often leads to bias results.
	
	\subsection{Theoretical implications and an illustrative example} \label{sec:theory_example}
	\subsubsection{Theoretical implications}\label{sec:theory}
		The inter-connections between the overlap weights, matching weights \eqref{eq:ow_ineq}, \eqref{eq:ow_mw} and entropy weights \eqref{eq:taylor_owent} along with the properties of the entropy weights lead to interesting theoretical implications and attractive features of selection functions $g(x)$ for clinical equipoise and the corresponding weights $w_z(x)$, $z=0, 1$.
	
  First, it is well-known that	OW combine nicely two opposing-effect operations: minimizing the influence of extreme weights while improving the efficiency of the treatment effect estimator.\cite{crump2006moving,li2018balancing} Therefore, as seen in practice, the overlap weight, matching weight,  entropy weight and beta weight estimators achieve high degree of covariate balance for the variables included in the propensity score model, prevent loss of information, and provide efficient analysis results. Nevertheless, this important feature also  means that cautious is warranted  in selecting variables to  include in the propensity score model. The goal is to adequately include all the confounding variables on the treatment-outcome relationship and prognostic variables (predictors of the outcome), using subject-matter knowledge of the study design and the outcome under study.  Once such a judicious choice is made, these weights help us obviate the cyclical process of "propensity score tautology" where researchers repeatedly change the propensity score model specifications until they reach covariate balance.\cite{imai2014covariate}
  
  Second, the above results reveal also an important fact: unless all the participants are such that $e_i(x)\leq 0.5$ (resp. $e_i(x)\geq 0.5$), the overlap weight, matching weight,  and entropy weight estimators  do not, in general, estimate the ATT (or ATC) per se.
	Because these estimators belong to the family of balancing weight estimators targeting the subpopulation of participants for whom there is equipoise, their estimates will be, in general, different from the ATE, ATT and ATC estimates whenever the treatment effect is heterogeneous.


	As indicated in Section \ref{sec:interpret},  $E(Y(z)|X)$, $z=0, 1$ can be estimated most reliably for values of the covariates $X$ in the vicinities 
	of $E(X|Z=z)$. Suppose $f({x})$ is the marginal distribution of the covariates ${X}$ and consider  $f_{x|Z}(x|z)=P(X=x|Z=z)$ the density of the covariates $X$ in the treatment group $Z=z$. From \eqref{eq:bayes_thm}, we have 
    \begin{align*}
    	\displaystyle {f_{X|z=1}(x)f_{X|z=0}(x)}&=\frac{e(x)(1-e(x))}{p(1-p)}f_X(x),
    \end{align*}
	which indicates that the contribution  of each subpopulation defined by the covariates $X=x$ into  $\widehat \tau_{g}$
	mirrors directly  the covariate joint density in the corresponding treatment $Z=z$ group. 		
	Whether $\tau_{g}$ is close to ATT (resp. ATC) can be anticipated by checking $p=P(Z=1)$  and  $r(x)= {\sigma_{1}^2(x)}\sigma_{0}(x)^{-2}$, where $\sigma_{z}^2(x)=\text{Var}[e_i(x)|Z=z], ~z=0, 1$.   Indeed, the conditional density functions $f_{{\!\!e(x)|Z=z}}$ are related to the marginal density $f_{{\!\!e(x)}}$ via (see, for instance, Shaikh et al.\cite{shaikh2009specification})   
	\begin{align}
		f_{{\!e(x)|Z=1}}(u)&=\frac{u}{p}f_{{\!e(x)}}(u); ~~
		f_{{\!e(x)|Z=0}}(u)= \frac{(1-u)}{(1-p)}f_{_{\!\!e(x)}}(u);~~\text{and thus}~~
		{f_{{\!e(x)|Z=1}}(u)}=\frac{(1-p)u}{p(1-u)}f_{{\!e(x)|Z=0}}(u)
	\end{align}  
	which means the weights OW, MW, and EW assign to ATT (resp. ATC) decrease (resp.  increase) in $p$ and $r(x)$. Thus, if  $p$ or $r(x)$  is large, OW, MW, and EW are close to ATC.   However, for small $r(x)$ and $p$,  OW, MW, and EW approximate ATT. In particular, when $r(x)\approx 1,$ whether we obtain a result close to an ATT or an ATC depends solely on whether the prevalence  $p$ is low or high. Using Rubin's  rule-of-thumb, we consider   $r(x)\approx 1$ if its estimate  $\widehat r(x)= {\widehat\sigma_{1}^2(x)}\widehat\sigma_{0}(x)^{-2}$ satisfies $0.5\leq \widehat r(x)\leq 2$; otherwise $r(x)$ is considered significantly different from 1.\cite{rubin2001using}

		
	Third, the simple case where $r(x)\approx 1$ sheds light on the difference between the weights assignment mechanisms of the overlap weight, matching, and entropy weight estimators compared to that of the IPW to estimate their respective estimator, with respect to the proportion $p$. 
	Just as the overlap weight, matching, and entropy weight estimators are combinations of the ATT and ATC,  so is the ATE estimator since $\text{ATE}=E[Y(1)-Y(0)]=pATT + (1-p)ATC$. However, how ATE weights ATT vs. ATT with respect to $p$ is in opposite direction of what the matching, overlap, or entropy weights estimators do. 
	If the treatment effect is heterogeneous and $p$ is small, we can expect the ATE to be close to the ATC whereas the overlap weight, matching weight,  and entropy weight estimands will be close to an ATT. On the other hand, if  $p$ is relatively large, then ATT receives more weights with the ATE estimator while it is the ATC is weighted heavily with the overlap weight, matching weight, and entropy weight estimators. In short, when there are more treated than control participants  overlap weight, matching weight, and entropy weight estimators give more weights to the ATC whereas the ATE estimator put more weights on the ATT instead, and vice-versa. How small (or large) should the proportion $p$ be for the difference between ATE estimated via IPW and the equipoise estimates to be noticeable? While an definitive answer to this question needs further investigations, in the case where  $r(x)\approx 1$ we conjecture that difference will be noticeable whenever $p$ is outside of interval [0.2, 0.8]   since $p(1-p)$ is fairly constant for $p\in$ [0.2, 0.8].
	
	Of course, when the treatment effect is homogeneous and there is sufficient overlap in the underlying distribution of the covariates, we have $\text{ATE}=ATT=ATC$. In that case, the estimators obtained through the overlap weight, matching weight, or entropy weight framework or any matching algorithm will be asympotically similar. Furthermore, under small or moderate sample size, the  overlap weight, matching weight, or entropy weight estimators will be less bias and more efficient than a pairwise PS matching estimator since the former estimation methods target and include a large sample of matchable participants and are more efficient.
	\subsubsection{Illustrative example}\label{sec:illust}
	We  consider a simple example using  $\nu=11$ and $\nu=81$, $X=(X_1,X_2)$ with $X_1\sim N(2,2)$, $X_2\sim N(1,1)$. Let $Z\sim \text{Bernoulli}(e(X))$, where $e(X)=[1+\exp\left\lbrace -(\alpha_{0}+\alpha_{1}X_1+\alpha_{2}X_2)\right\rbrace ]^{-1}$, $Y(1)=2+X_1+X_2+2X_1^2+0.5X_2^2+\varepsilon_1,$ with $\varepsilon_1\sim N(0,2)$, and $Y(0)=X_1+X_2+\varepsilon_2,$ with $\varepsilon_2\sim N(0,1)$. 
	We consider three different proportions $p=P(Z=1)$ of treatment participants,  under a limited overlap of the distributions of the propensity scores, by carefully choosing values of $\alpha = (\alpha_1, \alpha_2, \alpha_3)$ as follows: $\alpha = (-2.8, 0.2, 0.8)$, with $p\approx 20\%$ (Scenario A);  $\alpha = (-1.6, 0.45, 0.6)$, with $p\approx 45\%-50\%$ (Scenario B); and $\alpha = (0.2, 0.8, 0.2)$, with $p\approx 80\%$ (Scenario C).
	The "true" estimands, provided in Table \ref{tab:true_est}, are calculated using a "superpopulation" of size $10^7$ participants, based on the true parameters coefficients, covariates, and models. 
	
			\begin{table}[h!]
			\begin{center} 
				\begin{threeparttable}
					\caption{True estimands} \label{tab:true_est}
					\begin{tabular}{cccccccccccccccccccccc} 
						\toprule
									 & \multicolumn{8}{c}{Estimand} \\\cmidrule{2-9}
				Scenario & ATE &  ATT &  ATC & OW & MW & EW & BW(11) & BW (81) \\\cmidrule(lr){1-9}
					   A &  18.99 &  24.66 &  17.57 & 22.46 & 23.85  & 21.66 & 32.84 & 37.04 \\
					   B &  18.99 &  25.35 &  13.12 & 17.53 & 17.02  & 17.81 & 15.29 & 14.73\\
					   C &  18.99 &  21.61 &  8.41  & 8.52  & 7.50   & 9.79  & 3.73  & 3.43\\
						\bottomrule
					\end{tabular} 
				\end{threeparttable}
			\end{center} 
		\end{table} 


	%
	
		To investigate the above theoretical implications,  we generated $M=1000$ data replicates to estimate the treatment effects under the 3 scenarios for $N=1000$.
	%
		\begin{table} 
		\caption{Estimated treatment effects} \label{tab:est_trt}
		\begin{center} 
			\begin{threeparttable}[]
				\begin{tabular}{cccccccccccccccccccccc} 
					\toprule
				  && \multicolumn{3}{c}{Scenario A} &&  \multicolumn{3}{c}{Scenario B} &&  \multicolumn{3}{c}{Scenario C}\\
				  && \multicolumn{3}{c}{($p=0.20; ~r(x)=1.74$)} &&  \multicolumn{3}{c}{($p=0.48; ~r(x)=1.05$)} &&  \multicolumn{3}{c}{($p=0.80; ~r(x)=0.4$)}\\\cmidrule(lr){2-13}
				      &&   Est. & ARB & SE &&  Est.  & ARB & SE && Est.  & ARB & SE   \\\cmidrule(lr){3-5}\cmidrule(lr){7-9}\cmidrule(lr){11-13}
				Crude &&  25.95 & 36.65 & 1.76 && 27.24  & 43.44 &  1.12 &&  24.05  & 26.65 & 0.81 \\
				  ATE &&  18.94 & 0.26  & 1.15 &&  18.95 & 0.21  &   0.70 &&  19.00  & 0.05 &  0.76\\
				  ATT &&  24.54 & 0.49  & 1.64 &&  25.31 & 0.16  & 1.08 &&  21.65 & 0.18  & 0.88\\
				  ATC &&  17.54 & 0.17  & 1.29 &&  13.08 & 0.30  &  0.74 &&  8.29 & 1.43  & 0.74\\
				  OW  &&  22.36 & 0.44  & 1.20 &&  17.49 & 0.23  & 0.67 &&  8.47 & 0.57  & 0.54 \\
				  MW  &&  23.71 & 0.59  & 1.41 &&  16.99 & 0.18  & 0.72 &&  7.47 & 0.40  & 0.48 \\
				  EW  &&  21.57 & 0.42  & 1.10 &&  17.77 & 0.22  & 0.65 &&  9.73 & 0.61  & 0.58\\
			  BW(11)  &&  32.69 & 0.46  & 3.98 &&  15.32 & 0.20  & 1.10 &&  3.78 & 1.34  & 0.40\\
			  BW(81)  &&  36.99 & 0.13  & 7.83 &&  14.80 & 0.48  & 1.50 &&  3.50 & 2.04  & 0.71\\

					\bottomrule
				\end{tabular} 
								\begin{tablenotes}
								\footnotesize
								\item ARB: $100 \times $absolute relative bias
								\end{tablenotes}
			\end{threeparttable}
		\end{center} 
	\end{table}

	As opposed to the crude estimate, all the other methods (IPW for ATE and ATT, overlap weights, matching weights, entropy weights, and beta weights (for $\nu=11$ and $\nu=81$)) yield estimates that are close to their true values (see Table \ref{tab:est_trt}). Both scenarios A and B, correspond to $0.5\leq \widehat r(x)\leq 2$. We see that when $p$ is only 0.20 (scenario A),  the estimates of treatament effect for OW, MW, and EW are close the ATT estimated value and are all larger than  the ATE estimate. When $p\approx 0.5$, as in scenario B, the true estimand and their estimated values for ATE, OW, MW, and EW are in the same ballpark, but apart from the ATT or ATC estimates. Finally, in scenario C, because $p\approx 0.8$ on average, the ATE estimate is near the estimate of the ATT, whereas the estimates of the treatment effect estimated OW, MW, and EW are close the ATC estimate of 8.29. Overall, this example makes the case for the above theoretical implicatons: (1) MW does not estimate the analogue to the one-on-one matching, but really a standalone estimand; (2) OW, MW, and EW are data-adaptive and naturally depend on the proportion of treated (or control) participants in the sample; (3) MW, OW, and EW are not some knock-off versions the ATE that we happen to estimate due to the lack of positivity. Rather, they are estimand that estimate the treatment effect on a specific subpopulation of participants,i.e., the subpopulation of participants for whom there is sufficient overlap (or clinical equipoise, whenever such a context is more appealing) as we have made clear in this paper.

	\section{Improving efficiency via augmentation}	\label{sec:augmentation}
 	\subsection{Brief overview}	
	While the methods to estimate $\tau_{g}$ we have considered so far are based on estimating both the selection function $g$ and the propensity score $e(X),$ we can also use the selection function $g$ and outcome regression models as well as the combination of the selection function, propensity score, and regression models estimations.  The latter is motivated by the need to improve efficiency, provided that we can estimate the propensity scores approximately very well. In addition, augmented estimators tend to be more robust to model misspecification.
	
	Similar to the estimation the ATE,  we leverage the augmentation framework of Robins et al. \cite{robins1994estimation} 
	to combine the estimations of the selection function, the propensity score, and the regression models and  improve the efficiency of the estimator $\widehat \tau_g.$ 
	
	We consider the conditional mean models $m_z(X)=E(Y|Z=z, X)$, $z=0,~1.$ We have $E[g(X)m_1(X)]=E\left[g(X)Y(1)\right]$ and $E[g(X)m_0(X)]=E\left[g(X)Y(0)\right]$ (see Appendix \ref{appendix2}).
	Hence, 
	\begin{align*}
	\frac{ E[g(X)\left\lbrace m_1(X)-m_0(X)\right\rbrace ]}{E[g(X)]} =\frac{ E[g(X)\tau(X)]}{E[g(X)]} =\tau_g, ~~\text{where}~\tau(X)=E[Y(1)-Y(0)|X].
	\end{align*}
	 If we knew the true conditional expectations $m_z(X)$ and propensity scores $e(X)$, then the estimator  
	 \begin{align}\label{eq:regr}
 \displaystyle \left( {\displaystyle\sum_{i=1}^{N}  g(x_i)}\right)^{-1}{\displaystyle\sum_{i=1}^{N}   g(x_i)\left\lbrace  m_1(x_i)-m_0(x_i)\right\rbrace}
	 \end{align}
 would have been a natural candidate for an unbiased and consistent estimator of the above $\tau_g.$ Unfortunately, we don't usually know the true conditional expectation $m_z(X)$ or the true propensity score $e(X)$. Nevertheless, in addition to the propensity score model, we can also postulate a parametric model $m_z(X; \alpha_z)$ for $E(Y|Z=z, X)$	 and derive an estimator of $\tau_g$. 
	 
	It is important to mention that the parametric models $m_z(X; \alpha_z)$ are estimated separately and only on the subset of participants for which $Z=z$. Then, using the estimates $\widehat \alpha_z$ of the regression coefficients $\alpha_z$, $z=0, 1,$ and the covariate values, we calculate for each  participant in the sample (treated and control) the fitted values $\widehat m_z(x_i)=m(x_i; \widehat\alpha_z)$ for $z=0$ and $z=1.$ \cite{kang2007demystifying}	
	Although not apparent, the conditional mean models $m_z(X)$ are well-defined only under  the positivity assumption since $m_z(X)=\displaystyle  \int yf_{Y|X,Z}(y|x,z)f(x)dydx$.  Otherwise, whenever $e(x)=0$ or 1, we are conditioning on null event.  \cite{tsiatis2007semiparametric,schafer2008average} Unfortunately, since estimating $m_z(X)$ does not require estimating propensity scores,  the lack of adequate positivity or limited overlap may go unnoticed.  Nevertheless, when there is a limited overlap in the distribution of covariates between the two treatment groups, the models $\widehat m_z(x_i)= m_z(x_i;\widehat \alpha_z)$ extrapolate the relationship between the covariates $X$ and the outcome $Y$, from the overlap region of the distributions to the isolated regions of the distributions---where only observations from one  treatment group is present or data are sparse.\cite{ma2020robust,leger2022causal}  However, the success of this extrapolation will depends strongly on how well we specify the model $ m_z(x_i;\widehat \alpha_z)$, especially if there are nonlinear continuous variables.\cite{ma2020robust}  Such an extrapolation in regions of no overlap, which may be different from the true relationship, can be perilous and drastically affect the efficiency of the estimator $\widehat \tau_g^{m}$. Moreover, it may lead to a biased and inefficient estimator of $\tau_g.$\cite{schafer2008average,khan2010irregular,canavire2021outliers}
	
	  	\subsection{Augmented weighted average treatment effects}
	 For ATE, ATT, and ATC, we can write  the corresponding selection function as   $g(x)=a+b e(x)$ with, respectively,  $ a=1, b=0$; $  a=0, b=1$; and $  a=1, b=-1$. The augmentation framework allows us, with	$\widehat{w}_z(x)=(a + b\widehat e(x))\widehat e_i(x)^{-z}(1-\widehat e_i(x))^{z-1}$, $z=0, 1$, to combine both estimators $\widehat \tau_g$ from \eqref{eq:li_est} and $\widehat m_z(x_i)$  into a more robust   defined as 
	 \begin{align}\label{tau_dr}
	 	\widehat \tau_g^{DR}&=\sum_{i=1}^{N}{Z_i\widehat  W_1(x_i)\left\lbrace Y_i- \widehat m_1(x_i)\right\rbrace} - \sum_{i=1}^{N}{(1-Z_i)\widehat W_0(x_i)\left\lbrace Y_i- \widehat m_0(x_i)\right\rbrace} +\sum_{i=1}^{N} \widehat  W_3(x_i)\left\lbrace  \widehat m_1(x_i) -  \widehat m_0(x_i)  \right\rbrace
	 \end{align}
	 where ${\widehat  W_z(x_i)=\widehat w_z(x_i)}\Big/{\displaystyle\sum_{i=1}^{N}\widehat  w_z(x_i)}$,  $z=0, 1$, and ${\widehat  W_3(x_i)=(a+bZ_i)}\Big/{\displaystyle\sum_{i=1}^{N} (a+bZ_i)}.$
 	
 	Under the SUTVA, consistency, positivity, and unconfoundness assumptions, $\widehat \tau_g^{DR}$ is doubly robust, i.e., it is consistent if  $\widehat e(X)$ or $\widehat m_z(X)$ is correctly specified, but not necessarily both. When the two models are correctly specified, $\widehat \tau_g^{DR}$  achieves asymptotic efficiency  (see Tsiatis \cite{tsiatis2007semiparametric} and Moodie et al. \cite{moodie2014g}). 
 		
 	Nevertheless, the estimator $\widehat \tau_g^{DR}$ has some drawbacks: (1) it is less efficient than $\widehat \tau_g$ if $e(X)$ is correctly specified, but $m_z(X)$ is misspecified or both the propensity score and regression models are not correctly specified;  \cite{kang2007demystifying, robins2007comment}  (2) its estimate is also sensitive to the lack of adequate positivity, especially if it is in addition to misspecified propensity score or regression models. The lack of adequate positivity affects directly estimation of the weights $\widehat{w}_z(x)$ and the extrapolation of $\widehat m_z(X)$ in the regions where the data are sparse.  	
 	Let $\widetilde e(X)$ and $\widetilde m_z(X)$ denote, respectively, possibly misspecified propensity and regression models and consider $\widetilde \tau(X)=\widetilde  m_1(X)-\widetilde  m_0(X)$.  Following Zhou et al., \cite{zhou2020propensity}  we show in Appendix \ref{appendix3_models_misspecification} that the asymptotic bias $\text{ABias} (\widehat \tau_{g}^{DR})$ of the estimator $\widehat \tau_g^{DR}$ is  	
	\allowdisplaybreaks\begin{align}\label{eq:bias_unwgted}
		&E\left[ \frac{(a+b \widetilde e(X))}{E\left[a+b \widetilde e(X)\right]}\widetilde  \tau(X) - \frac{(a+b  e(X))}{E\left[a+b  e(X)\right]} \tau(X)\right] + E\left[\frac{ e(X)(a+b \widetilde e(X))}{\widetilde e(X)}\right]^{-1}E\left[\frac{e(X)(a+b \widetilde e(X)) }{\widetilde e(X)}\left\lbrace  m_1(X)-\widetilde  m_1(X)\right\rbrace\right]\nonumber\\
		& - E\left[\frac{(1- e(X))(a+b \widetilde e(X))}{1-\widetilde e(X)}\right]^{-1}E\left[ \frac{(1- e(X))(a+b \widetilde e(X))}{1-\widetilde e(X)}\left\lbrace   m_0(X)-\widetilde  m_0(X)\right\rbrace\right],~~\text{where}~~. 
	\end{align}
	This clearly highlights the detrimental impact the combination of lack of adequate overlap and misspecification of the regression model can have on the final estimation of the treatment effect. 
	
 As we have argued, there would be little scientific basis to estimate the ATE in such a condition as the treatment effect would be based in part on extrapolation. Thus, the imperative to estimate the treatment effect on the subgroup of participants in the regions of overlap.
	For OW and MW, the current literature proposes the following augmented estimator \cite{li2013weighting,mao2018propensity}
	 \begin{align}\label{tau_augment}
	  \widehat \tau_g^{aug}&=\sum_{i=1}^{N}{Z_i\widehat  W_1(x_i)\left\lbrace Y_i- \widehat m_1(x_i)\right\rbrace} - \sum_{i=1}^{N}{(1-Z_i)\widehat W_0(x_i)\left\lbrace Y_i- \widehat m_0(x_i)\right\rbrace} +\sum_{i=1}^{N} \widehat  W_4(x_i)\left\lbrace  \widehat m_1(x_i) -  \widehat m_0(x_i)  \right\rbrace
	 \end{align}
	 where $\widehat  w_z(x_i)=\widehat  g(x_i)\widehat e_i(x)^{-z}(1-\widehat e_i(x))^{z-1}$,  ${\widehat  W_z(x_i)=\widehat w_z(x_i)}\Big/{\displaystyle\sum_{i=1}^{N}\widehat  w_z(x_i)}$, , $z=0, 1$, and $\widehat  W_4(x_i)= \widehat  g(x_i)\Big/ \displaystyle\sum_{i=1}^{N} \widehat  g(x_i)$.
	 
	 The estimator $\widehat \tau_g^{aug}$ is not affected by the lack of adequate positivity. When the propensity score and the regression models are correctly specified and $\widehat \beta$ and $\widehat \alpha_z$ are consistent, the estimator $\widehat \tau_g^{aug}$ is consistent and can achieve asymptotic efficiency, i.e., have the smallest asymptotic variance. Moreover, when the propensity score model  is correctly specified and $\widehat \beta$ is consistent, the estimator  $\widehat \tau_g^{aug}$ is also consistent.
	 However, since the selection function $g(x)$ depends on  $e(x)$, when the propensity score model $\widehat e(X)$ is misspecified, $\widehat \tau_g^{aug}$ is no longer consistent, even if the regression models $ \widehat m_z(X)$ are correctly specified. Nevertheless, it tends to yield satisfactory results,\cite{mao2018propensity} since it is commonly known that issues related to both propensity score misspecification and extrapolation of regression models matter most in area of limited covariate distrubitions overlap. \cite{kang2016practice,ju2018adaptive,li2018addressing,khan2010irregular,canavire2021outliers,ma2005robust,ma2020robust} Its asymptotic bias $\text{ABias} (\widehat \tau_{g}^{aug})$, given by 
   \allowdisplaybreaks\begin{align}\label{eq:bias_wgted}
   	\text{ABias} (\widehat \tau_{g}^{aug})&=E\left[\frac{\widetilde g(X)}{E\left[ \widetilde g(X)\right]} \widetilde  \tau(X) -\frac{g(X)}{E\left[g(X)\right]} \tau(X)\right]   +E\left[\frac{ e(X)\widetilde g(X)}{\widetilde e(X)}\right]^{-1}E\left[\frac{e(X)\widetilde  g(X) }{\widetilde e(X)}\left\lbrace  m_1(X)-\widetilde  m_1(X)\right\rbrace\right] \\
   	& - E\left[\frac{(1- e(X))\widetilde g(X)}{1-\widetilde e(X)}\right]^{-1}E\left[ \frac{(1- e(X))\widetilde g(X)}{1-\widetilde e(X)}\left\lbrace   m_0(X)-\widetilde  m_0(X)\right\rbrace\right],\nonumber
   \end{align}	
	showcases how the selection function $\widetilde g(X)$ helps tremendously in the presence of observations with $\widetilde e(X)\approx 0$ or 1 since, in this case,  ${e(X)\widetilde  g(X) }/{\widetilde e(X)}$ and ${(1-e(X))\widetilde  g(X) }/{(1-\widetilde e(X))}$ can be negligible. 
	
	Models misspecifications are more likely to lead a poor estimation of the regression parameters in these regions of limited overlap since there is a sparsity of treated (or control) participants and the performances of these model parameters are based on extrapolation of the misspecified models. \cite{barsky2002accounting,kang2007demystifying} The equation \eqref{eq:bias_wgted} indicates that the impact of the misspecifications on the asymptotic bias $\text{ABias} (\widehat \tau_{g}^{aug})$ is minor whenever $\widetilde e(X)\approx$ 0 or 1 since $\widetilde g(X)\left\lbrace \widetilde  m_1(X)-\widetilde  m_0(X)\right\rbrace$, $\frac{e(X)\widetilde  g(X) }{\widetilde e(X)}\left\lbrace  m_1(X)-\widetilde  m_1(X)\right\rbrace,$ and $\frac{(1- e(X))\widetilde g(X)}{1-\widetilde e(X)}\left\lbrace   m_0(X)-\widetilde  m_0(X)\right\rbrace$ are all negligible and do not contribute much to the asymptotic bias.  Nevertheless, the impact of such misspecification in the asymptotic bias $\text{ABias} (\widehat \tau_{g}^{DR})$ of the augmented IPW estimator can be exacerbated by the presence of observations with $ \widetilde e(X)\approx$ 0 or 1, as it can be seen in equation \eqref{eq:bias_unwgted}.  For instance, for the ATE estimator with $a=1$ and $b=0$ the asymptotic bias \eqref{eq:bias_unwgted} simplifies to
	\allowdisplaybreaks\begin{align*}
		& E\left[\left(E\left[\frac{ e(X)}{\widetilde e(X)}\right]^{-1}\frac{e(X) }{\widetilde e(X)}-1\right) \left\lbrace  m_1(X)-\widetilde  m_1(X)\right\rbrace\right]- E\left[\left(E\left[\frac{1- e(X)}{1-\widetilde e(X)}\right]^{-1}\frac{1- e(X)}{1-\widetilde e(X)}-1\right) \left\lbrace   m_0(X)-\widetilde  m_0(X)\right\rbrace\right]. 
	\end{align*}

	\subsection{Why bother with augmentations if they are not doubly robust?}\label{sec:why.bother}
	In light of the above remarks, it becomes clear that regardless of the function used to reach the target subpopulation of patients from whom there is clinical equipoise, their corresponding augmented estimators are not doubly robust. 	
	Nevertheless, augmentation provides some additional benefits. Clearly, when the propensity score model is correctly specified, it leads to a correct specification of the selection function $g(x).$ The augmentation yields an unbiased estimator, even when regression models are misspecified. Moreover, if the regression models are also correctly specified, it leads to an efficient estimator. Finally, in practice, since the weights for equipoise estimators smoothly (and gradually) downweight observations outside the overlap regions, augmented estimators tend to be more efficient, if our estimation of the propensity score model is correct or nearly correct.  
	
	It is well-known that regression models are effective when the covariate distributions between treated and control participants are similar. \cite{schafer2008average,rubin2001using} By smoothly removing observations outside of the overlap regions through OW, MW, or EW, the application regression models  ${m}_z(X)$ to provide reliable and efficient treatment effects in the population of patients for whom there is clinical equipoise---provided some of the models are correctly specified, if not all. For instance, with misspecified propensity score model $\widetilde e(X))$  but correctly specified outcome models ${m}_z(X)$, the above asymptotic bias $\text{ABias} (\widehat \tau_{g}^{aug})$ in \eqref{eq:bias_wgted} simplifies to 
	\allowdisplaybreaks\begin{align*}
	E\left[\left( \frac{\widetilde g(X)}{E\left[ \widetilde g(X)\right]} -\frac{g(X)}{E\left[g(X)\right]}\right)  \tau(X)\right]   	
	\end{align*}	
compared to the asymptotic bias $\text{ABias} (\widehat \tau_{g})$ of the H\'ajek-type estimator \eqref{eq:li_est}
\allowdisplaybreaks\begin{align*}
	E\left[\left( E\left[\frac{ e(X)\widetilde g(X)}{\widetilde e(X)}\right]^{-1}\frac{e(X) }{\widetilde e(X)}m_1(X) - E\left[\frac{(1- e(X))\widetilde g(X)}{1-\widetilde e(X)}\right]^{-1}\frac{1- e(X)}{1-\widetilde e(X)}m_0(X)\right)\widetilde  g(X)\right]  -E\left[ \frac{g(X)\tau(X)}{E\left[g(X)\right]} \right].
\end{align*}	
Their difference \allowdisplaybreaks\begin{align*}
	E\left[\left( E\left[\frac{ e(X)\widetilde g(X)}{\widetilde e(X)}\right]^{-1}\frac{e(X) }{\widetilde e(X)}m_1(X) - E\left[\frac{(1- e(X))\widetilde g(X)}{1-\widetilde e(X)}\right]^{-1}\frac{1- e(X)}{1-\widetilde e(X)}m_0(X)-\frac{\tau(X)}{E\left[ \widetilde g(X)\right]} \right)\widetilde  g(X)\right].
\end{align*}
indicates clearly the behavior of $\widehat \tau_{g}$ vis-\`a-vis $\widehat \tau_{g}^{aug}$. The biases are the same only when $\widetilde e(X)\approx e(X)$; in which case $\widetilde g(X)\approx g(X)$, i.e., the estimators $\widehat \tau_{g}$ and $\widehat \tau_{g}^{aug}$  $\longrightarrow \tau_g.$ Otherwise, if the distribution of the ratio  $\widetilde e(X)/ e(X)$ or $(1-\widetilde e(X))/ (1-e(X))$ is skewed, this may drastically impact the overall difference between the two estimators. Therefore, combining OW, MW, and EW with regression models via augmentation can be advantageous.
	
	Furthermore, whenever the regression models are correctly specified, $\widehat\tau_{g}^{aug}$ converges in distribution to $E\left[ \widetilde g(X) (Y(1)-Y(0)\right] /E\left[ \widetilde g(X)\right],$ for a misspecified propensity score model. Therefore, $\widehat\tau_{g}^{aug}$ still retains a causal interpretation provided the outcome regression models are correctly specified, even though it will be different from true estimate of  $\tau_{g}$. It can be leverage to test for the (causal) "sharp" null hypothesis of no causal treatment effect. \cite{vansteelandt2014regression}
	
	Admittedly, it is important to lay out the asymptotic properties of estimators as we have done so far. Nevertheless, it is equally important (if not more important) to investigate how different estimators behave under a finite sample size. In practice  (where sample sizes are finite), combining the propensity score model and the outcome regression models to estimate augmented or doubly-robust estimators is better than dealing with either \eqref{eq:li_est} or \eqref{eq:regr} alone. It is important to remember that the outcome models $\widehat m_z(X)$ play a preeminent role, compared to the propensity score model $\widehat e(X)$, in reducing bias and improving efficiency.
	
	   The estimators  \eqref{tau_augment} improve over their counterparts in  \eqref{eq:li_est} by fine-tuning the estimators to substantially reduce  bias when the propensity score is misspecified, but the regression models are correctly specified. Moreover, when both models are moderately misspecified, these estimators maintain a good performance compared to the IPW estimator---this matters the most, in practice, since it is not always easy to guarantee correctly specified propensity or regression models.\cite{kang2007demystifying} Our simulations studies indicate that these improvements compare more favorably that the augmented estimators \eqref{tau_augment}.
	
	In the advent of a misspecified propensity score model, it has been our experience that the augmentation framework lead to high efficiency as long as (1) a careful consideration is used in the estimation of both the outcome regression and propensity score models and (2) the covariate balance is reached with the estimated propensity scores.   Such an advantage was also mentioned by Mao et al. \cite{mao2018propensity} as well as by Tao and Fu. \cite{tao2019doubly} This happens because  weighting by $\widetilde W_4(X)$  in \eqref{tau_augment}  still improves the performances of the augmented estimators for the overlap, "matching", and entropy weights by smoothly downweighting the influence of extreme weights at the tails of the distribution of the propensity scores, while the IPW doubly robust estimator may in fact unduly exacerbates the impact of these influential weights.   
	
	Insight into this surprising result can be found if we investigate the asymptotic behaviors of the estimators $\widehat \tau_g^{aug}.$	The asymptotic variance of $\widehat \tau_g^{aug}$ is the variance of its influence function $F_{\tau_{g}}$ (see Section \ref{appendix3_general}), 
	i.e., for $\mathcal{W}=(X, Y(1), Y(0)),$ 
	\begin{align}\allowdisplaybreaks\label{eq:asympt_variance}
\text{AVar}(\widehat  \tau_{g}^{aug})= 	& \, E_{\mathcal{W}}\left[ \frac{g(X)^{2}}{E\left[g(X)\right]^2} \left\lbrace\frac{\left(Y(1)- \tau_{g}^1\right)^2}{e(X)}+\frac{\left(Y(0)- \tau_{g}^0\right)^2}{1-e(X)}\right\rbrace \right] \\
 &	+\left[ \frac{g(X)^{2}}{E\left[g(X)\right]^2} \left(\sqrt{\frac{1-e(X)}{e(X)}}\left(m_1(X)- \tau_{g}^1\right) +\sqrt{\frac{e(X)}{1-e(X)}}\left(m_0(X)- \tau_{g}^0\right)\right)^2\right].  \nonumber
	\end{align}

	Whenever $e(X)\approx$ 0 or 1, the contrast between the asymptotic variance of the IPW estimator (where $g(X)=1$) and those of the  OW, MW, and EW  estimators can be seen through the last equation \eqref{eq:asympt_variance}.  The  four terms   have a negligible impact on the asymptotic variance $\text{AVar}(\widehat  \tau_{g}^{aug})$, since  $\allowdisplaybreaks g(X)^2 e(X)^{-1}\approx 0 $, $g(X)^2 (1-e(X))^{-1}\approx 0 $, $g(X)^2 \sqrt{(1-e(X))e(X)^{-1}}\approx 0 $, and $g(X)^2 \sqrt{e(X)(1-e(X))^{-1}} \approx 0 $ in the regions of the  propensity score spectrum where $e(X)\approx$ 0 or 1. Hence, the information leading to the estimation of  $ \tau_{g}^{aug}$ comes primarily from regions with sufficient overlap, i.e., where there is a good amount  	 of common information on the treatment effect, carried out by both treatment groups.
	 
	However,  for the IPW estimator, $\allowdisplaybreaks g(X)=1 $ and thus the terms  $\allowdisplaybreaks  e(X)^{-1}$, $\allowdisplaybreaks  (1-e(X))^{-1} $,  $\allowdisplaybreaks 
	 \sqrt{(1-e(X))e(X)^{-1}}$, and $\allowdisplaybreaks \sqrt{e(X)(1-e(X))^{-1}}$ 
	may take large values when  $e(X)\approx$ 0 or 1, even if the true treatment effect is constant.  Moreover, when $e(X)\approx$ 0 or 1, estimations of the regressions models $m_z(X),$ $z=0, 1,$ are susceptible to extrapolation or model misspecification bias  in these regions. Therefore, variance estimation for the IPW estimator depends on whether there is sufficient positivity. When the positivity assumption is satisfied, the estimator is consistent and has an asympotic normal distribution \cite{tsiatis2007semiparametric}. Otherwise, the central limit theorem may no longer apply since the variance of the the IPW estimator can be infinite and thus its asymptotic distribution may not be normally distributed.

	\section{Large sample properties}	\label{sec:variace estimation}
	To estimate the treatment effect $\tau_{g}$, the propensity score or the outcome regression models are first estimated and then plugged into via the estimator \eqref{eq:li_est} or \eqref{tau_augment}. As outlined in the previous section, the uncertainty inherent to the estimation of the propensity score or regression models sips through the derivation of the different estimators. In estimating their variances, we should account for the fact that these models have been estimated. Following the approach of Lunceford and Davidian, \cite{lunceford2004stratification} we can derive the asymptotic distribution of the estimators $\widehat{\tau}_{g}$, $\widehat\tau_{g}^{aug}$, and $\widehat\tau_{g,\,\text{REF}}^{aug}$  and subsequently estimate their large-sample variances, using the standard M-estimation theory. This allows us to incorporate the uncertainty associated with the estimation of the propensity scores and derive an empirical sandwich variance estimator by replacing all expected values by their corresponding empirical averages (see Stefanski and Boos \cite{stefanski2002calculus} or Wooldridge \cite{wooldridge2002inverse}). In this section, we show  how this can be done for $\widehat{\tau}_{g}$. The large-sample variances of  $\widehat\tau_{g}^{aug}$ and $\widehat\tau_{\mathds{_G}}$ can be derived similarly. We give the  detailed proofs for these estimators  in the Appendix \ref{appendix4}.
	
	To derive the large-sample variances, we  can view the estimator of interest (i.e., ${\tau}_{g}$, $\tau_{g}^{aug}$, or $\tau_{g,\,\text{REF}}^{aug}$) as linear combination $\tau_{g}^{aug} = c'\theta$ of the components of a vector $\theta$, for some $c=(c_1, c_2, \dots, c_J)'$. Using the solution $\widehat{\theta}$ 
	to an estimating equation $\displaystyle \displaystyle\sum_{i=1}^{N} \Psi_\theta({X}_i, Z_i, Y_i; \theta)=0$, with respect to ${\theta}$, we can determine the asymptotic behavior of the  estimator of interest. 	

 First, since $E[\Psi_\theta({X}_i, Z_i, Y_i; \theta)]=0$, this implies that $\widehat\theta\longrightarrow \theta$ in probability, as $N\longrightarrow \infty$, under some  regularity conditions. \cite{stefanski2002calculus} Therefore, by Slutsky's theorem, the estimator will also be consistent. 	 
	 Moreover,  $\sqrt{N}(\widehat{\theta}-{\theta})$ converges in distribution to the normal distribution $N({0}, \Sigma({\theta}))$ where $ \Sigma({\theta})=A(\theta)^{-1}B(\theta)\{A(\theta)'\}^{-1}$. 
	Therefore, a consistent estimator of $\Sigma({\theta})$ is then $\widehat \Sigma({\widehat\theta})=A_N(\widehat\theta)^{-1}B_N(\widehat\theta)\{A_N(\widehat\theta)'\}^{-1}$, where 
	\begin{align*}
	A_N(\widehat{\theta})&=N^{-1}\displaystyle\sum_{i=1}^{N}\left[ - \frac{\partial}{\partial{\theta'}}\Psi_\theta({X}_i, Z_i, Y_i)\right]_{\theta={\widehat\theta}}~;~~ A(\theta)=\lim_{N\rightarrow\infty}A_N(\widehat\theta)\\
	B_N(\widehat{\theta})&= N^{-1}\displaystyle\sum_{i=1}^{N}\Psi_\theta({X}_i, Z_i, Y_i)\Psi_\theta({X}_i, Z_i, Y_i)'\big|_{\theta=\widehat\theta}; ~~ B(\theta)=\lim_{N\rightarrow\infty}B_N(\widehat\theta)
	\end{align*} 
	 From  $ \widehat\Sigma({\widehat\theta})$, we derive  an estimator of the variance of the estimator of interest   as 
	\begin{equation}\label{eq:var.sandwich}
	\widehat{Var}({\widehat{\tau}_g})=N^{-1}\left[c'\widehat\Sigma({\widehat\theta})c\right].
	\end{equation} 
	For example, for  $\tau_{g}^{aug} =c'\theta_{aug}= \tau_{1g}^{m}- \tau_{0g}^{m}+ \mu_{1g}- \mu_{0g}$, we use the parameter vector $\theta = \theta_{aug}=(\beta',\alpha_1',  \alpha_0', \tau_{1g}^{m}, \tau_{0g}^{m},  \mu_{1g}, \mu_{0g})'$ and $c= (0,0,0,1,-1,1,-1)'$. Thus,  $\widehat\theta_{aug}=(\widehat\beta', \widehat\alpha_1',  \widehat\alpha_0', \widehat\tau_{1g}^{m}, \widehat\tau_{0g}^{m},\widehat\mu_{1g}, \widehat\mu_{0g})'$, is then the  solution to the estimating equation 
		\allowdisplaybreaks \begin{align*}
			\displaystyle \displaystyle\sum_{i=1}^{N}  \Psi_{\theta_{aug}}({X}_i, Z_i, Y_i)&
						= \displaystyle\sum_{i=1}^{N} 
			\begin{bmatrix}
				\psi_{\beta}({X}_i, Z_i)\\
				Z_i\psi_{\alpha_1}({X}_i, Y_i)\\
				(1-Z_i)\psi_{\alpha_0}({X}_i, Y_i)\\
				g(X_i)\{m_1(X_i)-\tau_{1g}^m\}\\
				g(X_i)\{m_0(X_i)-\tau_{0g}^m\}\\
				Z_iw_1({X}_i)(Y_i-m_1(X_i)-\mu_{1g})\\
				(1-Z_i)w_0({X}_i)(Y_i-m_0(X_i)-\mu_{0g})\\
			\end{bmatrix} =0
		\end{align*}
	 with respect to $\theta_{aug}=(\beta',\alpha_1',  \alpha_0', \tau_{1g}^{m}, \tau_{0g}^{m},  \mu_{1g}, \mu_{0g})',$ where  $\widehat{\tau}_{zg}= \displaystyle \displaystyle\sum_{i=1}^{N}\widehat W_4(x_i) \widehat m_z(x_i)$ and $\widehat{\mu}_{zg}= \displaystyle \displaystyle\sum_{i=1}^{N}Z_i^z(1-Z_i)^{1-z}\widehat W_z(x_i) \{Y_i-\widehat m_z(x_i)\}$,  for  $z=0,1.$ The functions	$\psi_{\beta}({X}_i, Z_i),$ and $\psi_{\alpha_{z}}({X}_i, Z_i, Y_i)$ are, respectively, the score functions of the propensity score and regression models, which we use to solve for  $\beta$ and $\alpha_z$. 
	 
	We have shown in the Appendix \ref{appendix4} how to derive estimators of these sandwich variances for the estimators ${\tau}_{g}$, $\tau_{g}^{aug}$, and $\tau_{g,\,\text{REF}}^{aug}$ when the propensity score is estimated by a logistic regression model and the outcome models $m_z(x)$ are linear regression models.
	Among the regularity conditions  (required to apply M-estimation theory), we need to have a smooth selection function $g$.  	Because the selection function  $u(x)=\min\left(e(x), 1-e(x)\right)$ for the "matching" weights is not differentiable at $e(x)=0.5$, Li et al. \cite{li2013weighting} use the smooth version $\widetilde u(x)$, which consists in replacing the portion of $u(x)$ around  $e(x)=0.5$ by a cubic polynomial to satisfy the regularity conditions for the sandwich variance estimation.

	Alternatively, when possible, the variance of $\widehat{\tau}_g$ can be estimated using a nonparametric bootstrap estimator. \cite{hastie2009elements} This is particularly important when the sample size $N$ is small or moderate, when one wants to estimate a confidence interval that does not rely on the asymptotic normality assumption, or when there is an inherent dependence between some of the observations such as in stratified, clustered, hierarchical or multilevel design. \cite{cheng2013cluster} In the latter case, strata or clusters of observations are resampled with repetition to preserve the correlation structure between dependent (or related) observations
	in each bootstrap sample. 
	
	Note,  however, that the bootstrap technique can become extremely time-consuming when performing numerical simulations, analyzing a large healthcare claims database or dealing with large electronic health records. Furthermore, for the inverse probability weight (IPW) estimator, standard bootstrap methods are known to fail when there is lack of positivity  as resampling with repetition can exacerbate issues related to propensity model convergence. \cite{romano1999subsampling,khan2010irregular}  Dealing with complex data structures in such a context and using standard bootstrap make inference more challenging, highly unreliable (possible high magnitude of data variations across bootstrap replicates), less optimal, and prone to errors.

	\section{Monte Carlo simulations}\label{sec:simulations}
	To assess the different estimators and evaluate the finite-sample performance, we considered a number of simulation studies based on a variety of settings via two specific data generating processes (DGPs). Our first DGP followed the design presented by  Li and Greene \cite{li2013weighting} under both homogeneous and heterogeneous treatment effect to illustrate the role of overlap weights as well as the similarities, in point estimates, between the different estimates compared to those from the matching weights, the Shannon's entropy weights and the beta weights. The chosen data generating process also helps  highlight the issues inherent to the lack of adequate positivity associated with the inverse probability weighting estimators, the advantage of the equipoise estimators via augmentation, and the efficient of the overlap weights vis-\`a-vis all the other weights, especially the beta weights. 
	
	The second DGP is based on the design developed by Li et Li \cite{li2018addressing}  and focuses on the impact of the prevalence of treatment.  The goal is to investigate the joint impact of random and structural violations of the positivity assumption while articulating the flexibility of overlap, matching, and Shannon's entropy weights and the limits of beta weights. Throughout the simulations, we calculated the true estimands for IPW, OW, MW, EW, and BW ($\nu=11$ and $81$) under heterogeneous treatment effect  using "super-populations" of size $10^7$ units, based on the true parameter coefficients, covariates, and models. For both DGPs, we run the simulations using 1000 replications for each scenario and we chose the sample size to be $N=2000$, to evaluate the finite-sample performance of the different estimators using the following measures: absolute relative bias percentage ARB, standard deviation (SD), standard error (SE), root mean squared error (RMSE) as well as the 95\% coverage probability (CP).
	To be concise, we only present  the first DGP and its results in this manuscript; the second DGP and the corresponding results are reported in the Appendix \ref{sec:additional_simulations}.   
%

	\subsection{First simulations}
	\subsubsection{Data generating process (DGP)}
	We  generated the covariates $\boldsymbol X=(X_1, \dots, X_4)$ and the treatment indicator $Z$ following the DGP of Li and Greene \cite{li2013weighting}, with $X_4 \sim Ber(0.5)$, $X_3\sim Ber(0.4+0.2X_4)$,
	\begin{flalign*}
		&\begin{pmatrix} X_{1}\\
			X_{2}
		\end{pmatrix}\! \sim  N\!\begin{bmatrix}\begin{pmatrix}
				X_{4}-X_{3}+0.5X_{3} X_{4}\\
				-X_{4}+X_{3}+X_{3} X_{4}
			\end{pmatrix}, \begin{pmatrix}
				2-X_{3} & 0.25(1+X_{3})\\
				0.25(1+X_{3}) & 2-X_{3}				
		\end{pmatrix} 	\end{bmatrix},
	\end{flalign*}
	and  $Z\sim Ber(e(\boldsymbol X))$, 	where $e(\boldsymbol X)=[1+\exp\{-(\beta_{0}+\beta_{1}X_{1}+\dots +\beta_{4}X_{4})\}]^{-1}.$
	
\begin{figure}[!htp]
		\begin{center}
			\includegraphics[trim=20 12 20 35, clip, width=0.9\linewidth]{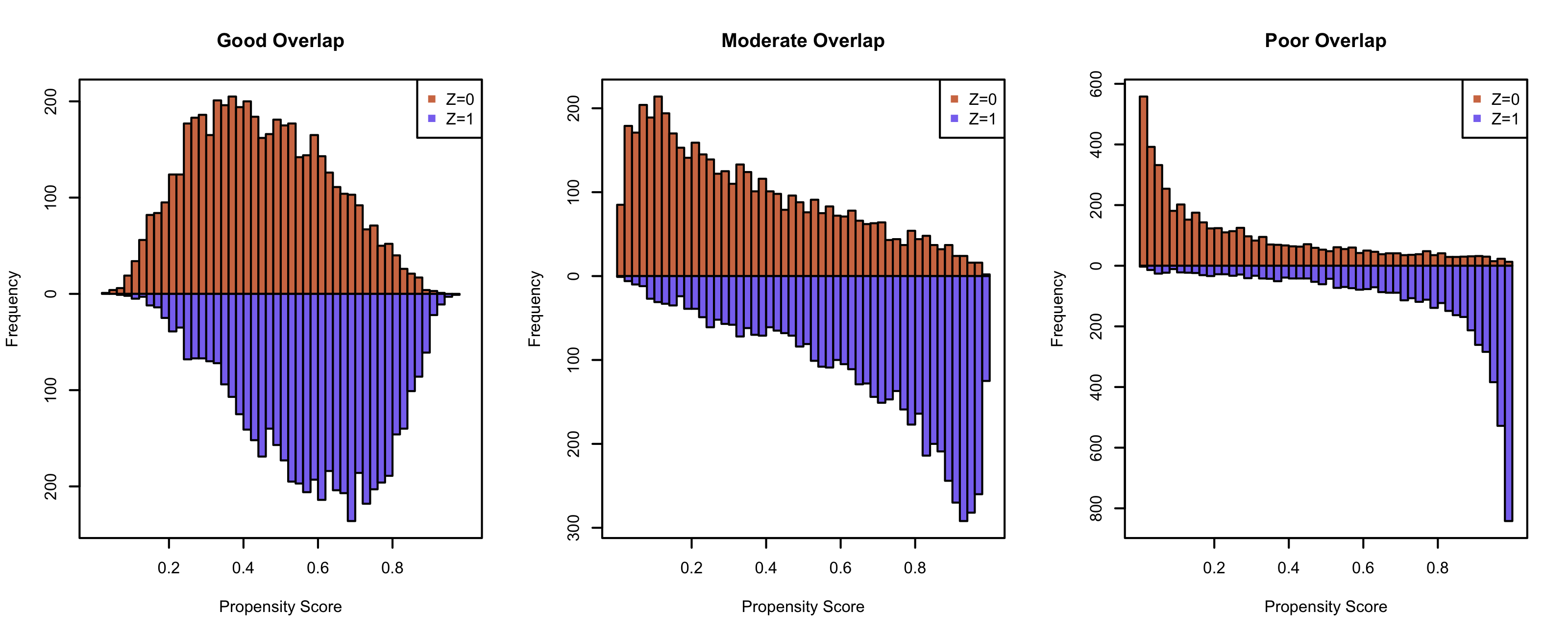}
		\end{center}
		\caption{Distribution of propensity scores---with good (left), moderate (middle), and poor (right) overlap. 
			\label{Mao_Overlap}}
	\end{figure}

	We considered, respectively, $(\beta_{0},\beta_{1}, \dots, $ $\beta_{4})=$ $\allowdisplaybreaks (-0.5,$ $0.3,$ $0.4,$ $0.4,$ $ 0.4)$, $(-1,$ $0.6,$ $0.8,$ $0.8,$ $0.8)$, and $(-1.5,$ $0.9,$ $1.2,$ $1.2,$ $1.2)$ for good, moderate, and poor overlap of PS distributions (with the average $p = $51.37\%, 53.05\%, and 52.77\%, respectively), as shown in Figure \ref{Mao_Overlap}. 
	The first plot shows a good overlap of the distributions of the propensity scores between the two treatment groups, the middle graph indicates a case where the overlap is moderate while the last graph shows a poor overlap, i.e., there is a strong separation in estimated propensity scores between the treated and control participants. Finally, we generated  the outcome $Y$  from the linear model $Y=0.5+\Delta Z+X_1+0.6X_2+2.2X_3+1.2X_4+\varepsilon$, where $\varepsilon\sim N(0,1)$ and $\Delta$ is the individual treatment effect. 
	We choose, respectively, $\Delta=3$ (homogeneous treatment effect) and  $-12e(\boldsymbol X)^2+12e(\boldsymbol X)+3$  (heterogeneous treatment effects), to compare the performance of the aforementioned methods. 
	
	We first assessed the  H\'ajek estimators $\widehat\tau_{g}$ \eqref{eq:li_est} of the weighted average treatment effect $\tau_{g}$ \eqref{eq:G-estimand}.  Then, we evaluated the performance of the augmented estimators $\widehat \tau_{g}^{aug}$ when both the propensity score and regression models are correctly specified and when at least one of them is misspecified. We misspecified the propensity score or the outcome model by omitting the covariate $X_1$ from the original correct model. 

	\begin{table}[h]
		\centering
				\begin{threeparttable} 
		\caption{Treatment effect estimation (without augmentation).\label{onlyweightingHo} { }} 
		\begin{tabular}{rrccccccccccccccccccccccccccccccccccccccccccccc}
			\toprule
				 &  & \multicolumn{6}{c}{Homogeneous treatment effect} &  &  \multicolumn{6}{c}{Heterogeneous treatment effect}\\\cmidrule{3-8} \cmidrule{10-15} 
			Weight & Overlap & True & ARB & RMSE & SD & SE & CP &  & True & ARB & RMSE & SD & SE & CP\\ 
			\midrule
		   IPW & Good & 3.00 & 0.07 & 6.54 & 6.54 & 6.39 & 0.94 & & 5.57 & 0.02 & 7.08 & 7.08 & 7.04 & 0.95 \\ 
			OW & Good & 3.00 & 0.05 & 4.72 & 4.72 & 4.83 & 0.95 & & 5.65 & 0.04 & 4.90 & 4.90 & 5.06 & 0.95\\ 
			MW & Good & 3.00 & 0.05 & 4.86 & 4.86 & 5.00 & 0.95 & & 5.70 & 0.06 & 5.07 & 5.07 & 5.14 & 0.94\\ 
			EW & Good & 3.00 & 0.05 & 4.80 & 4.80 & 4.89 & 0.94 & & 5.63 & 0.04 & 5.00 & 4.99 & 5.18 & 0.95\\ 
			BW(11) & Good & 3.00 & 0.08 & 7.29 & 7.29 & 7.40 & 0.96 & & 5.88 & 0.08 & 7.67 & 7.66 & 7.51 & 0.95 \\ 
			BW(81) & Good & 3.00 & 0.00 & 10.80 & 10.81 & 12.40 & 0.98 & &5.98 & 0.14 & 11.49 & 11.47 & 12.41 & 0.97 \\ \addlinespace
			
			IPW & Mod & 3.00 & 0.05 & 21.53 & 21.54 & 18.08 & 0.88 & & 4.94 & 0.10 & 24.78 & 24.79 & 20.99 & 0.87 \\ 
			OW & Mod & 3.00 & 0.06 & 5.55 & 5.55 & 5.56 & 0.95 & & 5.33 & 0.07 & 6.56 & 6.55 & 6.38 & 0.94 \\ 
			MW & Mod & 3.00 & 0.10 & 5.73 & 5.72 & 5.76 & 0.96 & & 5.44 & 0.04 & 6.46 & 6.46 & 6.35 & 0.95 \\ 
			EW & Mod & 3.00 & 0.06 & 5.89 & 5.89 & 5.83 & 0.94 & & 5.26 & 0.07 & 7.05 & 7.04 & 6.87 & 0.94 \\  
			BW(11) & Mod & 3.00 & 0.28 & 9.20 & 9.16 & 9.38 & 0.95 & &  5.86 & 0.07 & 9.18 & 9.18 & 9.51 & 0.96 \\ 
			BW(81) & Mod & 3.00 & 0.36 & 15.27 & 15.24 & 17.42 & 0.97 & &  5.98 & 0.09 & 15.19 & 15.18 & 17.51 & 0.98 \\ \addlinespace
			
			IPW & Poor & 3.00 & 1.28 & 52.00 & 51.89 & 36.69 & 0.77 & &  4.49 & 1.05 & 58.47 & 58.31 & 42.23 & 0.77 \\ 
			OW & Poor & 3.00 & 0.07 & 6.23 & 6.23 & 6.35 & 0.95 & &  5.18 & 0.03 & 7.40 & 7.41 & 7.64 & 0.96 \\  
			MW & Poor & 3.00 & 0.05 & 6.34 & 6.34 & 6.58 & 0.96 & & 5.32 & 0.06 & 7.33 & 7.32 & 7.56 & 0.95 \\  
			EW & Poor & 3.00 & 0.08 & 6.73 & 6.73 & 6.77 & 0.95 & & 5.07 & 0.01 & 8.05 & 8.05 & 8.38 & 0.96 \\ 
			BW(11) & Poor & 3.00 & 0.07 & 10.30 & 10.31 & 11.21 & 0.97& &  5.86 & 0.17 & 10.96 & 10.92 & 11.39 & 0.96 \\  
			BW(81) & Poor & 3.00 & 0.05 & 18.15 & 18.16 & 22.01 & 0.97& & 5.98 & 0.14 & 18.45 & 18.44 & 21.93 & 0.98 \\ 
			
			\bottomrule
		\end{tabular}
	
		\begin{tablenotes}\footnotesize		
		\item  These  results are based on 1000 repetitions of the simulated data. Mod: Moderate; BW($\nu$): BW with $\nu=11$ and $81$; ARB: absolute relative bias$\times 100$;
		\item   RMSE: root mean-squared error$\times 100$; SD:empirical standard deviation$\times 100$; SE: average estimated standard error$\times 100$; CP: 95\%  coverage probability. 
	\end{tablenotes}
			\end{threeparttable} 
	\end{table}
	
\subsubsection{Simulation results}
The treatment effect estimation using the propensity score weighting methods under comparison are shown in Table \ref{onlyweightingHo}, for the Hajek-type estimators \eqref{eq:li_est}. As expected, when the overlap of the distributions of the propensity scores is good, all estimators provide estimates with low bias and small standard errors. The coverage probability (CP) is between 0.94 and 0.96, except for BW(81) for which the CP is greater than or equal to 0.97. Nevertheless, BW(81) have a lower bias compared to BW (11). The differences among estimators are more noticeable instead in the variability of the estimates, with OW, MW, and EW being the most efficient (SE $\approx 5$), followed by  IPW, BW(11), and finally BW(81), for which SE $\approx 12.40$. 

Under moderate overlap,  IPW, OW, MW and EW have mostly smaller biases compared to BW(11) and BW (11). OW, MW and EW have the smaller variability ($SE< 6$) and adequate CPs, while BW (11) has SEs around 9 and CPs below or equal to 0.96 and BW(81) has  SEs around 17.40 and CP $> 96$. IPW has a lower ARB than BW(11) and BW (11) with a constant  treatment, but not when the treatment is heterogeneous. Both  BW(11) and BW (11) are more efficient than IPW. In addition, the CPs of IPW are below the 0.90.
The above trend in the performance of the estimators remain under poor overlap where the ARB of IPW is more than 16 times higher than those of the other estimators and its CP is now equal to 0.77. The only exception is that BW(81) is less bias than BW(11). Unfortunately, the gain in bias for BW(81) does not make up the loss in efficiency, as seen throughout all the scenarios we considered since its SEs are almost twice has much than those from BW(11). 

Overall, under moderate and poor overlap, IPW is the most sensitive estimator due to the lack of adequate positivity.  The corresponding relative bias and standard error of IPW estimates increase rapidly and can be much higher than those of the other methods as lack of overlap worsens from moderate to poor. In contrast, the family of balancing weights that target clinical equipoise perform stably, even under poor overlap, yielding treatment effect estimates with small biases and variances. However, the selected BW estimators (using the beta function with a relatively large $\nu=11$ and $81$) are shown to be less accurate and less efficient compared to OW, MW, and EW, which are special cases or approximations of the beta function with smaller $\nu$ $(1\leq\nu\leq4)$. The higher the value of $\nu$ (i.e., far away from 4), the less efficient the BW($\nu$) estimator.

The results for OW, MW, and EW are fairly consistent across treatment effects and different level of lack of positivity. Compared to IPW, the gains in RMSE are often substantial. For instance, in moderate  (resp. poor) overlap the RMSEs of OW, MW, and EW are about 3.7 (resp. 7) times lower than that of IPW, for both homogeneous and heterogeneous treatment effects. Oddly enough, even the worst case of beta weights (with  $\nu =81$) was doing better in terms of RMSE than the IPW under moderate and poor overlap.	
\begin{table}[!hbp]
	\begin{threeparttable} 
		\begin{center}
			\caption{Homogeneous treatment effect estimation (with and without model misspecification).\label{Mao_Ho} { }}
			\begin{tabular}{rrrrrrrrrrrrrrrrrrrrrrrrrrrrrrrrrrrrrrrrrrrrrrrrrrrrrrrrrrrrrrrrrrrrrrrrrrrrrrrrrrrrrrrr}
				\toprule
				&  &  & \multicolumn{10}{c}{Model misspecification} \\\cmidrule(lr){4-13}
				&  &  & \multicolumn{5}{c}{None }& \multicolumn{5}{c}{PS model misspecified} \\ \cmidrule(lr){4-8}\cmidrule(lr){9-13}
				Weight & Overlap & True & ARB & RMSE & SD & SE & CP & ARB & RMSE & SD & SE & CP \\
				\cmidrule(lr){1-13}

				IPW & Good & 3.00 & 0.05 & 5.00 & 5.00 & 4.95 & 0.96 & 0.05 & 4.93 & 4.93 & 4.90 & 0.96 \\ 
				OW & Good & 3.00 & 0.04 & 4.83 & 4.83 & 4.83 & 0.95 & 0.05 & 4.83 & 4.83 & 4.83 & 0.95 \\ 
				MW & Good & 3.00 & 0.04 & 4.86 & 4.86 & 4.89 & 0.95 & 0.04 & 4.83 & 4.83 & 4.88 & 0.95 \\ 
				EW & Good & 3.00 & 0.05 & 4.84 & 4.84 & 4.83 & 0.95 & 0.05 & 4.84 & 4.84 & 4.83 & 0.95 \\ 
				BW(11) & Good & 3.00 & 0.02 & 5.79 & 5.79 & 5.82 & 0.95 & 0.02 & 5.49 & 5.49 & 5.66 & 0.96 \\ 
				BW(81) & Good & 3.00 & 0.03 & 8.99 & 9.00 & 9.12 & 0.95 & 0.09 & 8.47 & 8.47 & 8.75 & 0.95 \\ \addlinespace
				
				IPW & Mod & 3.00 & 0.01 & 7.24 & 7.25 & 6.93 & 0.94 & 0.02 & 6.41 & 6.42 & 6.18 & 0.95 \\ 
				OW & Mod & 3.00 & 0.04 & 5.58 & 5.58 & 5.56 & 0.95 & 0.03 & 5.55 & 5.55 & 5.52 & 0.95 \\ 
				MW & Mod & 3.00 & 0.03 & 5.65 & 5.65 & 5.66 & 0.95 & 0.03 & 5.61 & 5.61 & 5.61 & 0.96 \\ 
				EW & Mod & 3.00 & 0.03 & 5.64 & 5.64 & 5.58 & 0.96 & 0.03 & 5.59 & 5.60 & 5.54 & 0.95 \\ 
				BW(11) & Mod & 3.00 & -0.01 & 7.80 & 7.81 & 7.77 & 0.95 & 0.00 & 7.33 & 7.33 & 7.37 & 0.95 \\ 
				BW(81) & Mod & 3.00 & -0.04 & 13.18 & 13.18 & 12.74 & 0.94 & 0.11 & 11.68 & 11.68 & 11.87 & 0.94 \\ \addlinespace
				
				IPW & Poor & 3.00 & 0.01 & 12.69 & 12.69 & 10.35 & 0.93 & -0.01 & 8.25 & 8.25 & 8.02 & 0.96 \\ 
				OW & Poor & 3.00 & 0.05 & 6.32 & 6.32 & 6.35 & 0.95 & 0.06 & 6.16 & 6.16 & 6.19 & 0.95 \\ 
				MW & Poor & 3.00 & 0.03 & 6.43 & 6.44 & 6.48 & 0.96 & 0.06 & 6.25 & 6.25 & 6.29 & 0.95 \\ 
				EW & Poor & 3.00 & 0.05 & 6.36 & 6.36 & 6.40 & 0.95 & 0.06 & 6.17 & 6.17 & 6.22 & 0.95 \\ 
				BW(11) & Poor & 3.00 & -0.04 & 9.32 & 9.32 & 9.42 & 0.95 & 0.06 & 8.57 & 8.57 & 8.64 & 0.95 \\ 
				BW(81) & Poor & 3.00 & -0.11 & 14.95 & 14.95 & 15.59 & 0.96 & -0.01 & 13.82 & 13.83 & 13.97 & 0.95 \\ \addlinespace

				&  &  &  \multicolumn{5}{c}{Outcome model misspecified} &   \multicolumn{5}{c}{Both models misspecified}  \\ \cmidrule(lr){4-8}\cmidrule(lr){9-13}
				Weight & Overlap & True & ARB & RMSE & SD & SE & CP & ARB & RMSE & SD & SE & CP\\
				\cmidrule(lr){1-13}

				IPW & Good & 3.00 & 0.07 & 5.24 & 5.24 & 5.22 & 0.95 & 13.90 & 42.35 & 7.39 & 7.53 & 0.00 \\ 
				OW & Good & 3.00 & 0.06 & 4.83 & 4.83 & 4.83 & 0.95 & 13.80 & 42.02 & 7.23 & 7.37 & 0.00 \\ 
				MW & Good & 3.00 & 0.05 & 4.95 & 4.95 & 4.95 & 0.95 & 13.76 & 41.92 & 7.31 & 7.43 & 0.00 \\ 
				EW & Good & 3.00 & 0.06 & 4.85 & 4.85 & 4.85 & 0.95 & 13.82 & 42.09 & 7.24 & 7.38 & 0.00 \\ 
				BW(11) & Good & 3.00 & 0.03 & 6.93 & 6.93 & 6.81 & 0.95 & 13.60 & 41.71 & 8.60 & 8.61 & 0.00 \\ 
				BW(81) & Good & 3.00 & 0.06 & 12.45 & 12.45 & 12.38 & 0.95 & 13.63 & 43.11 & 13.63 & 13.37 & 0.14 \\ \addlinespace
				
				IPW & Mod & 3.00 & 0.24 & 10.93 & 10.91 & 9.55 & 0.92 & 26.82 & 81.09 & 10.07 & 9.55 & 0.00 \\ 
				OW & Mod & 3.00 & 0.05 & 5.59 & 5.59 & 5.56 & 0.95 & 25.97 & 78.35 & 8.42 & 8.11 & 0.00 \\ 
				MW & Mod & 3.00 & 0.04 & 5.72 & 5.72 & 5.76 & 0.95 & 25.77 & 77.77 & 8.48 & 8.21 & 0.00 \\ 
				EW & Mod & 3.00 & 0.07 & 5.75 & 5.75 & 5.65 & 0.95 & 26.11 & 78.80 & 8.51 & 8.17 & 0.00 \\ 
				BW(11) & Mod & 3.00 & -0.00 & 9.73 & 9.74 & 9.57 & 0.94 & 24.97 & 75.72 & 11.05 & 10.86 & 0.00 \\ 
				BW(81) & Mod & 3.00 & -0.09 & 18.05 & 18.06 & 17.37 & 0.94 & 24.77 & 76.45 & 17.99 & 17.74 & 0.02 \\ \addlinespace
				
				IPW & Poor & 3.00 & 1.04 & 22.84 & 22.64 & 16.44 & 0.83 & 38.62 & 116.70 & 14.09 & 12.66 & 0.00 \\ 
				OW & Poor & 3.00 & 0.07 & 6.31 & 6.31 & 6.36 & 0.95 & 35.91 & 108.09 & 8.88 & 8.73 & 0.00 \\ 
				MW & Poor & 3.00 & 0.07 & 6.53 & 6.53 & 6.60 & 0.95 & 35.45 & 106.73 & 8.95 & 8.83 & 0.00 \\ 
				EW & Poor & 3.00 & 0.08 & 6.45 & 6.44 & 6.53 & 0.95 & 36.33 & 109.37 & 9.05 & 8.84 & 0.00 \\ 
				BW(11) & Poor & 3.00 & 0.11 & 11.56 & 11.56 & 11.64 & 0.95 & 33.69 & 101.85 & 12.46 & 12.30 & 0.00 \\ 
				BW(81) & Poor & 3.00 & 0.23 & 20.29 & 20.29 & 20.95 & 0.96 & 33.37 & 102.13 & 20.18 & 20.25 & 0.00 \\ 
				
				\bottomrule
			\end{tabular}
			\begin{tablenotes}\footnotesize		
				\item  These  results are based on 1000 repetitions of the simulated data. Mod: Moderate; BW($\nu$): BW with $\nu=11$ and $81$; ARB: absolute relative bias$\times 100$;
				\item   RMSE: root mean-squared error$\times 100$; SD:empirical standard deviation$\times 100$; SE: average estimated standard error$\times 100$; CP: 95\%  coverage probability. 
			\end{tablenotes}
		\end{center}
	\end{threeparttable}
\end{table}	

\begin{table}[htp!]
	\begin{threeparttable}
		\begin{center}
			\caption{Heterogeneous treatment effect (with and without model misspecification).\label{Mao_He} { }}
			\begin{tabular}{rrrrrrrrrrrrrrrrrrrrrrrrrrrrrrrrrrrrrrrrrrrrrrrrrrrrrrrrrrrrrrrrrrrrrrrrrrrrr}
				\toprule
				&  &  & \multicolumn{10}{c}{Model misspecification} \\\cmidrule(lr){4-13}
				&  &  & \multicolumn{5}{c}{None }& \multicolumn{5}{c}{PS model misspecified} \\ \cmidrule(lr){4-8}\cmidrule(lr){9-13}
				Weight & Overlap & True & ARB & RMSE & SD & SE & CP & ARB & RMSE & SD & SE & CP \\
				\cmidrule(lr){1-13}

				IPW & Good & 5.57 & 0.03 & 5.42 & 5.42 & 5.45 & 0.95 & 0.60  & 6.20 & 5.23 & 5.30 & 0.91 \\ 
				OW  & Good & 5.65 & 0.02 & 5.02 & 5.02 & 5.10 & 0.96 & 0.08  & 5.06 & 5.04 & 5.12 & 0.95 \\ 
				MW  & Good & 5.70 & 0.00 & 5.00 & 5.01 & 5.10 & 0.96 & -0.14 & 5.09 & 5.03 & 5.12 & 0.95 \\ 
				EW  & Good & 5.63 & 0.03 & 5.05 & 5.05 & 5.10 & 0.96 & 0.19  & 5.16 & 5.05 & 5.11 & 0.95 \\ 
				BW(11) & Good & 5.88 & -0.06 & 5.83 & 5.82 & 5.84 & 0.95 & -0.88 & 7.80 & 5.81 & 5.75 & 0.85 \\ 
				BW(81) & Good & 5.98 & -0.10 & 8.90 & 8.89 & 9.12 & 0.95 & -1.16 & 11.09 & 8.68 & 8.82 & 0.87 \\  \addlinespace
				
				IPW & Mod & 4.94 & 0.25 & 12.44 & 12.39 & 11.45 & 0.92 & 3.69 & 20.57 & 9.59 & 9.14 & 0.45 \\ 
				OW & Mod & 5.33 & 0.07 & 6.73 & 6.72 & 6.66 & 0.95 & 0.55 & 7.40 & 6.80 & 6.71 & 0.92 \\ 
				MW & Mod & 5.44 & 0.04 & 6.60 & 6.60 & 6.50 & 0.95 & -0.07 & 6.74 & 6.73 & 6.60 & 0.95 \\ 
				EW & Mod & 5.26 & 0.09 & 6.98 & 6.97 & 6.74 & 0.94 & 1.04 & 8.83 & 6.93 & 6.75 & 0.87 \\ 
				BW(11) & Mod & 5.86 & -0.08 & 8.20 & 8.19 & 7.83 & 0.94 & -2.36 & 16.06 & 8.17 & 7.87 & 0.57 \\ 
				BW(81) & Mod & 5.98 & -0.14 & 13.38 & 13.36 & 12.75 & 0.93 & -2.88 & 21.44 & 12.74 & 12.41 & 0.71 \\  \addlinespace
				
				IPW & Poor & 4.49 & 1.03 & 31.06 & 30.73 & 20.72 & 0.81 & 8.14 & 40.19 & 16.76 & 14.18 & 0.26 \\ 
				OW & Poor & 5.18 & 0.09 & 8.36 & 8.36 & 8.07 & 0.94 & 0.66 & 8.86 & 8.17 & 7.90 & 0.92 \\ 
				MW & Poor & 5.32 & 0.03 & 7.92 & 7.93 & 7.79 & 0.94 & -0.43 & 8.24 & 7.91 & 7.75 & 0.93 \\ 
				EW & Poor & 5.07 & 0.14 & 9.10 & 9.08 & 8.27 & 0.93 & 1.64 & 11.88 & 8.47 & 8.03 & 0.80 \\ 
				BW(11) & Poor & 5.86 & -0.20 & 9.62 & 9.56 & 9.49 & 0.94 & -4.54 & 28.29 & 9.64 & 9.57 & 0.21 \\ 
				BW(81) & Poor & 5.98 & -0.33 & 16.42 & 16.30 & 15.60 & 0.94 & -5.48 & 36.11 & 15.16 & 15.15 & 0.42 \\  \addlinespace

				&  &  &  \multicolumn{5}{c}{Outcome model misspecified} &   \multicolumn{5}{c}{Both models misspecified}  \\ \cmidrule(lr){4-8}\cmidrule(lr){9-13}
				Weight & Overlap & True & ARB & RMSE & SD & SE & CP & ARB & RMSE & SD & SE & CP\\
				\cmidrule(lr){1-13}

				IPW & Good & 5.57 & 0.03 & 5.80 & 5.80 & 5.81 & 0.95 & 7.68 & 43.44 & 7.63 & 7.81 & 0.00 \\ 
				OW & Good & 5.65 & 0.03 & 5.00 & 5.00 & 5.09 & 0.96 & 7.02 & 40.34 & 7.40 & 7.58 & 0.00 \\ 
				MW & Good & 5.70 & 0.01 & 5.06 & 5.07 & 5.14 & 0.96 & 6.72 & 39.05 & 7.48 & 7.60 & 0.00 \\ 
				EW & Good & 5.63 & 0.03 & 5.04 & 5.04 & 5.10 & 0.95 & 7.16 & 41.01 & 7.41 & 7.59 & 0.00 \\ 
				BW(11) & Good & 5.88 & -0.05 & 6.88 & 6.87 & 6.78 & 0.95 & 5.70 & 34.72 & 8.93 & 8.66 & 0.04 \\ 
				BW(81) & Good & 5.98 & -0.09 & 12.07 & 12.06 & 12.12 & 0.95 & 5.31 & 34.70 & 14.02 & 13.35 & 0.33 \\  \addlinespace
				
				IPW & Mod & 4.94 & 0.36 & 16.03 & 15.94 & 13.91 & 0.90 & 17.57 & 87.54 & 11.75 & 11.25 & 0.00 \\ 
				OW & Mod & 5.33 & 0.08 & 6.66 & 6.65 & 6.57 & 0.95 & 13.02 & 70.01 & 9.06 & 8.79 & 0.00 \\ 
				MW & Mod & 5.44 & 0.05 & 6.57 & 6.57 & 6.49 & 0.94 & 12.07 & 66.27 & 9.01 & 8.70 & 0.00 \\ 
				EW & Mod & 5.26 & 0.11 & 7.00 & 6.98 & 6.69 & 0.93 & 13.75 & 72.91 & 9.20 & 8.88 & 0.00 \\ 
				BW(11) & Mod & 5.86 & -0.07 & 9.57 & 9.56 & 9.23 & 0.94 & 8.59 & 51.65 & 11.32 & 10.79 & 0.01 \\ 
				BW(81) & Mod & 5.98 & -0.16 & 16.67 & 16.65 & 16.31 & 0.94 & 7.74 & 49.67 & 18.06 & 17.36 & 0.24 \\  \addlinespace
				
				IPW & Poor & 4.49 & 1.54 & 39.45 & 38.86 & 25.50 & 0.75 & 28.78 & 130.52 & 18.88 & 16.10 & 0.01 \\ 
				OW & Poor & 5.18 & 0.09 & 8.19 & 8.18 & 7.91 & 0.94 & 17.29 & 90.21 & 9.84 & 9.56 & 0.00 \\ 
				MW & Poor & 5.32 & 0.04 & 7.85 & 7.86 & 7.75 & 0.94 & 15.58 & 83.42 & 9.54 & 9.39 & 0.00 \\ 
				EW & Poor & 5.07 & 0.14 & 8.98 & 8.95 & 8.16 & 0.93 & 18.83 & 96.10 & 10.28 & 9.78 & 0.00 \\ 
				BW(11) & Poor & 5.86 & -0.13 & 11.11 & 11.09 & 11.01 & 0.95 & 9.32 & 55.95 & 12.19 & 11.95 & 0.01 \\ 
				BW(81) & Poor & 5.98 & -0.19 & 19.56 & 19.54 & 19.22 & 0.94 & 8.01 & 51.69 & 19.47 & 19.40 & 0.31 \\ 
				
				\bottomrule
			\end{tabular}
			\begin{tablenotes}
				\footnotesize
				\item  These  results are based on 1000 repetitions of the simulated data. Mod: Moderate; BW($\nu$): BW with $\nu=11$ and $81$; ARB: absolute relative bias$\times 100$;
				\item   RMSE: root mean-squared error$\times 100$; SD:empirical standard deviation$\times 100$; SE: average estimated standard error$\times 100$; CP: 95\%  coverage probability.  
			\end{tablenotes}
		\end{center}
	\end{threeparttable}
\end{table}

For the augmented estimators $\widehat \tau_{g}^{aug}$, the results are given in in Tables \ref{Mao_Ho} and \ref{Mao_He}. 
When at least one model is correctly specified and the overlap is good, augmentation tends to improve  bias and efficiency, overall. For instance, under homogeneous treatment, there is a substantial reduction in standard errors for IPW, BW(11), and BW(81) compared to the case without augmentation. The  OW, MW, and EW estimators are still the most efficient, while IPW is more efficient than BW(11) and BW(81) under good and moderate overlap. When the overlap is poor, IPW is only better than BW(81), but not BW(11). Compared to IPW, the gains in RMSE under homogeneous treatment effect are less substantial. For instance, in moderate  (resp. poor) overlap the RMSEs of OW, MW, and EW are about 1.2 (resp. 1.63) times lower than that of IPW, when both models or the outcome model is correctly specified. There are about 1.7 (resp. 2.5) when the propensity model is correctly specified.  When both models are misspecified, there is an excessive amount of bias for all the estimators and, judging by the their corresponding coverage probabilities, it is very difficult to pick the best performing estimators. 

Under heterogeneous treatment effect, the trends are similar to those from the homogeneous treatment as expected, with OW, MW, and EW being the most efficient, IPW is more efficient than  BW(11) and BW(81) only when the overlap is good. When the overlap is moderate or poor, IPW is more efficient than BW(81), but less efficient than BW(11). The gain in RMSE vis-\'a-vis IPW for OW, MW, and EW  for moderate (resp. poor) overlap, under heterogeneous treatment effect, is about 1.8 (resp. 2.6) when both models are correctly specified, 1.4 (resp. 1.8) when the outcome models are correctly specified, and 2.1 (resp. 3.3) when the propensity score model is correctly specified. Again, the RMSE of IPW is worse than that of the BW(81), in bias and RMSE, when the overlap is poor. This again highlight the inability of IPW estimator to handle violations of the positivity assumption. 

Overall, OW, MW, and EW are more efficient that IPW despite the fact that the latter is doubly-robust. Of the trio OW, MW, EW, the OW estimator has best RMSE in  most the scenarios we considered under homogeneous treatment effect, while MW is the best under heterogeneous treatment effect.  IPW is better than BW(81) almost everywhere, but not better than BW(11) in a good number of scenarios. 

	\subsection{Second simulations}
The second simulation  study also consider three degrees of propensity score distributions overlap (poor, moderate and good), but focuses on the impact of the proportion of treated participants (i.e., prevalence of treatment) in the sample, as explained in the Appendix \ref{sec:additional_simulations}, under both homogeneous and heterogeneous treatment effects. The results, summarized in Tables \ref{Li_high_Ho} and \ref{Li_low_Ho}, indicate that the prevalence of treatment heightens the differences in bias and efficiency across the methods; the smaller the prevalence of treatment the higher the bias and the variance of the estimators.
We see similar trend as in the first set of simulations, when we compare the different estimators, for a given prevalence of treatment and whether the treatment is homogeneous or not: OW, MW, and EW are the most efficient (with similar SEs) compared to IPW,  BW(11) and BW(81). There is not  a case where IPW has a small ARB and its RMSE was better than those from BW(11) and BW(81) only when the overlap was good. Otherwise, when the overlap was moderate or poor, both BW(11) and BW(81) have better RMSE. For instance, under constant treatment and median (resp. low) prevalence, the RMSE of BW(11) is equal to 15.25 (resp. 24.07), that of BW (81) equal to 24.79 (resp. 40.43) whereas IPW shows 55.46 (resp. 124.92). We observe a similar pattern under heterogeneous treatment effect. For more details, please, see  Tables \ref{Li_high_Ho} and \ref{Li_low_Ho} in the Appendix. 
%

	\section{Data illustrations} \label{sec:illustration}
	To illustrate the methods we have discussed, we present 2 examples with different levels of overlap in propensity scores between treatment groups. In the first example, we assess the effects of fish consumption on blood mercury level using data from the National Health and Nutrition Examination Survey (NHANES) 2013–2014 (see Zhao et al.\cite{zhao2019sensitivity}). The data set can be obtained through the R package \verb"CrossScreening". 
	In the second example, we evaluate the effects of children smoking on their pulmonary function using data from the Childhood Respiratory Disease Study (see Rosner \cite{rosner2016fundamentals}), which can be obtained through the R package \verb"covreg".
	For parsimony, we do not present the results from the beta selection function with $\nu = 81$ as the simulation studies indicate that for such a high of $\nu$ the performance of the methods is poor in these two contexts.
	\subsection{Fish consumption and blood mercury level} \label{sec:illustration_fish}
	The binary exposure $Z=1$  (high fish consumption) if the person had more than 12 servings of fish or shellfish in the previous month. The NHANES data consist 1107 adults with 234 (21\%) of them had high fish consumption. The data include participants' total blood mercury levels (in microgram per liter) as well as their age, sex, income (with missing values imputed by the median), the indicator of missing income, race, education, ever smoked, and number of cigarettes smoked last month. The outcome variable is defined as $\log_2$(total blood mercury level). 
	
		\begin{figure}[!h]	
		\hspace*{-1.6cm}
		\begin{subfigure}{.6\textwidth}	
						\begin{center}		
				\caption{Propensity scores}
				\label{fig:PS_dist_plot}
				\includegraphics[trim=80 25 45 200, clip, width=0.8\linewidth]{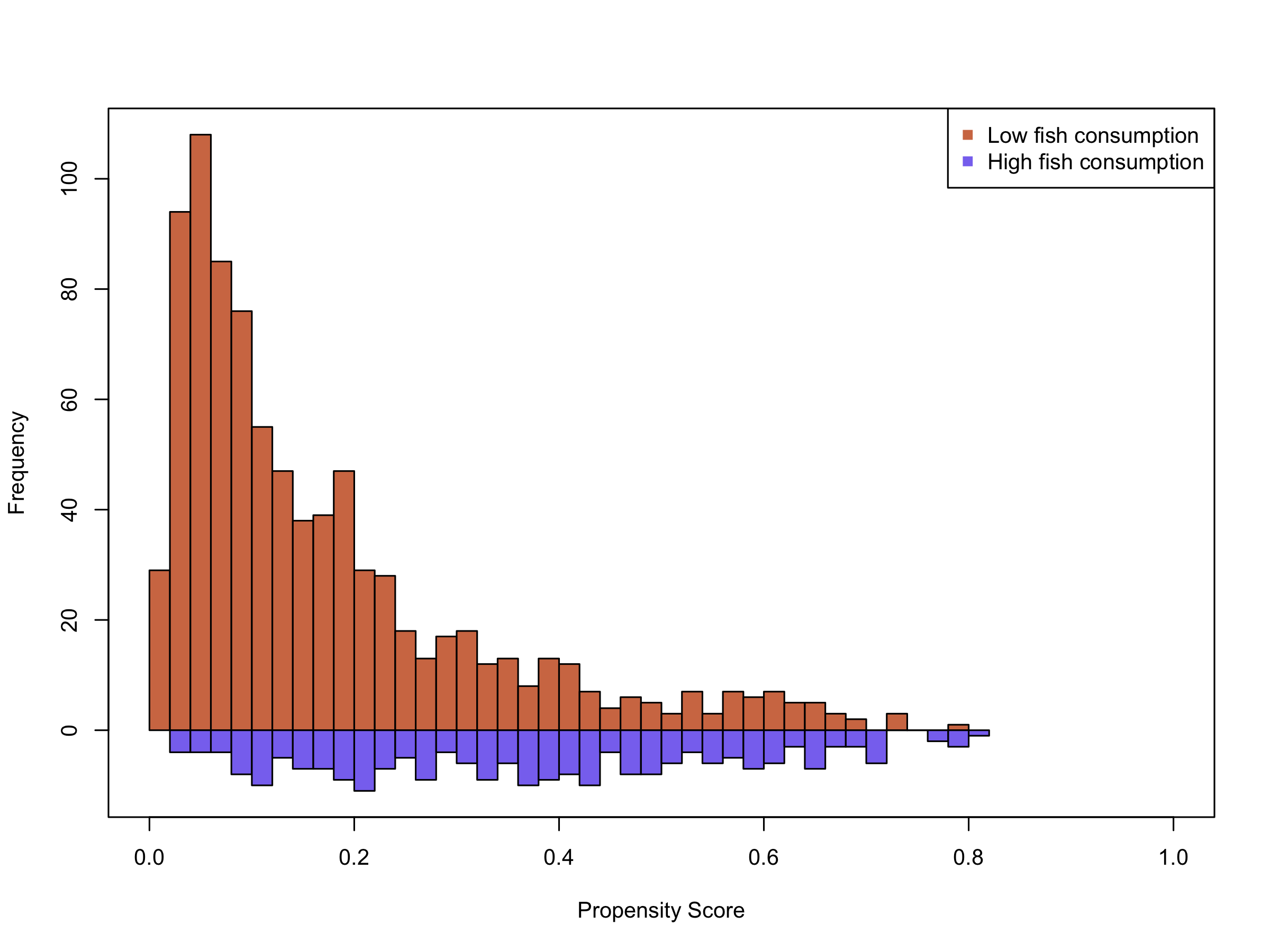}
				\tiny	\begin{tabular}{lllccccccccccccccccccc} 
					\toprule
					& $N$ & 	Min. &1st  Qu. & Median &  Mean  & 3rd Qu. &  Max.\\ \cmidrule(lr){1-8} 
					High fish consumption &   234 &  0.02 & 0.23 & 0.37 & 0.36 & 0.49 & 0.77  \\
					Low fish consumption &  873 &   0.004 & 0.05 & 0.12 & 0.17 & 0.24 & 0.75  \\
					\bottomrule
				\end{tabular} 
			\end{center}
		\end{subfigure}
		\hspace*{-1.3cm}
			\begin{subfigure}{.6\textwidth}
					\begin{center}
				\caption{Covariate balance}
				\label{fig:Fish_covariate_plot}
				\includegraphics[trim=19 25 45 90, clip, width=0.92\linewidth]{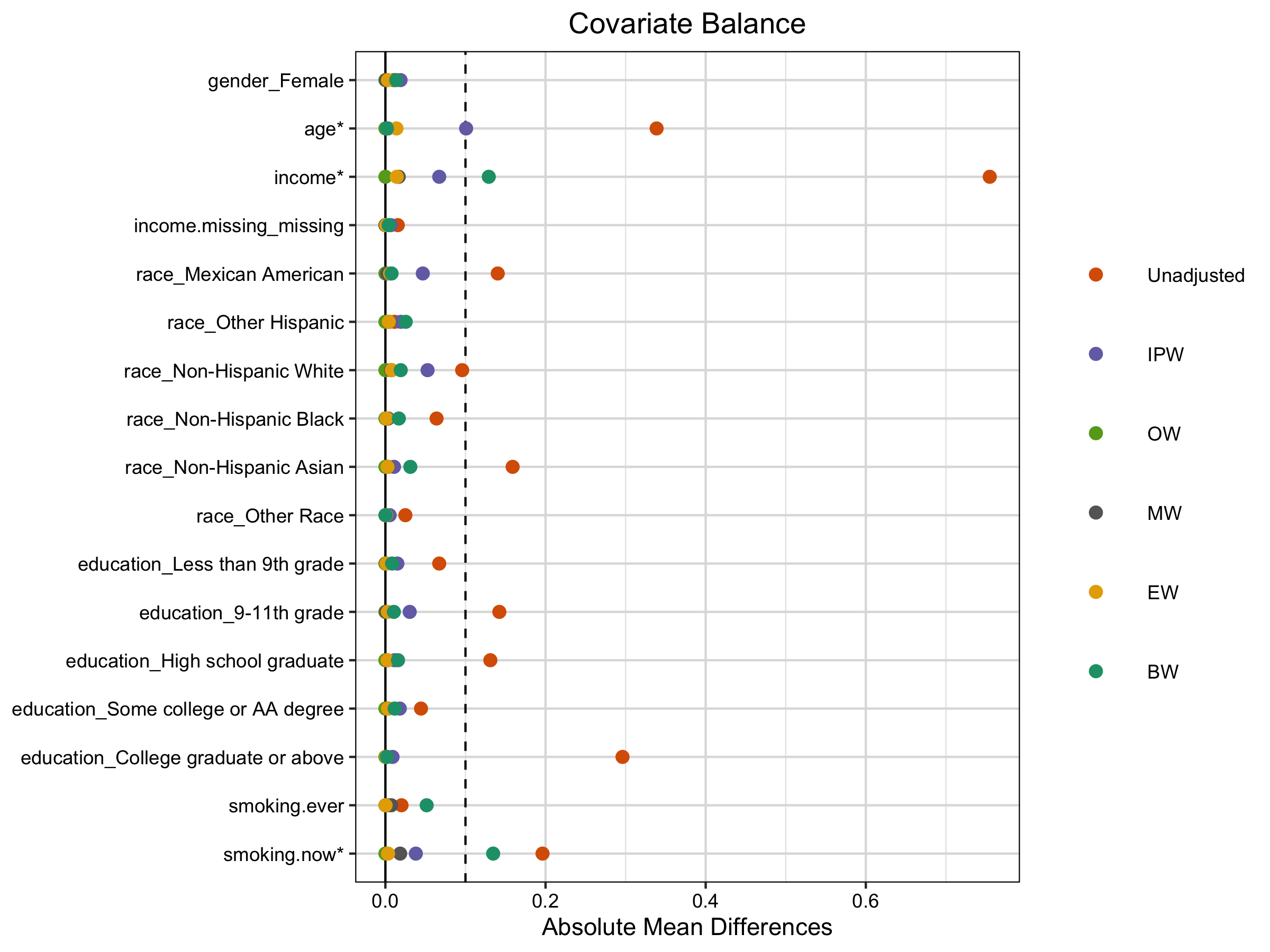}
			\end{center}
			\end{subfigure}
			
		\caption{NHANES data: Estimated propensity scores and covariate balance}
		\subcaption*{Covariate balance in the unweighted sample and the sample weighted by inverse probability weights (IPW), overlap weights (OW), matching weights (MW), entropy weights (EW), and beta weights (BW).}
	\end{figure}
	
	The estimated propensity scores range from 0.004 to 0.77 with moderate level of overlap between the treated and control groups (Figure \ref{fig:PS_dist_plot}). There is substantial imbalance in age, income, race, education, and number of cigarettes smoked, which can be mitigated by propensity score weighting (Figure \ref{fig:Fish_covariate_plot}). 
	The causal effects of fish consumption on blood mercury level, for the different weighting methods,  are summarized in Table \ref{tab:Fish_trt_est}. While the point estimates of OW (1.99), MW (2.04), and EW (1.94) are similar, those of IPW (1.77) and BW (2.24) are slightly different. OW, MW, and EW are more efficient than IPW and BW in terms of standard error, effective sample size, and variance inflation. Augmentation improves the point estimation and the efficiency of IPW and BW estimators. 
		\begin{table}[h!]
		\begin{center}
			\caption{NHANES data: Effect of fish consumption on $\log_2$(total blood mercury)}	\label{tab:Fish_trt_est}
			\begin{threeparttable}
				\begin{tabular}{crrrrrrrrrrrrrrrrrrrrrrrrrrrrrrrrrrrrrrrrrrrrrrrrrrrrrrrrrrrrrrrrrrrrrrrrrrrrrrrrrrrrrrrrrrrrrrrrrrrrrrrrrrrrrrrrrrrrrrrr}
					\toprule
						Estimator& Measure	& IPW & OW & MW & EW & BW(11)\\ \cmidrule(lr){1-7} 
					
	\multirow{1}{*}{$\widehat\tau_g$} & Estimate & 1.77 & 1.99 & 2.04 & 1.94 & 2.24 \\ 
									  & SE & 0.13 & 0.10 & 0.10 & 0.09 & 0.18  \\ \addlinespace
	\multirow{1}{*}{$\widehat\tau_g^{aug}$} & Estimate & 1.80 & 2.00 & 2.04 & 1.95 & 2.22  \\
					(augmentation) 			&SE  & 0.10 & 0.10 & 0.09 & 0.10 & 0.10  \\\addlinespace\cmidrule(lr){2-7} 
											& ESS & 304.99 & 595.03 & 574.16 & 581.17 & 282.31  \\ 
											& VI & 2.42 & 1.24 & 1.29 & 1.27 & 2.61  \\
					\bottomrule
				\end{tabular}				
				\begin{tablenotes}
					\footnotesize
					\item {SE: standard error; ESS: effective sample size; VI: variance inflation function}
				\end{tablenotes}		
			\end{threeparttable}
		\end{center}
	\end{table}

	\subsection{Smoking and force expiratory volume among children} \label{sec:illustration_fish}
	
	The data consist of 654 children, aged 3--19, and include the children's forced expiratory volume (FEV), age, sex, height, and smoking status. In total, 65 children (about 10\%) were smokers, with the youngest smoker being 9 years old. 
	
			\begin{figure}[!h]
					\hspace*{-1.5cm}	
		\begin{subfigure}{.6\textwidth}
		\begin{center}
			\caption{Propensity scores}
			\label{fig:PS_dist_FEV}
			\includegraphics[trim=95 25 13 200, clip, width=0.8\linewidth]{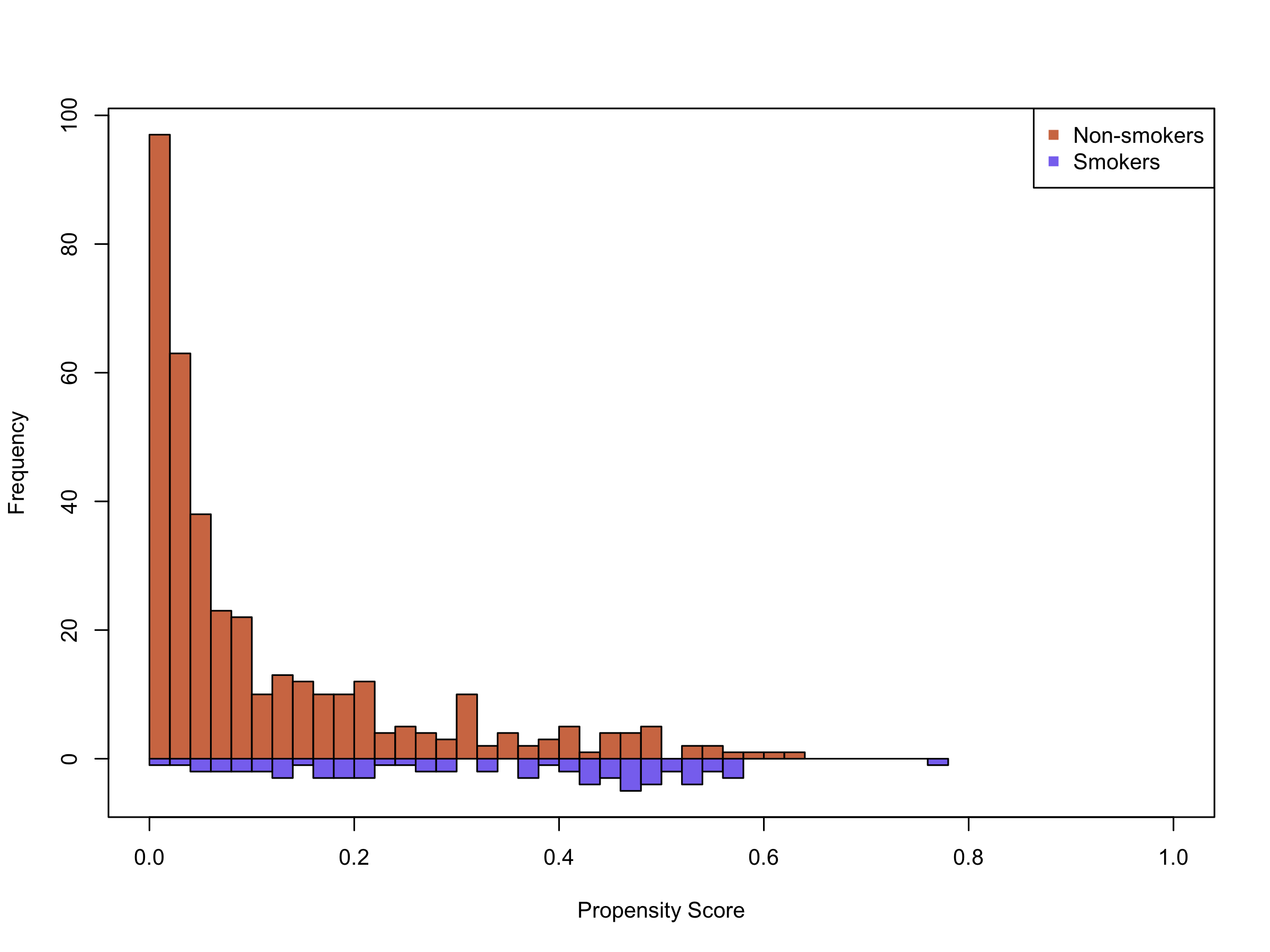}
			\tiny	\begin{tabular}{rcccccccccccccccccccccccccccccccccccccccccccccccccccccccccccccccccccccccccccccc} 
				\toprule
				& $N$ & 	Min. &1st  Qu. & Median &  Mean  & 3rd Qu. &  Max.\\ \cmidrule(lr){1-8} 
				Smokers &   65 &  0.0022  & 0.20 &  0.34  &  0.32  &  0.46 &  0.63  \\
				Non-smokers &  374 &   0.0009 &   0.01 &   0.06   &  0.12 &   0.19  &   0.68  \\
				\bottomrule
			\end{tabular} 
		\end{center}
	\end{subfigure}
		\hspace*{-1.0cm}
		\begin{subfigure}{.52\textwidth}
			\vspace{-1.4cm}
			\begin{center}
				\caption{Covariate balance}
				\label{fig:FEV_covariate_plot}
				\includegraphics[trim=15 0 5 90, clip, width=1\linewidth]{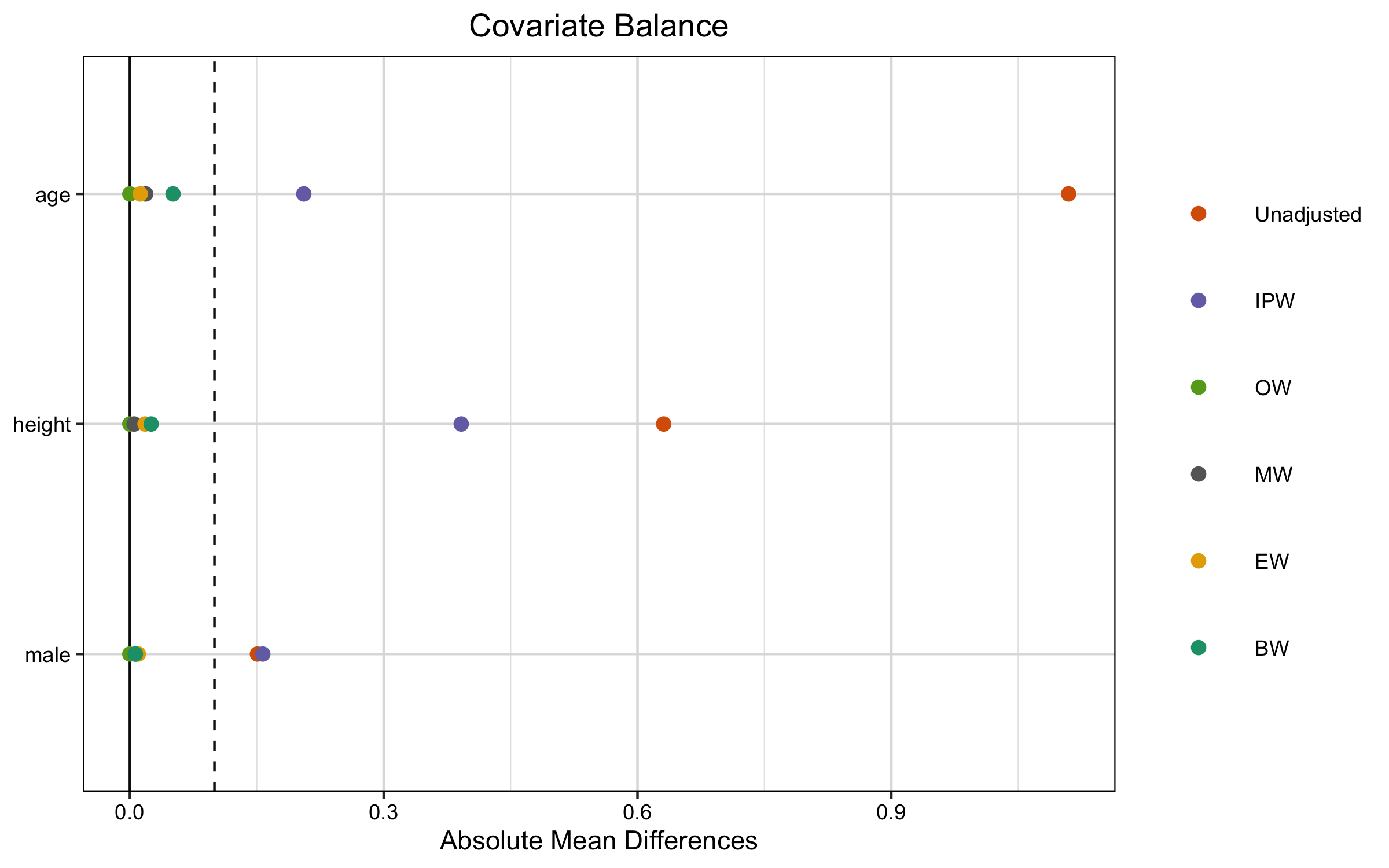}
			\end{center}
		\end{subfigure}
		\caption{FEV data: Estimated propensity scores and covariate balance.}
		\subcaption*{Covariate balance in the unweighted sample and the sample weighted by inverse probability weights (IPW), overlap weights (OW), matching weights (MW), entropy weights (EW), and beta weights (BW).}
	\end{figure}

We present the results for the subgroup of 439 children who are 9 years old or older. The estimated propensity scores range from 0.0009 to 0.68 with moderate to poor overlap between treatment groups (Figure \ref{fig:PS_dist_FEV}). Standard inverse probability weighting (IPW) fails to bring satisfactory balance in the three covariates (Figure \ref{fig:FEV_covariate_plot}).  Causal effect estimation for FEV data are summarized in Table \ref{tab:FEV_trt_est}. Due to the existence of extreme weights and poor overlap, the point estimates of IPW ($-0.58$) and BW ($-0.30$) are substantially different compares to those from  OW ($-0.10$), MW ($-0.13$), and EW ($-0.12$)  estimators, i.e., the class of weights targeting treatment equipoise. In addition, IPW and BW result in large standard error around the point estimates, 0.81 and 0.13 respectively compared to 0.07 for the equipoise estimators. Augmentation improves efficiency of IPW ($0.07$) and BW ($0.08$), and also pulls the point estimate of IPW (($-0.13$)) away from the non-parametric estimate ($-0.58$). The magnitude of the point estimate for BW decreased but remain high (more than twice the estimate of AW, MW, and EW). Nevertheless, the augmentation has drastically improve the variance estimation, going from 0.13 to 0.08. The point estimates and variances of the equipoise estimators for the augmented estimators barely budged from the non-augmented inference. Across both examples, OW and the class of weights targeting treatment equipoise maintain their efficiency compared to IPW and BW, regardless of the level of overlap.	

\begin{table}[t!]
	\begin{center}
		\caption{FEV data: Effect of smoking on FEV}	\label{tab:FEV_trt_est}
		\begin{threeparttable}[]
			\begin{tabular}{crrrrrrrrrrrrrrrrrrrrrrrrrrrrrrrrrrrrrrrrrrrrrrrrrrr}
				\toprule
			Estimator& Measure	& IPW & OW & MW & EW & BW(11)\\ \cmidrule(lr){1-7} 
				
\multirow{2}{*}{$\widehat\tau_g$} &	Estimate & -0.58 & -0.10 & -0.13 & -0.12 & -0.30 \\ 
								  &   SE & 0.81 & 0.07 & 0.07 & 0.07 & 0.13 \\ \addlinespace
\multirow{1}{*}{$\widehat\tau_g^{aug}$} &   Estimate  & -0.13 & -0.11 & -0.13 & -0.11 & -0.28 \\ 
		(augmented)			  &   SE & 0.07 & 0.08 & 0.08 & 0.07 & 0.08 \\ \addlinespace\cmidrule(lr){2-7}
								  &   ESS & 10.88 & 177.73 & 170.93 & 167.92 & 72.79 \\ 
								  &   VI & 20.37 & 1.25 & 1.30 & 1.32 & 3.04 \\ 
				\bottomrule
			\end{tabular}
			\begin{tablenotes}
				\footnotesize
				\item {SE: standard error; ESS: effective sample size; VI: variance inflation function}
			\end{tablenotes}
		\end{threeparttable}
	\end{center}
\end{table}

	\section{Conclusion} \label{sec:conclusion}
	This paper acknowledges that structural lack of positivity assumption is ubiquitous  and should be handled appropriately.  Whether lack of adequate positivity is expected or not, IPW with (or without) stabilized weights,  trimming, or truncation can lead to bias and unstable results.\cite{ma2020robust,hong2020inference} Furthermore, we have provided a number of examples where lack of adequate positivity is expected and sometimes welcomed. Our  statistical analyses of these data structures cannot  ignore lack of positivity, treat it as random, or even worse just use simple conventional methods. Handling structural lack of positivity often requires appropriate methods to yield accurate estimates and robust inference. These estimators have now risen in popularity, sparked lot of curiosity among a wide range of stakeholders,  been implemented in many studies, and even have software packages made freely accessible for whomever want to use them. \cite{thomas2020overlap,zhou2020propensity,zhou2020psweight}  In this paper, we have presented  alternative methods that do not require the positivity assumption per se and handle treatment estimation appropriately, provided we focus our inference  on the population of participants for whom there clinical equipoise. Therefore, the central role of this paper is to shed  light on the role of the overlap weight estimator along with the three other related weighting estimators, i.e., the matching weight, the entropy weight, and the beta weight estimators. 
	
	Using mathematical derivations and large-sample properties, we have demonstrated that overlap weights, matching weights, Shannon's entropy weights, and beta weights target indeed the same underlying estimand of interest: the equipoise average treatment effect, i.e., the treatment effect on the participants for whom there is clinical equipoise. The beta weights are a family of weights that target a wide range of estimands and  that have, as special cases,  the average treatment effect (ATE), the average treatment effect on the treated (ATT), the average treatment effect on the controls (ATC), and the average treatment effect of the overlap (ATO). The family can also approximate the matching weights (MW) and entropy weights (EW) very well. In this paper, we selected two specific beta weights, notably BW(11) and BW(81) to prove a point and answer questions we have been asked in our daily practices by clinical investigators: if $e(x)=0.5$ represent equipoise and since overlap weights focuses on participants whose propensity is around 0.5, could we make the windows where we select the participants in the neighborhood of 0.5? We have demonstrated that while such an inquiry is legitimate, there is a bias-variance trade-off we need to take into consideration. The choice of BW($\nu$) with $\nu$ far and far away from 4, achieves that desired neighborhood of $e(x)=0.5$, but often with biased and less efficient results.
	
	The implementation of the methods presented in this paper is straightforward. What is challenging is the assessment of the underlying assumptions, which aid with correct estimation and inference of treatment effect. The use of the inverse probability weighting (IPW) methods (with truncation, trimming, or not) in the presence of lack of adequate positivity makes additional assumptions about the data that are not always elucidated or target an estimand, the ATE, when it is no longer valid since the assumptions are no longer valid---especially the positivity assumption.\cite{ma2020robust, zhou2020propensity} The paucity of these key assessments of the assumptions creates lots of confusion,  leads to controversy, and undermines the appreciation of innovative method, often deemed too complicated and fancy, with less motivation to adopt them.  We have emphasized that a best practice to implementing the methods presented in this paper is to formulate  the scientific questions of interest appropriately, define the goals of the study clearly, understand  the structure of  data pretty well, and elicit subjet-matter knowledge if necessary. In some cases, just drawing a directed acyclic graph (DAG) to decipher what variables to include in the propensity score or outcome regression models can help tremendously. More importantly, DAGs can aid to identify variables we  should not include in our regression models (e.g., instrumental variables, mediators, and colliders).\cite{digitale2022tutorial, diemer2021more} As shown by the simulation study, variable selection and model specification are primordial to obtain reliable estimates of the treatment effect and reach sensible statistical inference. 
	
	While the methods presented in this paper focus on studies where a large number participants are eligible to either treatment, we recognize that the proportion of participants assigned to either treatment groups can influence directly the estimated treatment effect. Whether the proportion $p$ of participants assigned to the treatment group is small (resp. large) in the population, we  know that the ATE is close to the average treatment effect on the control (ATC) (resp. the average treatment effect on the treated (ATT)) since $ATE = pATT + (1-p)ATC$. In this paper, we have demonstrated that the equipoise estimators would yield point estimates that are close the ATT (resp. ATC). We have illustrated such a phenomenon through simulation studies, which confirms the heuristic proof provided by Li et al.\cite{li2018balancing}. Nevertheless, further investigations are warranted to assess the performance of different methods under finite sample size and maybe explain such a stark contrast in targeting specific (sub)population. Nevertheless, as we have demonstrated in this paper, deciding whether we need to estimate ATE or if it is even appropriate should not be limited to making sure that the SUTVA, uncounfoundness, and poistivity assumptions    are satisfied, we need also to ensure that we have a balanced number of participants in both treatment groups. Otherwise, in addition to the issues inherent to propensity score methods at large, comparing different propensity score methods (matching, weighting, stratification) may not afford us to expect the results to be similar, especially when the underlying treatment  effect we want to estimate  is heterogeneous. We recommend the use of equipoise estimators as we gaining deeper understanding of their statistical properties---they are adaptive, data-sensible, flexible, and lead to efficient estimates. In addition, they always balance the distributions of the covariates included in the propensity score model. 
	
	We hope this paper will help those who would like to expand their statistical toolsets as the equipoise estimators (OW, MW, EW, and BW) offer an unprecedented opportunity to explore complex studies where lack of adequate positivity is not just by happenstance. Rather it is an integral part of the data underlying structure which should be acknowledged and handled appropriately. Whenever possible assessment of the assumptions and model specifications should always be conducted, but this cannot be done at the price of ignoring important tenets of causal inference: run observational studies that emulate randomized clinical trials and elicit experts' opinion when necessary.  Although we mainly focus on continuous outcomes, the methods presented in this paper can be extended to other types of outcomes and can easily be implemented using the PSweight R-package.\cite{zhou2020psweight}
	
\section{Acknowledgments}
The authors are indebted to Yi Liu who read later versions of the manuscript and made valuable comments and suggestions, which helped improve the content of this paper. 

\newpage	\appendix
    \section{Technical details}\label{appendix}
    \newcounter{Appendix}[section]
    \numberwithin{equation}{subsection}
    \renewcommand\theequation{\Alph{section}.\arabic{subsection}.\arabic{equation}}
    \numberwithin{table}{subsection}
    \numberwithin{figure}{subsection}
	 \subsection{The weighted estimator}
    \subsubsection{Inverse weighted mean estimation}\label{appendix1}
	\allowdisplaybreaks \begin{align*}\allowdisplaybreaks
	E[Zw_{1}(X)]&=E[Zg(X)e(X)^{-1}]=E[E\{Zg(X)e(X)^{-1}|X\}]\nonumber\\
	&=E[g(X)e(X)^{-1}E(Z|X)]=E[g(X)]
		\end{align*} 
		Similarly, $E[(1-Z)w_{0}(X)]=E[(1-Z)g(X)(1-e(X))^{-1}]
		=E[E\{(1-Z)g(X)(1-e(X))^{-1}|X\}]=E[g(X)].$\\	
	Combining these results, lead to 
	\allowdisplaybreaks \begin{align}\label{eq:weight_sum}
	E[Zw_{1}(X)+(1-Z)w_{0}(X)]&=2E[g(X)]
	\end{align} 
Moreover,
	\allowdisplaybreaks \begin{align*}\allowdisplaybreaks
	E[Zw_{1}(X)Y]&=E[Zw_{1}(X)Y(1)]=E[Zg(X)e(X)^{-1}Y(1)]\\
	&=E\left[E\left\lbrace Zg(X)e(X)^{-1}Y(1)\big|X,  Y(1)\right\rbrace \right] =E[g(X)e(X)^{-1}Y(1)E(Z|X, Y(1))]\\
	&=E[g(X)e(X)^{-1}Y(1)E(Z|X)]\\
	&=E[g(X)Y(1)]\\
	E[(1-Z)w_{0}(X)Y]&=E[(1-Z)w_{0}(X)Y(0)]=E[(1-Z)g(X)(1-e(X))^{-1}Y(0)]\\
	&=E\left[ E\left\lbrace (1-Z)g(X)(1-e(X))^{-1}Y(0)\big|X,  Y(0)\right\rbrace\right] \\
	&=E[g(X)(1-e(X))^{-1}Y(0)E(1-Z|X, Y(0))]=E[g(X)(1-e(X))^{-1}Y(0)E(1-Z|X)]
	\\&=E[g(X)Y(0)]
	\end{align*} 
		\subsubsection{Estimation via regression models}\label{appendix2}
By the SUTVA, $m_z(X)=E(Y|Z=z, X)=E[Y(z)|Z=z, X]$ while by  the unconfoundness assumption $E(Y(z)|X)=E[Y(z)|Z=z, X],~ z=0, 1$. Therefore, 
 \allowdisplaybreaks \begin{align*}
  E[g(X)m_1(X)]&=E\left[g(X)E(Y|Z=1, X)\right] =E\left[g(X)E(Y(1)|Z=1, X)\right] \\
  &=E\left[E(g(X)Y(1)|Z=1, X)\right] =E\left[E(g(X)Y(1)|X)\right] \\
  &=E\left[g(X)Y(1)\right]. 
  \end{align*}
  Similarly,
  \allowdisplaybreaks \begin{align*}
  E[g(X)m_0(X)]&=E\left[g(X)E(Y|Z=0, X)\right] =E\left[g(X)E(Y(0)|Z=0, X)\right] \\
  &=E\left[E(g(X)Y(0)|Z=0, X)\right] =E\left[E(g(X)Y(0)|X)\right] \\
  &=E\left[g(X)Y(0)\right]. 
  \end{align*}
  Thus, \allowdisplaybreaks \begin{align} \label{eq:selectm1_m0}
  \frac{ E[g(X)\left\lbrace m_1(X)-m_0(X)\right\rbrace ]}{E[g(X)]} =\frac{E\left[g(X)\left\lbrace Y(1)-Y(0)\right\rbrace \right]}{E[g(X)]}=\tau_g
  \end{align}
  Finally, 
 \allowdisplaybreaks \begin{align} \label{eq:weight_m1}
  E\left[\left( Zw_1(X) + (1-Z)w_0(x)\right) m_1(X)\right]&=E\left[\displaystyle \left\lbrace  Zg(X)e(X)^{-1}+(1-Z)g(X)(1-e(X))^{-1}\right\rbrace  m_1(X)\right]\nonumber\\
  &=E\left[E\left\lbrace \displaystyle  Zg(X)e(X)^{-1} m_1(X)+\displaystyle (1-Z)g(X)(1-e(X))^{-1}m_1(X)|X\right\rbrace \right]\nonumber\\
  &=E\left[\displaystyle g(X)m_1(X)\left\lbrace e(X)^{-1}E\left( Z|X\right) + \left( 1-e(X)\right) ^{-1} \displaystyle E\left((1-Z)|X\right)\right\rbrace \right]\nonumber\\
  &=E\left[\displaystyle g(X)m_1(X)\left\lbrace e(X)^{-1}E\left( Z|X\right) + (1-e(X))^{-1} \displaystyle E\left((1-Z)|X\right)\right\rbrace \right]\nonumber\\
  &=2E\left[\displaystyle g(X)m_1(X)\right]
   \end{align}
  and 
  \allowdisplaybreaks\begin{align}  \label{eq:weight_m0}
  E\left[\left( Zw_1(X) + (1-Z)w_0(x)\right) m_0(X)\right]&=E\left[\displaystyle \left\lbrace  Zg(X)e(X)^{-1}+(1-Z)g(X)(1-e(X))^{-1}\right\rbrace  m_0(X)\right]\nonumber\\
  &=E\left[E\left\lbrace \displaystyle  Zg(X)e(X)^{-1} m_0(X)+\displaystyle (1-Z)g(X)(1-e(X))^{-1}m_0(X)|X\right\rbrace \right]\nonumber\\
  &=E\left[\displaystyle g(X)m_0(X)\left\lbrace e(X)^{-1}E\left( Z|X\right) + \left( 1-e(X)\right) ^{-1} \displaystyle E\left((1-Z)|X\right)\right\rbrace \right]\nonumber\\
  &=E\left[\displaystyle g(X)m_0(X)\left\lbrace e(X)^{-1}E\left( Z|X\right) + (1-e(X))^{-1} \displaystyle E\left((1-Z)|X\right)\right\rbrace \right]\nonumber\\
  &=2E\left[\displaystyle g(X)m_0(X)\right]
  \end{align}
  The results \eqref{eq:weight_sum}, \eqref{eq:weight_m1}, and \eqref{eq:weight_m0} put together imply that 
 \allowdisplaybreaks \begin{align} \label{eq:weightm1_m0}
 \frac{ E[\left( Zw_1(X) + (1-Z)w_0(x)\right)\left\lbrace m_1(X)-m_0(X)\right\rbrace ]}{E[\left( Zw_1(X) + (1-Z)w_0(x)\right)]} = \frac{ E[g(X)\left\lbrace m_1(X)-m_0(X)\right\rbrace ]}{E[g(X)]}= \frac{ E[g(X)\tau(X)]}{E[g(X)]} =\tau_g
  \end{align}
		\subsection{The augmented estimators }\label{appendix3}
 To demonstrate the results related to the augmentation of the estimator $\widehat \tau_{g}$, we consider that the SUTVA, consistency, positivity, and unconfoundness assumptions are satisfied. 
 
 Consider 
\allowdisplaybreaks \begin{align}
  &\widehat \tau_g^{aug}= 
  \displaystyle\sum_{i=1}^{N}  Z_i\widehat W_1(x_i)\left\lbrace Y_i - \widehat m_1(x_i)\right\rbrace
  -\displaystyle\sum_{i=1}^{N} (1-Z_i) \widehat  W_0(x_i)\left\lbrace Y_i-\widehat m_0(x_i)\right\rbrace + \displaystyle\sum_{i=1}^{N} \widehat  g(x_i)\left\lbrace \widehat m_1(x_i)-\widehat m_0(x_i)\right\rbrace\big/\displaystyle\sum_{i=1}^{N} \widehat  g(x_i).\label{eq_gdr1}
 \end{align}

Note that $\widehat \tau_g^{aug}$ is similar to the double robust estimator for normalized inverse probability weights given by Robins et al. \cite{robins2007comment} (see the second version of their equation (1)).

If the selection function is affine, i.e.,  $g(x)=a+b e(x)$,  $a, b \in \mathbb{R},$ we use the following estimator instead  
\begin{align}
\widehat \tau_g^{DR}&=
\displaystyle\sum_{i=1}^{N}  Z_i\widehat W_1(x_i)\left\lbrace Y_i - \widehat m_1(x_i)\right\rbrace
-\displaystyle\sum_{i=1}^{N} (1-Z_i) \widehat  W_0(x_i)\left\lbrace Y_i-\widehat m_0(x_i)\right\rbrace + \displaystyle\sum_{i=1}^{N} \left(W_3(x_i)\left\lbrace  \widehat m_1(x_i)- \widehat m_0(x_i)\right\rbrace\right)
\end{align}
where $\allowdisplaybreaks\widehat  w_z(x_i)= (a+b\widehat e(x))\left[z_i\widehat e_i(x)^{-1}+ (1-z_i)(1-\widehat e_i(x))^{-1}\right]$ and $ W_3(x_i) = {(a+bZ_i)}\big/{\displaystyle\sum_{i=1}^{N}(a+b Z_i)}.$

Notice that the ATE, ATT, and ATC correspond to the affine functions $g(x)=a+b e(x)$ with, respectively,  $b=0$; $a=0$; and $  a=-b$. Their respective estimators via $\widehat \tau_g^{aug}$ are doubly robust and can achieve asymptotic efficiency (see Tsiatis \cite{tsiatis2007semiparametric}, Tan \cite{tan2007comment}, and Mercatanti and Li \cite{mercatanti2014debit}). We will demonstrate later that $\widehat \tau_g^{DR}$ is doubly robust.

 \subsubsection{General case}\label{appendix3_general}   
 
 In general, how good of an estimator is $\widehat \tau_g^{aug}$ depends on how we specify the parametric models for $\widehat e(x), \widehat  g(x)$ and $\widehat m_z(x)$, $z=0, 1.$

 First, suppose that  the selection function $\widehat  g(x)$  and the propensity score  $\widehat  e(x)$  are correctly specified, but the regression models $\widehat m_z(x)$ are misspecified, i.e., $\widehat m_z(X)\longrightarrow \widetilde  m_z(X)\neq  m_z(X)$. 
Using the results from Appendix \ref{appendix1}, we have 
 \allowdisplaybreaks\begin{align}\label{eq_regres0}
 N^{-1}\displaystyle\sum_{i=1}^{N} Z_i\widehat w_1(x_i) & \longrightarrow  E(Zw_1(X))=E\left[g(X)\right]~~~\text{and}~~
N^{-1}\displaystyle\sum_{i=1}^{N} (1-Z_i)\widehat w_0(x_i)  \longrightarrow  E((1-Z)w_0(X))=E\left[ g(X)\right].
 \end{align}
Also,  
\begin{align}
	N^{-1}\displaystyle\sum_{i=1}^{N} Z_i\widehat w_1(x_i)\left\lbrace Y_i - \widehat m_1(x_i)\right\rbrace  \longrightarrow  E\left[ g(X)\left\lbrace Y(1)-\widetilde  m_1(X)\right\rbrace \right]
 \end{align}
since  $N^{-1}\displaystyle\sum_{i=1}^{N}Z_i\widehat w_1(x_i)\left\lbrace Y_i - \widehat m_1(x_i)\right\rbrace = N^{-1}\displaystyle\sum_{i=1}^{N} \frac{Z_i\widehat g(X_i)}{\widehat e(X)}\left\lbrace Y_i - \widehat m_1(x_i)\right\rbrace$ converges in probability to 
\begin{align*}	
	E\left[\frac{Zg(X)}{e(X)}\left\lbrace Y(1)- \widetilde  m_1(X)\right\rbrace \right] &=E\left[ E\left[\frac{Zg(X)}{ e(X)}\left\lbrace Y(1)- \widetilde  m_1(X)\right\rbrace\Big|Y(1), X
\right] \right]= E\Bigg[g(X) \{ Y(1)- \widetilde  m_1(X)\}  \frac{E(Z|Y(1), X)}{e(X)} \Bigg]\\
&= E\left[ g(X)\left\lbrace Y(1)-\widetilde   m_1(X)\right\rbrace \frac{E(Z| X)}{e(X)}\right]  =E\left[ g(X)\left\lbrace Y(1)-\widetilde  m_1(X)\right\rbrace \right],
  \end{align*}
 since by the unconfoundness assumption, $E(Z|Y(1), X)=E(Z|X)=e(X).$
 
 Similarly, 
 \begin{align}
N^{-1}\displaystyle\sum_{i=1}^{N}(1- Z_i)\widehat w_0(x_i)\left\lbrace Y_i - \widehat m_0(x_i)\right\rbrace\rightarrow 
E\left[g(X)\left\lbrace Y(0)-\widetilde   m_0(X)\right\rbrace \right]\label{eq_regres2}
 \end{align}
 In addition, 
 \begin{align}
 N^{-1}\displaystyle\sum_{i=1}^{N} \widehat  g(x_i)\left\lbrace \widehat m_1(x_i)-\widehat m_0(x_i)\right\rbrace\longrightarrow & E(g(X)\left\lbrace  \widetilde m_1(x_i)-\widetilde m_0(x_i)\right\rbrace)\label{eq_regres3}
 \end{align}
 We combine the equations \eqref{eq_regres0}--\eqref{eq_regres3} to conclude that $\widehat \tau_{g}^{aug}\longrightarrow \tau_{g}$, by the Slutsky's Theorem \cite{gut2012probability}. 
 
 Therefore, whenever $ g(X)$ is correctly specified, $\widehat \tau_{g}^{aug}\longrightarrow \tau_{g}$ if either the propensity score or the selection function are correctly specified. This proves that  $\widehat \tau_{g}^{aug}$ is doubly robust, whenever $ g(X)$ is correctly specified.

 Furthermore, when the selection function, the propensity score model, and the regression models are all correctly specified, we can use the efficient influence function of the weighted average treatment effect  to derive the asymptotic variance of $\widehat  \tau_{g}^{aug}$ (as in Newey \cite{newey1994asymptotic}) given by
 $
 A\mathbb{V}(\widehat  \tau_{g}^{aug})=\mathbb{V}(F_{\tau_{g}}(X,Z,Y))
 $
 where  $F_{\tau_{g}}(X,Z,Y)$ is the efficient influence function. \\Using semiparametric theory \cite{tsiatis2007semiparametric}, the class of all influence functions of estimators of $\tau_{g}$ that are regular asymptotically linear (RAL) are given by $\displaystyle \frac{g(X)}{E\left[g(X)\right]}\left\lbrace \bigg[ \frac{Z(Y-\tau_{g}^1)}{e(X)}
 \left. ~-\frac{(1-Z)(Y-\tau_{g}^0)}{1-e(X)}\right] + \left\{Z-e(X)\right\}h(X)\right\rbrace $ for any arbitrary function $h(X).$   
 Therefore, following Theorem 13.1 (page 333) of Tsiatis \cite{tsiatis2007semiparametric}, there exist a unique efficient influence function $F_{\tau_{g}}(X,Z,Y)$, i.e., the one with the smallest variance among all the influences functions of RAL  estimators of $\tau_{g}$ given by 
 \begin{align}\label{eq:eif}
 F_{\tau_{g}}(X,Z,Y)&=\displaystyle \frac{g(X)}{E\left[g(X)\right]}\bigg[\frac{Z(Y-\tau_{g}^1)}{e(X)}-\frac{(1-Z)(Y-\tau_{g}^0)}{1-e(X)}\bigg]-\displaystyle \frac{g(X)}{E\left[g(X)\right]}\Bigg[(Z-e(X))\left\lbrace \frac{m_1(X)-\tau_{g}^1}{e(X)}+\frac{m_0(X)-\tau_{g}^0}{1-e(X)}\right\rbrace\Bigg] \nonumber \\
 &=\displaystyle \displaystyle \frac{g(X)}{E\left[g(X)\right]}\bigg[\frac{Z(Y-m_1(X))}{e(X)}-\frac{(1-Z)(Y-m_0(X))}{1-e(X)} + \left\lbrace m_1(X)- m_0(X)\right\rbrace  -\tau_g  \Bigg].
 \end{align}
 We can calculate  the asymptotic variance of $\widehat  \tau_{g}^{aug}$ using the law of iterative variance, conditional with respect to  $\mathcal{W}=(X, Y(1), Y(0))$ as $\text{AVar}(\widehat  \tau_{g}^{aug})=Var_{\mathcal{W}}\left( E(F_{\tau_{g}}(X,Z,Y)|\mathcal{W})\right)+E_{\mathcal{W}}\left(Var(F_{\tau_{g}}(X,Z,Y)|\mathcal{W})\right),$ which is equal to
  \begin{align*}
& E_{\mathcal{W}}\left( \frac{g(X)^2}{\left(E\left[g(X)\right]\right)^2}\left\lbrace {\frac{\left(Y(1)- \tau_{g}^1\right)^2}{e(X)}}+{\frac{\left(Y(0)- \tau_{g}^0\right)^2}{1-e(X)}}\right\rbrace \right)\\
 &-E_{\mathcal{W}}\left[ \left( \frac{g(X)}{E\left[g(X)\right]}\left\lbrace \sqrt{\frac{1-e(X)}{e(X)}}\left(m_1(X)- \tau_{g}^1\right) +\sqrt{\frac{e(X)}{1-e(X)}}\left(m_0(X)- \tau_{g}^0\right)\right\rbrace \right)^2 \right]. 
 \end{align*}
It achieves the minimum variance when the regression models are correctly specified. 
 However, if the selection function depends on the propensity score, it is crucial to have it correctly specified. Unless the selection function $g$ is an affine function of the propensity score, i.e., it is of the form $ g(x)=a+b  e(x)$,  $a, b \in \mathbb{R}$ (see \ref{appendix3_affine} ). Otherwise, whenever the propensity score is misspecified, i.e., $\widehat  e(X)\longrightarrow \widetilde e(X)\neq e(x)=P(Z=1|X)$, the selection function is also misspecified, i.e., $\widehat  g(X)\longrightarrow \widetilde g(X)\neq g(X)$. Therefore, $\widehat \tau_{g}^{aug}$ does not converges to $\tau_{g}$, even if the conditional regression models  $\widehat m_z(X)$ are correctly specified. 
 
 \subsubsection{Bias under model misspecifications}\label{appendix3_models_misspecification}
 Suppose that  $g(x)$ is not an affine function of $ e(x)$ and that  $\widehat  e(X)\longrightarrow \widetilde e(X)$, possibly different from the true propensity score $e(X)=  P(Z=1|X)$ and $\widehat m_z(X)\longrightarrow \widetilde  m_z(X)$ and possibly different from $m_z(X)$. By the law of large numbers,  $N^{-1}\displaystyle\sum_{i=1}^{N} Z_i\widehat w_1(x_i)\left\lbrace Y_i- \widehat m_1(x_i)\right\rbrace$ converges  to $E\left[\frac{Z\widetilde g(X)}{\widetilde e(X)}\left\lbrace Y(1)-\widetilde m_1(X)\right\rbrace
 \right]=E\left[ E\left[\frac{Z\widetilde g(X) }{\widetilde e(X)}\left\lbrace Y(1)-\widetilde m_1(X)\right\rbrace
 \Big|X\right] \right]$. 
 
 Thus, $
 N^{-1}\displaystyle\sum_{i=1}^{N}  Z_i\widehat w_1(x_i)\left\lbrace Y_i- \widehat m_1(x_i)\right\rbrace  \longrightarrow 
 E\left[\frac{e(X)\widetilde  g(X) }{\widetilde e(X)}\left\lbrace  m_1(X)-\widetilde  m_1(X)\right\rbrace\right] 
$
  by the SUTVA and the unconfoundness assumption. 
 Similarly, 
 $
 N^{-1}\displaystyle\sum_{i=1}^{N}(1- Z_i)\widehat w_0(x_i) \left\lbrace Y_i- \widehat m_0(x_i)\right\rbrace   \longrightarrow 
 E\left[\frac{(1- e(X))\widetilde  g(X)}{1-\widetilde e(X)}\left\lbrace  m_0(X)- \widetilde  m_0(X)\right\rbrace\right].
 $
 
 Finally, by the law of large numbers,  we also have the following results:
 \allowdisplaybreaks\begin{align*}
& N^{-1}\displaystyle\sum_{i=1}^{N} \widehat  g(x_i)\longrightarrow  E\left[ \widetilde g(X)\right];~~
 N^{-1}\displaystyle\sum_{i=1}^{N} Z_i\widehat w_1(x_i)  \longrightarrow  E\left[ Z\frac{\widetilde g(X)}{\widetilde e(X)}\right] = E\left[\frac{ e(X)\widetilde g(X)}{\widetilde e(X)}\right];\\
& N^{-1}\displaystyle\sum_{i=1}^{N} (1-Z_i)\widehat w_0(x_i)  \longrightarrow   E\left[(1- Z)\frac{\widetilde g(X)}{1-\widetilde e(X)}\right] = E\left[\frac{(1- e(X))\widetilde g(X)}{1-\widetilde e(X)}\right];\\
& N^{-1}\displaystyle\sum_{i=1}^{N} \widehat  g(x_i)\left\lbrace \widehat m_1(x_i)-\widehat m_0(x_i)\right\rbrace  \longrightarrow E(\widetilde g(X)\left\lbrace \widetilde  m_1(X)-\widetilde  m_0(X)\right\rbrace)=E(\widetilde g(X) \widetilde  \tau(X)), ~~\text{with}~~ \widetilde  \tau(X)= E(\widetilde g(X)\left\lbrace \widetilde  m_1(X)-\widetilde  m_0(X)\right\rbrace).
 \end{align*}
  Therefore, putting all together, we have 
   \allowdisplaybreaks\begin{align}\label{eq:aug_tau_g}
  \widehat \tau_{g}^{aug}- \tau_{g}& \longrightarrow E\left[\frac{\widetilde g(X)}{E\left[ \widetilde g(X)\right]}\widetilde  \tau(X)-\frac{ g(X)}{E\left[  g(X)\right]}\tau(X)\right] +E\left[\frac{ e(X)\widetilde g(X)}{\widetilde e(X)}\right]^{-1}E\left[\frac{e(X)\widetilde  g(X) }{\widetilde e(X)}\left\lbrace  m_1(X)-\widetilde  m_1(X)\right\rbrace\right] \\
  & - E\left[\frac{(1- e(X))\widetilde g(X)}{1-\widetilde e(X)}\right]^{-1}E\left[ \frac{(1- e(X))\widetilde g(X)}{1-\widetilde e(X)}\left\lbrace   m_0(X)-\widetilde  m_0(X)\right\rbrace\right], ~~\text{with}~~\tau_{g} = E\left[\frac{ g(X)}{E\left[  g(X)\right]}\tau(X)\right], \nonumber
  \end{align}
where $\widetilde \tau(X)= \widetilde  m_1(X)-\widetilde  m_0(X)$ and $\tau(X)=E[Y(1)-Y(0)|X] =  m_1(X)-  m_0(X).$ 
  \subsubsection{Special cases: affine selection functions}\label{appendix3_affine}  
 When the selection function $g$ is an affine function of the propensity score, i.e., $ g(x)=a+b  e(x)$,  $a, b \in \mathbb{R},$ however, we can re-define  the augmentation au $\widehat \tau_{g}$ to obtain a doubly robust estimator.
To do this, first notice that the sample average $N^{-1}\displaystyle\sum_{i=1}^{N} (a+ bZ_i)$ converges in probability to 
$E[(a+bZ)]=a + bE[E(Z|X)]=E[g(X)].$ 
Similarly,  for any function $\zeta(.)$, the average $N^{-1}\displaystyle\sum_{i=1}^{N} (a+ bZ_i)\zeta(X_i)$ also converges in probability to 
$E[(a+bZ)\zeta(X)]= a + bE[\zeta(X)E(Z|X)] =E[g(X)\zeta(X)].$ 

Thus, we can replace $\widehat g(x)=a+b \widehat e(x)$ in the first term of $\tau_{g}^{aug}$ by $a+bZ$ to obtain, for $a, b \in \mathbb{R}$, 
\begin{align*}
\widehat \tau_g^{DR}&=\displaystyle\sum_{i=1}^{N}(a+bZ_i)\left\lbrace  \widehat m_1(x_i)- \widehat m_0(x_i)\right\rbrace\Big/\displaystyle\sum_{i=1}^{N}( a+bZ_i) + \displaystyle\sum_{i=1}^{N} \left[\frac{Z_i \widehat  w_1(x_i)}{ N_{\widehat w_1}}\left\lbrace Y_i -m_1(x_i)\right\rbrace- \frac{(1-Z_i) \widehat  w_0(x_i)}{ N_{\widehat w_0}}\left\lbrace Y_i-\widehat m_0(x_i)\right\rbrace \right] 
\end{align*}
As defined, $\widehat \tau_g^{DR}$ is doubly robust as it converges to $\tau_{g}$ whenever the propensity model or the regression models are correctly specified.
To prove this, first suppose that the regression models $\widehat m_z(X)$ are correctly specified, but the propensity score is not. In that case, the first term of  $\widehat \tau_g^{DR}$  converges to $\tau_{g}.$  
In addition, the second term in $\widehat \tau_g^{DR}$ converges to 0, using an approach similar to those leading to \eqref{eq:aug_tau_g}, even though  $\widetilde  g(x)\longrightarrow a+b\widetilde e(x)$. 
On the other hand, if the propensity model $\widehat e(X)$  is correctly specified, but the regression models $\widehat m_z(x)$ are not, 
both $N^{-1}\displaystyle\sum_{i=1}^{N}  Z_iw_1(x_i)$ and  $N^{-1}\displaystyle\sum_{i=1}^{N} (1- Z_i)w_0(x_i)$ converge to $E[g(X)].$ 

Plus, $N^{-1}\displaystyle\sum_{i=1}^{N}  (a+bZ_i)\widehat m_z(x_i)\longrightarrow 
E\left[E\left[(a+bZ) \widetilde m_z(X)|X\right]\right]=E\left[(a+bE(Z|X)) \widetilde m_z(X) \right]=E\left[g(X)\widetilde m_z(X) \right]$ as well as
 
\begin{align}\allowdisplaybreaks\label{eq:dreq1}
N^{-1}\displaystyle\sum_{i=1}^{N} \widehat  g(x_i)\left[\frac{Z_i}{\widehat e_i(x)}\left\lbrace Y_i- \widehat m_1(x_i)\right\rbrace\right]
&\longrightarrow E\left[\frac{Z}{e(X) g(X)}\left\lbrace Y - \widetilde m_1(X)\right\rbrace \right] =
E\left[E\left[\frac{Zg(X)}{e(X)} \left\lbrace Y(1) -\widetilde m_1(X) \right\rbrace\Big|X\right] \right]\\
&=E\left[g(X)\left\lbrace\frac{E(Z|X)}{e(X)} E(Y(1)|X)-\frac{E(Z|X)}{e(X)} \widetilde m_1(X) \right\rbrace\right]\nonumber\\
&=E\left[g(X) \left\lbrace E[Y(1)|X]- \widetilde m_1(X) \right\rbrace\right]=E\left[ E[g(X) Y(1)|X]\right]- E\left[g(X) \widetilde m_1(X) \right]\nonumber\\
&=E\left[g(X)Y(1)\right]-E\left[g(X)\left\lbrace \widetilde m_1(X) \right\rbrace\right]=E\left[ g(X)\left\lbrace Y(1)- \widetilde m_1(X) \right\rbrace\right]\nonumber
\end{align}
\begin{align}\allowdisplaybreaks\label{eq:dreq2}
N^{-1}\displaystyle\sum_{i=1}^{N} \widehat  g(x_i)\left[\frac{(1-Z_i)Y_i}{1-\widehat e_i(x)}\left\lbrace  Y_i- \widehat m_0(x_i)\right\rbrace\right]
&\longrightarrow E\left[ g(X)\left\lbrace \frac{(1-Z)Y}{1-e(X)}-\frac{(1-Z)}{1-e(X)}\widetilde  m_0(X)\right\rbrace \right]\\
&= E\left[g(X) \left\lbrace \frac{(1-Z)Y(0)}{1-e(X)} -\frac{(1-Z)}{1-e(X)}\widetilde  m_0(X)\right\rbrace\right] \nonumber\\
&=E\left[g(X)E\left\lbrace\frac{(1-Z)Y(0)}{1-e(X)}-\frac{(1-Z)}{1-e(X)}\widetilde  m_0(X)\Big|X\right\rbrace\right]\\
&=E\left[g(X)\left\lbrace\frac{1-E(Z|X)}{1-e(X)}E(Y(0)|X)-\frac{1-E(Z|X)}{1-e(X)}\widetilde  m_0(X)\right\rbrace\right]\nonumber\\
&=E\left[E\left\lbrace g(X)(Y(0)-\widetilde  m_0(X)|X \right\rbrace\right]=E\left[ g(X)\left\lbrace Y(0)-\widetilde  m_0(X)\right\rbrace\right]
\end{align}
Therefore, if we combine the above results, we conclude that  $\widehat \tau_{g}^{DR}\longrightarrow \tau_{g}.$   
  

\section{Variance estimation}\label{appendix4}                                                                                                                                                                                                                                                                                                                                                                     


Following Lunceford and Davidian, \cite{lunceford2004stratification} we estimate the variance of $\widehat{\tau}_{g}$ using the standard theory of M-estimation (see Stefanski and Boos \cite{stefanski2002calculus} or Wooldridge \cite{wooldridge2002inverse}).  
In general, the sandwich variance any of the estimators we considered in this paper is derived using an estimating equation of the form $\displaystyle 0=\displaystyle\sum_{i=1}^{N} \Psi_\theta({X}_i, Z_i, Y_i)$ for which $\theta$ is solution to the equation and where the estimator is a linear combination of the components of the component of $\theta$, for some matrix $ \Psi_\theta({X}_i, Z_i, Y_i).$ 

Using $E[\Psi_\theta({X}_i, Z_i, Y_i)]=0$ $\Longrightarrow \widehat\theta\overset{p}{\longrightarrow}  \theta $ when  $N\longrightarrow \infty$, under some  regularity conditions,  \cite{stefanski2002calculus} we can conclude that the estimator $ \widehat\theta$ is consistent, by Slutsky's theorem. In addition,  $\sqrt{N}(\widehat{\theta}-{\theta})\overset{d}{\longrightarrow}N({0}, \Sigma({\theta})),$ with $ \Sigma({\theta})=A(\theta)^{-1}B(\theta)\{A(\theta)'\}^{-1}$. A consistent estimator of variance $\Sigma({\theta})$ of $\widehat{\theta}$ is  $ \widehat \Sigma({\widehat\theta})=A_N(\widehat\theta)^{-1}B_N(\widehat\theta)\{A_N(\widehat\theta)'\}^{-1}$, where  $A(\theta), B(\theta),$ $ A_N(\widehat\theta)$ and $B_N(\widehat\theta)$ are the following matrices: 
\begin{align*}\label{eq:matrices_appendix}
	A_N(\widehat{\theta})&=N^{-1}\displaystyle\sum_{i=1}^{N}\left[ - \frac{\partial}{\partial{\theta'}}\Psi_\theta({X}_i, Z_i, Y_i)\right]_{\theta={\widehat\theta}}; ~~
	A(\theta)=\lim_{N\rightarrow\infty}A_N(\widehat\theta)\\
	B_N(\widehat{\theta})&= N^{-1}\displaystyle\sum_{i=1}^{N}\Psi_\theta({X}_i, Z_i, Y_i)\Psi_\theta({X}_i, Z_i, Y_i)'\big|_{\theta=\widehat\theta}; ~\text{and}~~
	B(\theta)=\lim_{N\rightarrow\infty}B_N(\widehat\theta)=E\left[ \Psi_\theta({X}_i, Z_i, Y_i)\Psi_\theta({X}_i, Z_i, Y_i)'\right] 
\end{align*} 
In this section, we describe the corresponding matrices $ \Psi_\theta({X}_i, Z_i, Y_i),$ $A_N(\widehat{\theta})$, and $B_N(\widehat{\theta})$ respectively for $\widehat{\tau}_g$, $\widehat{\tau}_g$, $\widehat\tau_{g, \,\text{REF}}^{aug}$. We assume that the propensity score model $e({x_i};{\beta})=P(Z=1|{X_i=x_i}; {\beta}))$ and the regression models  $\widehat m_z(x_i)=m(x_i; \widehat\alpha_z)$ for $z=0$ and $z=1$ are estimated by maximum likelihood using, respectively, the logistic and linear regression models. 

Finally, as we specified in Section \ref{sec:balancing_overview}, the notation $X$ represents the baseline covariates as well as their possible transformations including quadratic terms and interactions terms, if needed. To be more granular, we can consider that different combinations (or subsets) of covariates $X$  enter the (treatment) logistic and (outcome) regression models, which we denote  $V$ and $W$ respectively.

\subsection{Sandwich variance for the weighted mean estimator }\label{appendix4_sandwich}   
The propensity score parameters $\widehat{\beta}$ and the estimator $(\widehat{\mu}_{1g}, ~\widehat{\mu}_{0g}$)  are derived as solutions to the estimating equation	
\begin{align*}
\displaystyle 0=\displaystyle\sum_{i=1}^{N} \Psi_\theta({X}_i, Z_i, Y_i)
= \displaystyle\sum_{i=1}^{N} 
\begin{bmatrix}
\psi_{\beta}({X}_i, Z_i)\\
\psi_{\mu_{1g}}({X}_i, Z_i, Y_i) \\
\psi_{\mu_{0g}}({X}_i, Z_i, Y_i) \\
\end{bmatrix} 
= \displaystyle\sum_{i=1}^{N} \begin{bmatrix}
\psi_{\beta}({X}_i, Z_i)\\
Z_iw_1({X}_i)(Y_i-\mu_{1g})\\
(1-Z_i)w_0({X}_i)(Y_i-\mu_{0g})\\
\end{bmatrix} 
\end{align*}
with respect to  ${\theta}=({\beta}',\mu_{1g}, \mu_{0g})'$ where $\widehat{\tau}_g=c'{\theta}=\widehat{\mu}_{1g}-\widehat{\mu}_{0g}$ and $c=(0,1,-1)'$. 
The matrices $A(\theta)$, $A_N(\widehat\theta)$, $B(\theta)$ and $B_N(\widehat\theta)$ are 
\begin{align*}
A_N(\widehat{\theta})&=N^{-1}\displaystyle\sum_{i=1}^{N}\left[ - \frac{\partial}{\partial{\theta'}}\Psi_\theta({X}_i, Z_i, Y_i)\right]_{\theta={\widehat\theta}}=
\begin{bmatrix} 
\widehat A_{11} & 0 &0  \\
\widehat A_{21} &\widehat A_{22} &0\\
\widehat A_{31} & 0 &\widehat A_{33}
\end{bmatrix};~~
A(\theta)=\lim_{N\rightarrow\infty}A_N(\widehat\theta)=\begin{bmatrix} 
A_{11} & 0 &0  \\
A_{21} & A_{22} &0\\
A_{31} & 0 &A_{33}
\end{bmatrix}\\
B_N(\widehat{\theta})&= N^{-1}\displaystyle\sum_{i=1}^{N}\Psi_\theta({X}_i, Z_i, Y_i)\Psi_\theta({X}_i, Z_i, Y_i)'\big|_{\theta=\widehat\theta}; ~\text{and}~
B(\theta)=\lim_{N\rightarrow\infty}B_N(\widehat\theta)=E\left[ \Psi_\theta({X}_i, Z_i, Y_i)\Psi_\theta({X}_i, Z_i, Y_i)'\right] 
\end{align*} 
where 
\allowdisplaybreaks\begin{align*}
A_{11}&=-E\left( \frac{\partial}{\partial{\beta'}}\psi_\beta({X}_i, Z_i)\right); \\
A_{21}&=-E\left( \frac{\partial}{\partial{\beta'}}\psi_{\mu_{1g}}({X}_i, Z_i, Y_i) \right) =-E\left( \frac{\partial w_1({X}_i)}{\partial{\beta'}}Z_i(Y_i-\mu_{1g}) \right);~
A_{22}=-E\left( \frac{\partial}{\partial{\mu_{1g}}}\psi_{\mu_{1g}}({X}_i, Z_i, Y_i) \right) =E\left( Z_iw_1({X}_i) \right)\\
A_{31}&=-E\left( \frac{\partial}{\partial{\beta'}}\psi_{\mu_{0g}}({X}_i, Z_i, Y_i) \right)=-E\left( \frac{\partial w_0({X}_i)}{\partial{\beta'}}(1-Z_i)(Y_i-\mu_{0g}) \right);~
A_{33}=-E\left( \frac{\partial}{\partial{\mu_{0g}}}\psi_{\mu_{0g}}({X}_i, Z_i, Y_i) \right)=E\left( (1-Z_i)w_0({X}_i) \right). 
\end{align*}   
If we estimate the propensity scores via a logistic regression model $e(X_i)=[1+\exp(-V_{i}'\beta)]^{-1}$,   we have $\psi_{{\beta}}({X}_i, Z_i)=[Z_i-e(V_i;{\beta})]V_i$. The components of the matrix $A_N$ are given by 
\allowdisplaybreaks\begin{align*}
\widehat A_{11}&=N^{-1}\displaystyle\sum_{i=1}^{N} \widehat e_i(\mathrm{v})(1-\widehat e_i(\mathrm{v}))V_iV_i'; ~~
\widehat A_{21}=-N^{-1}\displaystyle\sum_{i=1}^{N} Z_i\left[\frac{\partial w_1(V_i)}{\partial{\beta}}\right]_{\beta=\widehat\beta} \left(\widehat Y_i-\widehat\mu_{1g}\right);~~
\widehat A_{22}=N^{-1}\displaystyle\sum_{i=1}^{N} Z_i\widehat w_1(\mathrm{v}_1);\\
\widehat A_{31}&=-N^{-1}\displaystyle\sum_{i=1}^{N} (1-Z_i)\left[\frac{\partial w_0(V_i)}{\partial{\beta}}\right]_{\beta=\widehat\beta} \left( \widehat Y_i-\widehat\mu_{0g}\right);  ~~
\widehat A_{33} =N^{-1}\displaystyle\sum_{i=1}^{N}(1- Z_i)\widehat w_0(\mathrm{v}_i).
\end{align*} 
The partial derivatives  $\displaystyle\left[\displaystyle {\partial w_z(V_i)}/{\partial{\beta}}\right]_{\beta=\widehat\beta}$  are given in Table \ref{weight_derivatives}.  The selection function $u(x)=\min\{e(x), 1-e(x)\}$ is not differentiable at $e(x) = 0.5.$ To calculate the corresponding partial derivatives $\displaystyle\left[\displaystyle {\partial w_z(V_i)}/{\partial{\beta}}\right]_{\beta=\widehat\beta},$ we replace the portion of $u(x)$ in the neighborhood of 0.5 with a cubic polynomial function that connects smoothly with $u(x)$ at boundaries of the neighborhood. \cite{li2013weighting}
An estimator of the variance of $\widehat{\tau}_g$  is thus $\widehat{Var}({\widehat{\tau}_g})=N^{-1}c'\widehat\Sigma({\widehat\theta})c$, where $c=(0,1,-1)'$.
\begin{table}[h]
	\begin{center}
		\begin{threeparttable}

	\caption{Partial derivatives of the weights and selection function by estimand}\label{weight_derivatives}
	\begin{tabular}{cccccccccccccccccccccccc}
		\toprule
		& ATE & OW & \multicolumn{3}{c}{MW} & EW \\\cmidrule(lr){1-7}
		&   &   & $e(\mathrm{v})<0.5-\delta $ & $0.5-\delta\leq e(\mathrm{v})\leq 0.5+\delta$ & $e(\mathrm{v})>0.5+\delta$ & \\\cmidrule(lr){4-6}
		$\displaystyle \frac{\partial w_1(V_i)}{\partial{\beta}}$ & $\eta_1(\mathrm{v}_i)V_i'$ & $-\eta_2(\mathrm{v}_i)V_i'$ &  0 & $\eta_3(\mathrm{v})\eta_2(\mathrm{v}_i)V_i'$ & $\eta_1(\mathrm{v}_i)V_i'$& $\ln(1-e(\mathrm{v}))\eta_1(\mathrm{v}_i)V_i'$ \\ \addlinespace
		$\displaystyle \frac{\partial w_0(V_i)}{\partial{\beta}}$ & $\eta_0(\mathrm{v}_i)V_i'$ & $\eta_2(\mathrm{v}_i)V_i'$ &   $\eta_0(\mathrm{v}_i)V_i'$ & $\eta_4(\mathrm{v}_i)\eta_2(\mathrm{v}_i)V_i'$ & 0 & $\ln(e(\mathrm{v}))\eta_0(\mathrm{v}_i)V_i'$   \\ \addlinespace
		$\displaystyle \frac{\partial g(V_i)}{\partial{\beta}}$ & 0 &   $[1-2e_i(\mathrm{v}_i)]\eta_2(\mathrm{v}_i)V_i'$ & $ \eta_2(\mathrm{v}_i)V_i'$ & $\eta_5(\mathrm{v}_i)\eta_2(\mathrm{v}_i)V_i'$ & $-\eta_2(\mathrm{v}_i)V_i'$  & $\log(\eta_0(\mathrm{v}_i))\eta_2(\mathrm{v}_i)V_i' $\\ 
		\bottomrule\addlinespace
		\end{tabular}
			\begin{tablenotes}
	\scriptsize
	\item $\eta_0(\mathrm{v}_i)= \displaystyle \frac{e_i(\mathrm{v})}{1-e_i(\mathrm{v})} $; ~$\eta_1(\mathrm{v}_i) = -\displaystyle \frac{1-e_i(\mathrm{v})}{e_i(\mathrm{v})} $;  ~$\eta_2(\mathrm{v}_i) =  b_2(\mathrm{v}) =\displaystyle {e_i(\mathrm{v})}{[1-e_i(\mathrm{v})]} $; ~$\eta_3(\mathrm{v})= a_{11} + 2a_{12}e_i(\mathrm{v}) + 3a_{13}e_i(\mathrm{v})^2$; ~$\eta_4(\mathrm{v})= a_{21} + 2a_{22}e_i(\mathrm{v}) + 3a_{23}e_i(\mathrm{v})^2$; ~$\eta_5(\mathrm{v})= a_{10} + 2a_{11}e_i(\mathrm{v}) + 3a_{12}e_i(\mathrm{v})^2+ 4a_{13}e_i(\mathrm{v})^3$, where the vectors $(a_{10}, a_{11}, a_{12}, a_{13})' = \mathbf{D}^{-1}\left(1, 0, \frac{1-2\delta}{1+2\delta},  \frac{-4}{(1+2\delta)^2}\right)' $ and $(a_{20}, a_{21}, a_{22}, a_{23})' = \mathbf{D}^{-1}\left(\frac{1-2\delta}{1+2\delta},  \frac{4}{(1+2\delta)^2}, 1, 0 \right)',$ with the matrix $\mathbf{D}=
	\begin{pmatrix}
		1 & 0.5-\delta &(0.5-\delta)^2&(0.5-\delta)^3\\	
		0 & 1& 2(0.5-\delta) &3(0.5-\delta)^2\\	
		1 & 0.5+\delta &(0.5-\delta)^2&(0.5-\delta)^3\\	
		0 & 1 &2(0.5+\delta) &3(0.5+\delta)^2
	\end{pmatrix}$ and $\delta = 0.002.$\cite{li2013weighting}  
		\end{tablenotes}
			\end{threeparttable}
				\end{center}
\end{table}

\subsection{Sandwich variance for the augmented estimator}\label{appendix4_sandwich_augmentation}  
For the augmented estimator $\widehat\tau_{g}^{aug}$, we also consider the estimating functions $\psi_{\alpha_z}(X)$ for the regression models $m_z(X)=m_z(X; \alpha_z),$ $z=0,1$. Let $c=(0,0,0,1,-1,1,-1)'$;  we  can derive the estimator $\widehat\tau_{g}^{aug}=c'\widehat\theta_{aug}=\widehat\tau_{1g}^{m}-\widehat\tau_{0g}^{m}+\widehat\mu_{1g}-\widehat\mu_{0g}$ through $\widehat\theta_{aug}=(\widehat\beta', \widehat\alpha_1',  \widehat\alpha_0', \widehat\tau_{1g}^{m}, \widehat\tau_{0g}^{m},\widehat\mu_{1g}, \widehat\mu_{0g})'$, the  solution to the estimating equation $	\displaystyle \displaystyle\sum_{i=1}^{N} \Psi_{\theta_{aug}}({X}_i, Z_i, Y_i)=0$ with respect to $\theta_{aug}=(\beta',\alpha_1',  \alpha_0', \tau_{1g}^{m}, \tau_{0g}^{m},  \mu_{1g}, \mu_{0g})',$ where 
\allowdisplaybreaks \begin{align*}
\Psi_{\theta_{aug}}({X}_i, Z_i, Y_i)&= 
\begin{bmatrix}
\Psi_{\beta}({X}_i, Z_i)\\
\Psi_{\alpha_{1}}({X}_i, Z_i, Y_i) \\
\Psi_{\alpha_{0}}({X}_i, Z_i, Y_i) \\
\Psi_{\tau_{1g}^m}({X}_i, Z_i, Y_i)\\
\Psi_{\tau_{0g}^m}({X}_i, Z_i, Y_i)\\
\Psi_{\mu_{1g}}({X}_i, Z_i, Y_i) \\
\Psi_{\mu_{0g}}({X}_i, Z_i, Y_i) \\
\end{bmatrix} = 
\begin{bmatrix}
\psi_{\beta}({X}_i, Z_i)\\
Z_i\psi_{\alpha_1}({X}_i, Y_i)\\
(1-Z_i)\psi_{\alpha_0}({X}_i, Y_i)\\
g(X_i)\{m_1(X_i)-\tau_{1g}^m\}\\
g(X_i)\{m_0(X_i)-\tau_{0g}^m\}\\
Z_iw_1({X}_i)(Y_i-m_1(X_i)-\mu_{1g})\\
(1-Z_i)w_0({X}_i)(Y_i-m_0(X_i)-\mu_{0g})\\
\end{bmatrix} 
\end{align*}
We also consider the matrices,
\begin{align*}
&A_N(\widehat{\theta}_{aug})=N^{-1}\displaystyle\sum_{i=1}^{N}\left[ - \frac{\partial}{\partial{\theta}}\Psi_{\theta_{aug}}({X}_i, Z_i, Y_i)\right]_{{\theta_{aug}}={\widehat{\theta}_{aug}}}\!\!;
~~ A({\theta_{aug}})=\lim_{N\rightarrow\infty}A_N(\widehat{\theta}_{aug})\\ 
&B_N({\widehat\theta_{aug}})= N^{-1}\displaystyle\sum_{i=1}^{N}\Psi_{\theta_{aug}}({X}_i, Z_i, Y_i)\Psi_{\theta_{aug}}({X}_i, Z_i, Y_i)'\big|_{{\theta_{aug}}=\widehat{\theta}_{aug}} 
~~ B(\theta)=\lim_{N\rightarrow\infty}B_N(\widehat\theta)=E\left[ \Psi_{\theta_{aug}}({X}_i, Z_i, Y_i)\Psi_{\theta_{aug}}({X}_i, Z_i, Y_i)'\right]
\end{align*} 
When we estimate the propensity score $e(X)$ and the regression models $m_z(X)$ using, respectively, a logistic regression model and a linear regression model, i.e., $e(V_i)=[1+\exp(-V_i'\beta)]^{-1}$ and the regression models $m_z(W_i)=W_i'\alpha_z$, $z=0,1$. Hence, $\psi_{{\beta}}({X}_i, Z_i)=[Z_i-e({V_i};{\beta})]V_i$ and $\psi_{{\alpha}_z}({X}_i, Z_i)=W_i(Y_i-W_i'{\alpha}_z)$. 
Assuming that the same covariates appear as predictors in the regression models $m_z(W)$, the non-zero components $\widehat A_{ij}$ of the matrix $ A_N$ are given by 
\allowdisplaybreaks\begin{align*}
\widehat A_{11}&=N^{-1}\displaystyle\sum_{i=1}^{N} \widehat e_i(\mathrm{v})(1-\widehat e_i(\mathrm{v}))V_iV_i'; ~
\widehat A_{22}=N^{-1}\displaystyle\sum_{i=1}^{N} Z_iW_iW_i';~~ 
\widehat A_{33}=N^{-1}\displaystyle\sum_{i=1}^{N}(1- Z_i)W_iW_i';\\ 
\widehat A_{41}&=-N^{-1}\displaystyle\sum_{i=1}^{N} \left[\frac{\partial g(V_i)}{\partial{\beta}}\right]_{\beta=\widehat\beta}\{\widehat m_1(W_i)- \widehat \tau_{1g}^m\}; ~~
\widehat A_{42}=\widehat A_{53}=-N^{-1}\displaystyle\sum_{i=1}^{N} \widehat g(V_i)W_i'; ~\widehat A_{44}= \widehat A_{55}=N^{-1}\displaystyle\sum_{i=1}^{N} \widehat g(V_i)\\
\widehat A_{51}&=-N^{-1}\displaystyle\sum_{i=1}^{N} \left[\frac{\partial g(V_i)}{\partial{\beta}}\right]_{\beta=\widehat\beta}\{\widehat m_0(W_i)- \widehat \tau_{0g}^m\}; \\
\widehat A_{61}&=-N^{-1}\displaystyle\sum_{i=1}^{N} Z_i\left[\frac{\partial w_1(V_i)}{\partial{\beta}}\right]_{\beta=\widehat\beta} \left( \widehat Y_i-\widehat m_1(W_i)-\widehat\mu_{1g}\right);~~
\widehat A_{62}=N^{-1}\displaystyle\sum_{i=1}^{N} Z_i\widehat w_1(\mathrm{v})W_i';~~
\widehat A_{66}=N^{-1}\displaystyle\sum_{i=1}^{N} Z_i\widehat w_1(\mathrm{v});\\
\widehat A_{71}&=- N^{-1}\displaystyle\sum_{i=1}^{N} (1-Z_i)\left[\frac{\partial w_0(V_i)}{\partial{\beta}}\right]_{\beta=\widehat\beta} ( \widehat Y_i-\widehat m_0(W_i)-\widehat\mu_{0g});~~
\widehat A_{73}=N^{-1}\displaystyle\sum_{i=1}^{N} (1-Z_i)\widehat w_0(\mathrm{v})W_i'; ~~ \widehat A_{77}=N^{-1}\displaystyle\sum_{i=1}^{N}(1- Z_i)\widehat w_0(\mathrm{v}).
\end{align*}   
An estimator of  $\Sigma({\theta}_{aug})$ is then $ \widehat\Sigma({\theta}_{aug})=A_N(\widehat\theta_{aug})^{-1}B(\widehat\theta_{aug})\{A(\widehat\theta_{aug})'\}^{-1},$ from which we can derive the variance of $\widehat\tau_{g}^{aug}=c'\theta_{aug}$ as $\widehat{Var}(\widehat{\tau}_g^{aug})=N^{-1}c'\widehat\Sigma({\widehat{\theta}_{aug}})c.$
\section{Additional simulations}\label{sec:additional_simulations}

\subsection{Data generating process (DGP)}
We generated covariates $\boldsymbol X=(X_1, \dots, X_6)$ and treatment assignment $Z$ following the data generating process of Li et al. \cite{li2018addressing}. We first generated $X_1,\dots, X_6$ from a multivariate normal distribution with zero mean, unit marginal variance, and pairwise covariance of $0.5$. Then, we dichotomized $X_4, \dots, X_6$ at $0$. Finally, we generated the treatment assignment indicator $Z\sim Ber(e(\boldsymbol X))$, where $e(\boldsymbol X)=[1+\exp\{-(\beta_{0}+\beta_{1}X_{1}+\dots +\beta_{6}X_{6})\}]^{-1}.$

\begin{figure}[h]
	\begin{center}
		\includegraphics[trim=35 12 15 35, clip, width=0.9\linewidth]{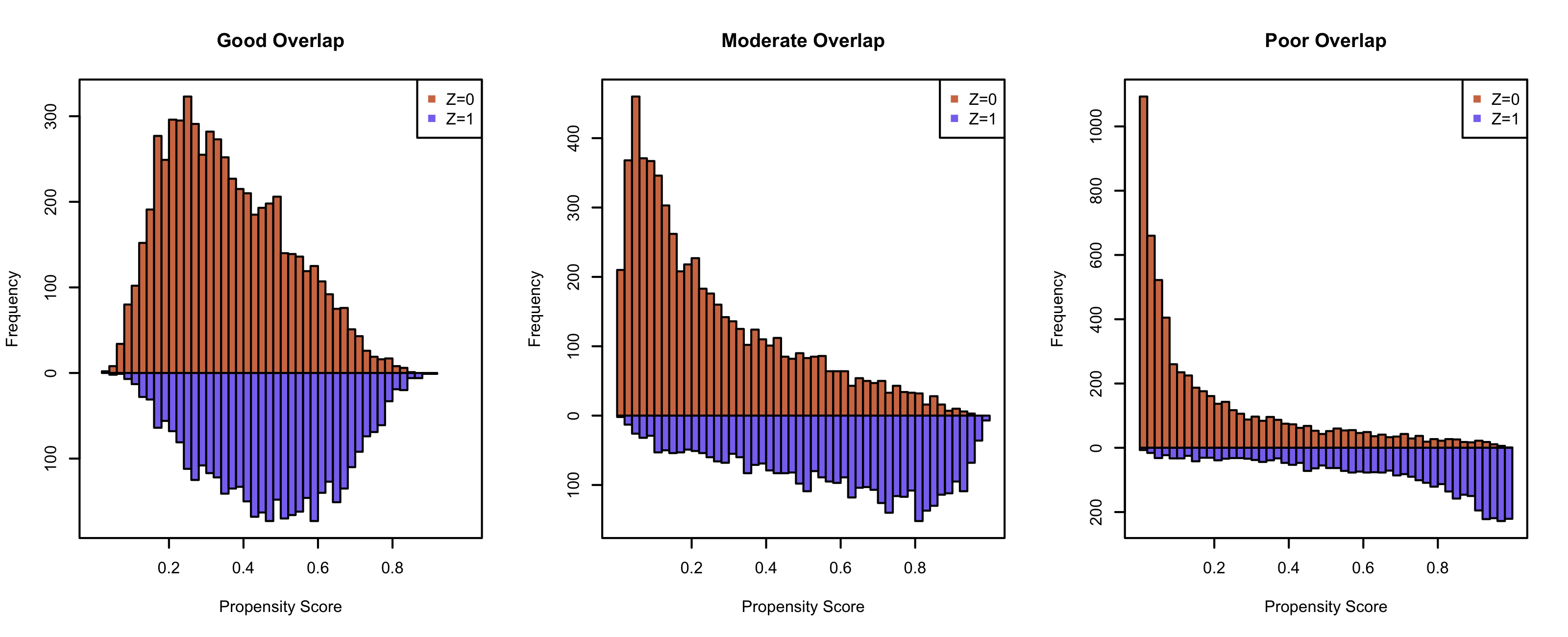}
	\end{center}
	\caption{Distributions of propensity scores under medium treatment prevalence, with good (left), moderate (middle), and poor (right) overlap. 
		\label{Li_PS_highprev}}
\end{figure}


\begin{figure}[!h]
	\begin{center}
		\includegraphics[trim=20 12 15 35, clip, width=0.9\linewidth]{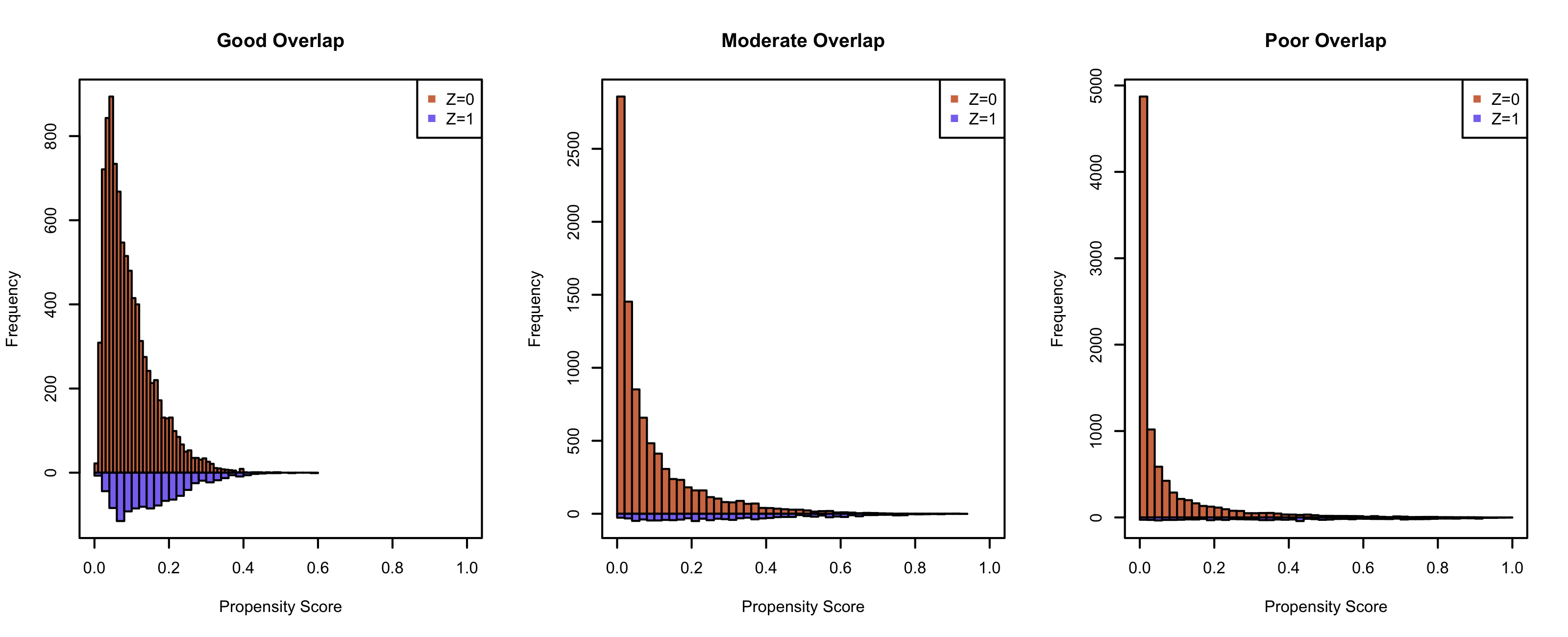}
	\end{center}
	\caption{Distributions of propensity scores under low prevalence of treatment, with good (left), moderate (middle), and poor (right) overlap. 
		\label{Li_PS_lowprev}}
\end{figure}

We considered $(\beta_{1}, \dots, $ $\beta_{6})=$ $\allowdisplaybreaks (0.15\gamma,$ $0.3\gamma,$ $0.3\gamma,$ $-0.2\gamma,$ $-0.25\gamma,$ $-0.25\gamma)$ and varied $\gamma$ from $1$ to $3$ for good, moderate, and poor overlap of PS distributions, as shown in Figure \ref{Li_PS_highprev}. The intercept $\beta_0$ was chosen to reflect the prevalence of treatment as shown in Table \ref{trt_prevalence}. We selected $\beta_0$ such that the prevalence is around $0.10$ for low prevalence ($\beta_0=$ -2.1, -2.2, and -2.8 for good, moderate, and poor overlap respectively under low treatment prevalence), and the prevalence is around 0.40 for medium prevalence ($\beta_0=$ -0.1, 0, and 0.2 for good, moderate, and poor overlap respectively under medium treatment prevalence). The corresponding distributions of propensity scores is shown in Figure \ref{Li_PS_lowprev}.  We generate $Y$  from the linear model $Y=0+\Delta Z-0.5X_1-0.5X_2-1.5X_3+0.8X_4+0.8X_5+X_6+\varepsilon$, where $\varepsilon\sim N(0,1.5)$ and $\Delta$ is the individual treatment effect. 
We choose, respectively, $\Delta=0.75$ (homogeneous treatment effect) and  $e(\boldsymbol X)^2+2e(\boldsymbol X)+1$  (heterogeneous treatment effects), to compare the performance of the aforementioned methods.  	
The results are summarized in Table \ref{Li_high_Ho}  (resp. Table \ref{Li_low_Ho}) for medium (resp. low) prevalence of treatment.

		\begin{table}[h]
	\caption{Average treatment prevalence for different levels of PS overlap.}\label{trt_prevalence}
	\begin{center}
		\begin{tabular}{cccccccccccccc}
			\toprule
			& \multicolumn{3}{c}{Overlap}  \\ \cmidrule(lr){2-4}
			Prevalence   & Good & Moderate & Poor   \\
			\cmidrule(lr){1-4}			
			Medium & 40.08\% & 39.81\% & 39.63\%  \\
			Low  & 10.26\% & 10.79\% &	10.24\%  \\
			\bottomrule
		\end{tabular}
	\end{center}
\end{table}	
		\begin{table}[!htp]
	\begin{threeparttable}
		\centering
		\caption{Treatment effect estimation (medium  prevalence of treatment).\label{Li_high_Ho} { }}
		\begin{tabular}{rrrrrrrrrrrrrrrrrrrrrrrrrrrrrrrrrrrrrrrrrrrrrrrrrrrrrrrrrrrrrrrrrrrrrrrrrrrrrrrrrrrrrrrrrrrrrrrrrrrr}
			\toprule
			&  & \multicolumn{6}{c}{Homogeneous treatment effect} &  &  \multicolumn{6}{c}{Heterogeneous treatment effect}\\\cmidrule{3-8} \cmidrule{10-15} 
			Weight & Overlap & True & ARB & RMSE & SD & SE & CP & & True & ARB & RMSE & SD & SE & CP \\ 
			\midrule
			IPW & Good & 0.75 & 1.18 & 9.08 & 9.05 & 8.96 & 0.94 &  & 2.00 & 0.12 & 9.05 & 9.05 & 8.79 & 0.94 \\
			OW & Good & 0.75 & 0.46 & 7.35 & 7.35 & 7.31 & 0.94  &  & 2.06 & 0.07  & 7.55 & 7.55 & 7.42 & 0.95 \\ 
			MW & Good & 0.75 & 0.32 & 7.60 & 7.60 & 7.53 & 0.95  &   & 2.09 & 0.08 & 7.67 & 7.67 & 7.65 & 0.95 \\  
			EW & Good & 0.75 & 0.59 & 7.40 & 7.39 & 7.37 & 0.94  &   & 2.04 & 0.05 & 7.61 & 7.61 & 7.46 & 0.95 \\  
			BW(11) & Good & 0.75 & 0.06 & 10.73 & 10.74 & 10.74 & 0.95 &   & 2.20 & 0.05 & 10.40 & 10.40 & 10.85 & 0.96 \\ 
			BW(81) & Good & 0.75 & 0.29 & 15.72 & 15.73 & 18.03 & 0.98 &   & 2.24 & 0.02 & 15.81 & 15.81 & 18.25 & 0.98 \\ \addlinespace
			
			IPW & Mod & 0.75 & 1.54 & 24.19 & 24.17 & 21.16 & 0.90 &  & 2.00 & 0.57 & 21.84 & 21.83 & 19.29 & 0.90 \\
			OW & Mod & 0.75 & 1.05 & 8.36 & 8.33 & 8.40 & 0.94  & & 2.13 & 0.05 & 8.92 & 8.93 & 8.73 & 0.94 \\ 
			MW & Mod & 0.75 & 1.12 & 8.61 & 8.57 & 8.66 & 0.95 & & 2.15 & 0.03 & 9.09 & 9.10 & 8.98 & 0.94 \\ 
			EW & Mod & 0.75 & 0.91 & 8.57 & 8.54 & 8.67 & 0.95 & & 2.11 & 0.05 & 9.18 & 9.18 & 8.94 & 0.94 \\ 
			BW(11) & Mod & 0.75 & 1.43 & 13.08 & 13.04 & 13.28 & 0.96 & & 2.24 & 0.08 & 13.14 & 13.15 & 13.57 & 0.96 \\
			BW(81) & Mod & 0.75 & 0.15 & 20.87 & 20.88 & 24.83 & 0.98 & & 2.25 & 0.25 & 20.82 & 20.82 & 25.15 & 0.98 \\ \addlinespace
			
			IPW & Poor & 0.75 & 6.49 & 55.46 & 55.27 & 41.21 & 0.79 & & 2.05 & 2.34 & 49.06 & 48.85 & 36.51 & 0.81 \\  
			OW & Poor & 0.75 & 0.11 & 9.44 & 9.44 & 9.51 & 0.95 & & 2.20 & 0.18 & 10.20 & 10.20 & 10.05 & 0.95 \\
			MW & Poor & 0.75 & 0.36 & 9.74 & 9.75 & 9.79 & 0.95 & & 2.21 & 0.22 & 10.51 & 10.50 & 10.31 & 0.95 \\ 
			EW & Poor & 0.75 & 0.16 & 10.01 & 10.01 & 9.97 & 0.95 & & 2.19 & 0.16 & 10.47 & 10.47 & 10.44 & 0.95 \\  
			BW(11) & Poor & 0.75 & 1.17 & 15.25 & 15.23 & 15.58 & 0.95 & & 2.25 & 0.40 & 16.03 & 16.02 & 15.99 & 0.94 \\ 
			BW(81) & Poor & 0.75 & 1.16 & 24.79 & 24.79 & 30.92 & 0.98 & & 2.25 & 0.68 & 25.76 & 25.73 & 31.05 & 0.98 \\ 
			
			\bottomrule
		\end{tabular}
		
		\begin{tablenotes}
			\footnotesize
			\item  These  results are based on 1000 repetitions of the simulated data. Mod: Moderate; BW($\nu$): BW with $\nu=11$ and $81$; ARB: absolute relative bias$\times 100$;
			\item   RMSE: root mean-squared error$\times 100$; SD:empirical standard deviation$\times 100$; SE: average estimated standard error$\times 100$; CP: 95\%  coverage probability. 
		\end{tablenotes}
	\end{threeparttable}
\end{table}

\begin{table}[]
	\centering
	\begin{threeparttable}
	\caption{Treatment effect estimation (low prevalence of treatment).\label{Li_low_Ho} { }}
	\begin{tabular}{rrrrrrrrrrrrrrrrrrrrrrrrrrrrrrrrrrrrrrrrrrrrrrrrrrrrrrrrrrrrrrrrrrrrrrrrrrrrrrrrrrrrrrrrrrrrrrrrrrrr}
		\toprule
		&  & \multicolumn{6}{c}{Homogeneous treatment effect} &  &  \multicolumn{6}{c}{Heterogeneous treatment effect}\\\cmidrule{3-8} \cmidrule{10-15} 
		Weight & Overlap & True & ARB & RMSE & SD & SE & CP & & True & ARB & RMSE & SD & SE & CP \\ 
		\midrule
		IPW & Good & 0.75 & 1.18 & 25.59 & 25.59 & 24.55 & 0.93 & & 1.21 & 0.91 & 25.54 & 25.53 & 24.31 & 0.92 \\  
		OW & Good & 0.75 & 0.59 & 11.28 & 11.27 & 11.47 & 0.96 & & 1.31 & 0.06 & 11.83 & 11.83 & 11.57 & 0.94 \\ 
		MW & Good & 0.75 & 0.59 & 11.72 & 11.72 & 11.99 & 0.96  & & 1.34 & 0.03 & 12.34 & 12.34 & 12.00 & 0.95 \\  
		EW & Good & 0.75 & 0.52 & 11.87 & 11.87 & 12.02 & 0.96  & & 1.29 & 0.08 & 12.35 & 12.36 & 12.14 & 0.95 \\  
		BW(11) & Good & 0.75 & 1.35 & 50.17 & 50.18 & 52.18 & 0.96  & & 1.78 & 0.44 & 48.93 & 48.95 & 49.40 & 0.96 \\ 
		BW(81) & Good & 0.75 & 6.74 & 167.78 & 167.79 & 151.60 & 0.80  & & 2.16 & 2.21 & 162.19 & 162.20 & 142.24 & 0.80  \\ \addlinespace
		
		IPW & Mod & 0.75 & 11.33 & 70.11 & 69.63 & 51.24 & 0.78 & & 1.26 & 7.40 & 64.79 & 64.15 & 48.29 & 0.76 \\ 
		OW & Mod & 0.75 & 0.86 & 11.98 & 11.97 & 11.87 & 0.96  & & 1.57 & 0.25 & 12.12 & 12.12 & 12.12 & 0.94 \\  
		MW & Mod & 0.75 & 0.68 & 12.51 & 12.50 & 12.50 & 0.95  & & 1.64 & 0.25 & 12.87 & 12.87 & 12.61 & 0.94 \\  
		EW & Mod & 0.75 & 1.06 & 13.38 & 13.36 & 12.92 & 0.95 & & 1.51 & 0.36 & 12.92 & 12.92 & 13.09 & 0.95 \\  
		BW(11) & Mod & 0.75 & 0.14 & 24.48 & 24.49 & 24.75 & 0.95  & & 2.10 & 0.16 & 25.22 & 25.23 & 25.04 & 0.94 \\ 
		BW(81) & Mod & 0.75 & 0.11 & 41.02 & 41.04 & 51.10 & 0.97  & & 2.23 & 0.02 & 43.93 & 43.95 & 52.03 & 0.97 \\  \addlinespace
		
		IPW & Poor & 0.75 & 69.11 & 124.92 & 113.72 & 63.01 & 0.51 & & 1.25 & 40.01 & 118.23 & 107.22 & 59.63 & 0.51 \\  
		OW & Poor & 0.75 & 0.50 & 13.54 & 13.54 & 13.58 & 0.94 &  & 1.71 & 0.46 & 13.88 & 13.86 & 13.98 & 0.96 \\ 
		MW & Poor & 0.75 & 0.51 & 13.99 & 13.99 & 14.14 & 0.94 & & 1.79 & 0.56 & 14.19 & 14.17 & 14.46 & 0.95 \\ 
		EW & Poor & 0.75 & 0.17 & 14.75 & 14.76 & 14.72 & 0.94 &  & 1.64 & 0.52 & 15.14 & 15.12 & 15.04 & 0.95 \\ 
		BW(11) & Poor & 0.75 & 0.24 & 24.07 & 24.09 & 25.45 & 0.96  &  & 2.16 & 1.06 & 24.89 & 24.80 & 26.04 & 0.96 \\ 
		BW(81) & Poor & 0.75 & 2.08 & 40.43 & 40.42 & 53.75 & 0.99  & & 2.24 & 1.52 & 41.84 & 41.72 & 54.04 & 0.98 \\  
		
		\bottomrule
	\end{tabular}
	
	\begin{tablenotes}
		\footnotesize
		\item  These  results are based on 1000 repetitions of the simulated data. Mod: Moderate; BW($\nu$): BW with $\nu=11$ and $81$; ARB: absolute relative bias$\times 100$;
		\item   RMSE: root mean-squared error$\times 100$; SD:empirical standard deviation$\times 100$; SE: average estimated standard error$\times 100$; CP: 95\%  coverage probability. 
	\end{tablenotes}
\end{threeparttable}
\end{table}

\newpage
\bibliography{ow_augmentation_sim}
\end{document}